\documentclass[twocolumn]{aastex63}

\usepackage{amsmath}
\usepackage{amssymb}
\usepackage{graphics}
\usepackage{enumerate}
\usepackage{soul}

\usepackage{savesym}
\savesymbol{tablenum}
\usepackage{siunitx}
\restoresymbol{SIX}{tablenum}
\usepackage{natbib}

\newcommand{\del}{\partial}

\DeclareRobustCommand{\delsf}{\bgroup\markoverwith{\textcolor{red}{\rule[.5ex]{2pt}{0.4pt}}}\ULon}

\newcommand{\isotope}[2]{{\ensuremath{{}^{#1}\mathrm{#2}}}}
\newcommand{\Yisotope}[2]{{\ensuremath{Y_{{}^{#1}\mathrm{#2}}}}}

\shorttitle{Collapses and explosions of rotating VMSs with $\alpha$-chain network}

\shortauthors{Fujibayashi et al.}

\begin{document}

\title{Explosions from Rotating Very Massive Star Collapses to Black Holes: Effects of Nuclear Burning}

\author[0000-0001-6467-4969]{Sho Fujibayashi}
\affiliation{Frontier Research Institute for Interdisciplinary Sciences, Tohoku University, Aramaki aza Aoba 6-3, Aoba-ku, Sendai 980-8578, Japan}
\affiliation{Astronomical Institute, Graduate School of Science, Tohoku University, Sendai 980-8578, Japan}
\affiliation{Max-Planck-Institut f\"ur Gravitationsphysik (Albert-Einstein-Institut), Am M\"uhlenberg 1, D-14476 Potsdam-Golm, Germany}

\author[0000-0002-1307-1401]{Alan Tsz-Lok Lam}
\affiliation{Institute for Gravitation and the Cosmos, The Pennsylvania State University, University Park, PA 16802, USA}
\affiliation{Department of Physics, The Pennsylvania State University, University Park, PA 16802, USA}
\affiliation{Max-Planck-Institut f\"ur Gravitationsphysik (Albert-Einstein-Institut), Am M\"uhlenberg 1, D-14476 Potsdam-Golm, Germany}

\author[0000-0002-2648-3835]{Yuichiro Sekiguchi}
\affiliation{Center for Gravitational Physics and Quantum Information, Yukawa Institute for Theoretical Physics, Kyoto University, Kyoto, 606-8502, Japan}
\affiliation{Department of Physics, Toho University, Funabashi, Chiba 274-8510, Japan}

\author[0000-0002-4979-5671]{Masaru Shibata}
\affiliation{Max-Planck-Institut f\"ur Gravitationsphysik (Albert-Einstein-Institut), Am M\"uhlenberg 1, D-14476 Potsdam-Golm, Germany}
\affiliation{Center for Gravitational Physics and Quantum Information, Yukawa Institute for Theoretical Physics, Kyoto University, Kyoto, 606-8502, Japan}

\date{\today}

\correspondingauthor{Sho Fujibayashi}
\email{sho.fujibayashi@astr.tohoku.ac.jp}

\begin{abstract}
We investigate the collapse of rotating very massive and supermassive stellar cores using numerical relativity simulations including an alpha-chain nuclear reaction network and neutrino cooling. Our main survey focuses on newly constructed models with initial core masses of $2 \times 10^3$--$5\times 10^4M_\odot$. The collapse is triggered either by pair instability in lower-mass cores or by general-relativistic instability in higher-mass cores. We find that higher-mass cores undergo a nearly homologous collapse, whereas lower-mass cores show a more runaway-like collapse because neutrino cooling becomes more efficient at their higher densities and temperatures. Consequently, the black hole formed in lower-mass models initially contains a smaller fraction of the core mass, and disk formation occurs while a larger amount of matter remains outside the black hole. The lower compactness of pair-unstable cores also allows larger dimensionless angular momentum, favoring the formation of rapidly rotating black holes and massive disks. The disk bounce drives mass ejection with ejecta masses of order $10$--$10^3M_\odot$ and kinetic energies of order $10^{53}$--$10^{55}\,\mathrm{erg}$. Significant \isotope{56}{Ni} production in the disk-bounce ejecta occurs only in the lowest-mass models. For selected models, we further follow the viscous evolution of the disk and find that viscosity enhances the ejecta mass and kinetic energy. In models with $\lesssim10^4M_\odot$, the viscosity-driven ejecta can originate from disk matter that has reached nuclear statistical equilibrium and can therefore become rich in \isotope{56}{Ni}. These results suggest that rotating very massive star collapses can produce massive, energetic ejecta and, for sufficiently low core masses, substantial iron-group elements.

\end{abstract} 



\section{Introduction}

Stars with initial masses larger than $\sim 10M_\odot$ undergo gravitational collapse after forming an iron core. Further massive stars with initial masses $\gtrsim 130M_\odot$ become gravitationally unstable due to the production of electron-positron pairs (e.g.,~\citealt{2001ApJ...550..372F, 2012A&A...542A.113Y, Takahashi2018}). Such stars are referred to as \textit{very massive stars}. The most massive class of stars with masses $\gtrsim 10^4M_\odot$ undergoes collapse due to the general relativistic instability~\citep{1964ApJ...140..417C, 1971reas.book.....Z}. They are called \textit{supermassive stars} (see, e.g., \citealt{Woosley2002a} and \citealt{Langer2012sep} for reviews).

The very massive and supermassive stars are likely to be born as the high-mass end of the stars formed in low-metallicity environments. Theoretical works have suggested that the typical mass of the first generation (Population~III; PopIII) stars is 10--100$M_\odot$ (e.g., \citealt{Hirano2014feb,Hirano2015mar,Susa2014sep}). However, further works reported a possibility for the formation of very massive and supermassive stars with masses $\gtrsim10^3M_\odot$ (\citealt{Li2023jun,Toyouchi2023jan}). Such stars are likely to leave behind black holes with masses similar to their progenitor star's cores, which can then be a seed black hole that could grow to supermassive black holes later on (see, e.g., \citealt{Inayoshi2020aug}). 

Explosions in the collapses of very massive and supermassive stars have been studied in several previous works. One of the explosion scenarios is the one driven by energy generation due to nuclear burning. This possibility was
investigated in early studies by \citet{Fricke1973aug} and was later
revisited by \citet{Fuller1986aug}, focusing on explosions of metal-rich stars powered by hydrogen burning through the CNO cycle and related processes
(see also \citealt{Montero2012apr}). Explosions powered by helium burning
were reported by \citet{Chen2014aug} for a $\sim\SI{5e4}{}M_\odot$
supermassive star and later by \citet{Nagele2022dec} for somewhat lighter
stars with masses of $\sim\SI{3e4}{}M_\odot$.

The effects of the potential rotation of the supermassive stars are investigated by \cite{Uchida2017oct} and \cite{Fujibayashi2025mar} (see also \citealt{Shibata:2002br, Liu2007oct}). In these studies, they showed that a massive disk is formed around the natal black hole for the collapse of rapidly rotating supermassive stars. The disk then bounces violently at its formation, leading to the explosion through the propagation of a strong shock.

The observational properties of transients associated with the explosions of supermassive stars have also been discussed in previous studies. \citet{Whalen2013ApJ.778.17}, \citet{Moriya2021may}, and \citet{Nagele2023mar} modeled their emission as arising from the expansion and cooling of massive hydrogen-rich ejecta, in analogy with
scaled-up Type IIP supernovae. On the other hand, \citet{Jockel2026jan} considered a more realistic formation environment, in which the ejecta interact with the pristine atomic gas that accretes onto and feeds the supermassive star. In this case, the conversion of kinetic energy into internal energy at the optically thick shock powers the emission, making the transient analogous to a scaled-up Type IIn or circumstellar-matter-interacting supernovae. The transients resulting from the explosion of supermassive stars are characterized by their long timescale and large luminosity. They are promising targets for deep near-infrared observations and surveys with facilities such as \textit{James Webb Space Telescope}, \textit{Euclid}, and \textit{Nancy Grace Roman Space Telescope}.

In this paper, we model explosions from rotating very massive stellar cores as a result of the collapse triggered by pair instability or general relativistic instability. Our main survey covers initial core masses of $2 \times 10^3$--$5 \times 10^4M_\odot$, corresponding to very massive stars and relatively low-mass supermassive stars. For such objects, the density and temperature of the collapsing matter are higher than those for high-mass supermassive stars studied in \citet{Uchida2017oct, Fujibayashi2025mar}. As a result, energy generation by nuclear burning and energy loss by neutrino emission are more efficient. Therefore, these microphysical processes have to be incorporated to model the collapse and explosion of these stars.

This paper is organized as follows: In \S~\ref{sec:method}, we summarize the numerical method employed to simulate the core collapse of rotating very massive and supermassive stars. In \S~\ref{sec:results}, the results of our numerical simulations are presented. We focus especially on the ejecta properties. In \S~\ref {sec:discussion}, we discuss the observational properties and detection prospects, the longer-term evolution of the formed disk, and the possibility of heavy-element synthesis. \S~\ref{sec:summary} is devoted to a summary. Throughout this paper, $c$, $G$, and $k$ are the speed of light, gravitational constant, and Boltzmann constant, respectively.

\section{Method} \label{sec:method}
The solvers of Einstein's equation and the radiation hydrodynamics equation are the same as those adopted in \cite{Fujibayashi2025mar} and described in detail in \cite{Sekiguchi2010a,fujibayashi2017a,Fujibayashi2020c}. In this section, we describe the additional features to model the core collapse of very massive stars, which necessitate a more detailed microphysical treatment.

\subsection{Nuclear burning}
In addition to the usual hydrodynamical variables, mass fractions of several nuclear species are evolved in this work. Here, we summarize the basic equations that govern their evolution. In the following, $\rho$, $n_\mathrm{b}$, $n_I$, $u^\mu$, and $T$ denote the rest-mass density, baryon number density, number density of $I$th species, four velocity, and temperature, respectively.

We construct an alpha-chain approximate nuclear network consisting of 13 nuclear species from \isotope{4}{He} to \isotope{56}{Ni} following \cite{Timmes2000jul} (see also Appendix~\ref{app:alpha-network}). The time evolution of the abundance of $I$th nuclear species, $Y_I := n_I/n_\mathrm{b}$, is governed by
\begin{align}
\frac{dY_I}{dt} &= \frac{dY_I}{dt}\bigg|_\mathrm{reac}\notag\\
&=\sum_J \lambda_{I,J} Y_J - \sum_J \lambda_{J,I} Y_I\notag\\
&+ \rho \sum_{JK} \lambda_{I,JK} Y_J Y_K - \rho \sum_{JK} \lambda_{K,JI} Y_J Y_I \notag\\
& + \rho^2 \sum_{JKL} \lambda_{I,JKL} Y_J Y_K Y_L - \rho^2 \sum_{JKL} \lambda_{K,LJI} Y_L Y_J Y_I, \label{eq:nuc}
\end{align}
where $\lambda_{I,J}$, $\lambda_{I,JK}$, and $\lambda_{I,JKL}$ are the reaction rates of one, two, and three-body reactions ($J\rightarrow I$, $J+K\rightarrow I$, and $J+K+L\rightarrow I$). These reaction rates depend on the temperature. We note that $\lambda$'s include the factors introduced to avoid double-counting the identical incident and reactant nuclei. To evaluate the reaction rates, we use fitting formulae for the reaction rates provided by JINA REACLIB v2.2 \citep{Cyburt2010a}.

The nuclear burning is solved in an operator-splitting way \citep{Uchida2017oct}, in which we first update the abundance of each species explicitly, only with advection terms, as
\begin{align}
\del_\mu(\sqrt{-g} \rho A_I Y_I u^\mu) = 0, \label{eq:adv}
\end{align}
where $A_I$ is the mass number of $I$th species, and $g$ denotes the determinant of the spacetime metric $g_{\mu\nu}$. For the Greek subscripts appearing above and below, it is implicitly assumed that the sum is taken over the space-time components. Given the abundance $Y_I^*$ updated by Eq.~\eqref{eq:adv}, we then update it with the source term due to nuclear burning as
\begin{align}
\frac{Y_I^{(n+1)}-Y_I^*}{\Delta t} = \frac{dY_I}{dt}\bigg|_\mathrm{reac}^{(n+1)},
\end{align}
where $dY_I/dt|_\mathrm{reac}^{(n+1)}$ is the right-hand side of Eq.~\eqref{eq:nuc} evaluated with the abundance of $(n+1)$th timestep, $Y_I^{(n+1)}$, and $\Delta t$ denotes the time interval in a numerical simulation. We note that the density and temperature used in evaluating Eq.~\eqref{eq:nuc} are those in $n$th timestep.  We solve Eq.~\eqref{eq:nuc} implicitly to obtain $Y_I^{(n+1)}$. 

For $T>\SI{1e10}{K}$ (10\,GK), the evolution equation is not solved because the nuclear statistical equilibrium (NSE) is achieved for such high temperature in the density range considered in this paper. Thus, we update the nuclear abundance based on the local temperature and density assuming the NSE. 

\subsection{Neutrino cooling}
We adopt the fitting formula of \cite{Itoh1996feb} for neutrino emission due to the electron-positron pair annihilation, Bremsstrahlung, and plasmon decay processes.
We do not include charged-current weak-interaction channels in the dynamical simulations. The additional channels, such as electron and positron captures by nuclei and free nucleons, which we do not take into account in this paper, can enhance neutrino cooling and, in electron-degenerate matter, can also change the electron fraction. Their possible effects are discussed in \S~\ref{subsec:cc}.

We also do not consider the trapping of neutrinos emitted in the central region of collapsing cores, although the emitted neutrinos can be trapped, reducing the cooling rate in a high-density region~\citep{1975PThPh..54.1325S}. Neutrino trapping can modify the collapse dynamics before the black hole formation, and hence the initial black hole properties. Its direct effect after the black hole formation is, however, expected to be limited, because the optically thick central matter is rapidly swallowed by the black hole. According to the spherically symmetric models with masses of order $10^4M_\odot$ in \cite{Nagele2021nov}, the density at the neutrino sphere in the collapsing supermassive stars is $10^7$--$\SI{e8}{g.cm^{-3}}$. This is much lower than that in the usual stellar core collapse. After the neutrino sphere formation, the cooling efficiency of the central region of the collapsing core drops. Hence, the collapse proceeds rather adiabatically after the neutrino trapping sets in. Especially for the low-mass very-massive stars, the central density before the black-hole formation can become high enough for the neutrino optical depth to exceed unity, and thus, the emitted neutrinos are trapped and decrease the cooling efficiency. To check how the collapse dynamics are affected by the trapping, we perform simulations suppressing the neutrino cooling by multiplying the cooling function by the following factor, $\exp(-\rho/\rho_\mathrm{crit})$, where $\rho_\mathrm{crit}$ is the critical density above which the cooling is artificially suppressed. In this paper, $\rho_\mathrm{crit}$ is chosen to be $\SI{e10}{}$ and \SI{e8}{g.cm^{-3}} (see \S~\ref{subsec:nuc-nu}).

\subsection{Grid setup and regridding procedure}
We adopt the same grid structure and regridding strategy as in \citet{Fujibayashi2025mar}, to which readers may refer for further details. For the standard-resolution runs, the finest grid spacing is $dx_0 = 0.025 GM_0/c^2$, where $M_0$ denotes the total mass of the stellar core. This resolution corresponds to the high-resolution setup in \citet{Fujibayashi2025mar}. The lower- and higher-resolution runs, labeled by ``L'' and ``H'', respectively, have $dx_0 = 0.040 GM_0/c^2$ and $0.015 GM_0/c^2$.

We also follow the same regridding procedure as in \citet{Fujibayashi2025mar}. During the simulation, when the collapsing core becomes sufficiently compact, we stop the run, map the hydrodynamical quantities onto a finer grid structure, and restart the simulation from the remapped data.

\subsection{Initial condition}
For the initial conditions of the present work, we prepare rotating equilibrium configurations that represent the cores of progenitor stars at the onset of collapse triggered by either the pair instability or the general-relativistic instability. They are constructed in the same manner as in \cite{Shibata2025jan}. In this method, we approximate the equation of state as a polytropic one $P= K\rho^\Gamma$ where $P$ denotes the pressure. The polytropic index $\Gamma$ and the constant $K$ are determined by the entropy per baryon and the composition, which are assumed to be constant in the entire stellar core. We obtain rotating solutions of the stellar cores by imposing the following conditions: (1) the core is either marginally stable to the pair instability or to the general-relativistic instability, (2) the mass and the ratio of rotational kinetic energy to gravitational binding energy, $T_\mathrm{rot}/|W|$, have the desired values, and (3) the energy generation rate by nuclear burning is equal to the Eddington luminosity of the core (see \cite{Shibata2025jan} for details on the computation of the energy generation rate).

\subsection{Models}
Table~\ref{tab:key-result} summarizes the models employed in this study. The main survey consists of rotating stellar cores with initial core masses $M_0=2\times10^3M_\odot$--$5\times10^4M_\odot$. The initial rotation is characterized by the ratio of rotational kinetic energy to gravitational binding energy, $T_\mathrm{rot}/|W|$. 
These main models are named according to their initial core mass and rotation parameter. For example, the model 2E3-09 has $M_0=2\times10^3M_\odot$ and $T_\mathrm{rot}/|W|\approx 0.009$, while the model 5E4-09 has $M_0=5\times10^4M_\odot$ and $T_\mathrm{rot}/|W|\approx 0.009$. Unless otherwise stated, these models are initialized with pure \isotope{16}{O} for baryons.

For selected core masses, we also perform simulations with slower rotation models, adopting $T_\mathrm{rot}/|W| \approx 0.004$. These models are labeled in the same way, with the suffix ``04''; for example, 2E3-04 denotes a model with $M_0=2\times10^3M_\odot$ and $T_\mathrm{rot}/|W|=0.004$. We also consider a test model 5E4-09-He, which has the same mass and rotation as 5E4-09 but is initialized with pure \isotope{4}{He}. This model is used to examine the dependence of the collapse and ejecta composition on the initial abundance of alpha particles.

In addition to the main survey models, we include selected higher-mass supermassive-star models to connect the present results to our previous study. The H2 and H4 models are taken from \cite{Fujibayashi2025mar} and are shown for comparison. They correspond to hydrogen-burning supermassive-star cores with primordial composition. The He2 and He4 models are also based on the initial conditions of \cite{Fujibayashi2025mar}, but in the present work we recompute their evolution with the alpha-chain reaction network and thermal neutrino cooling, in order to evaluate their ejecta composition in the same framework as the lower-mass models.

For several models, we follow the viscous evolution after the formation of a massive disk around the black hole. For these simulations, viscosity is switched on after the disk has settled into a quasi-steady state and the disk-bounce ejecta mass has saturated. The viscous models are labeled by the suffixes ``v03'' and ``v10'', corresponding to $\alpha_\mathrm{vis}=0.03$ and 0.10, respectively. Here $\alpha_\mathrm{vis}$ denotes the so-called alpha parameter for the shear viscous coefficient (see \cite{Fujibayashi2025mar} for details). For example, 2E3-09-v03 denotes the viscous evolution of the 2E3-09 model with $\alpha_\mathrm{vis}=0.03$.

\section{Results} \label{sec:results}

\subsection{Collapse profile differences in different core masses}

\begin{figure}
    \centering
    \includegraphics[width=0.5\textwidth]{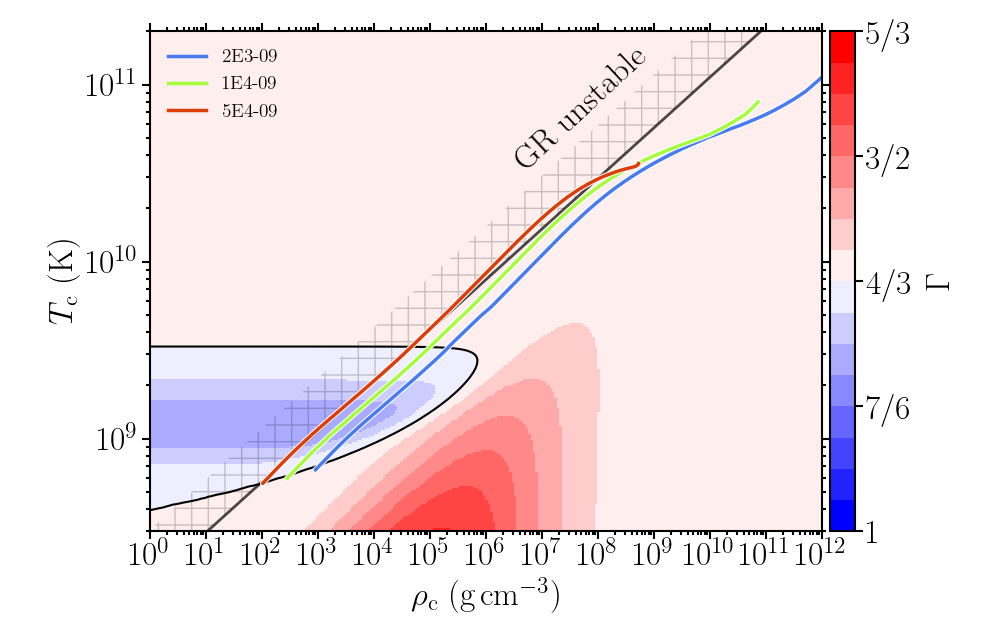}
    \caption{Evolution tracks of the central density and temperature for the 5E4-09, 1E4-09, and 2E3-09 models. The color indicates the adiabatic index $\Gamma$. The top-left region above the diagonal line shows where the corresponding non-rotating equilibrium configuration is unstable to the general-relativistic instability.}
    \label{fig:rho-T-cen}
\end{figure}

\begin{figure*}
    \centering
    (a)\includegraphics[width=0.47\textwidth]{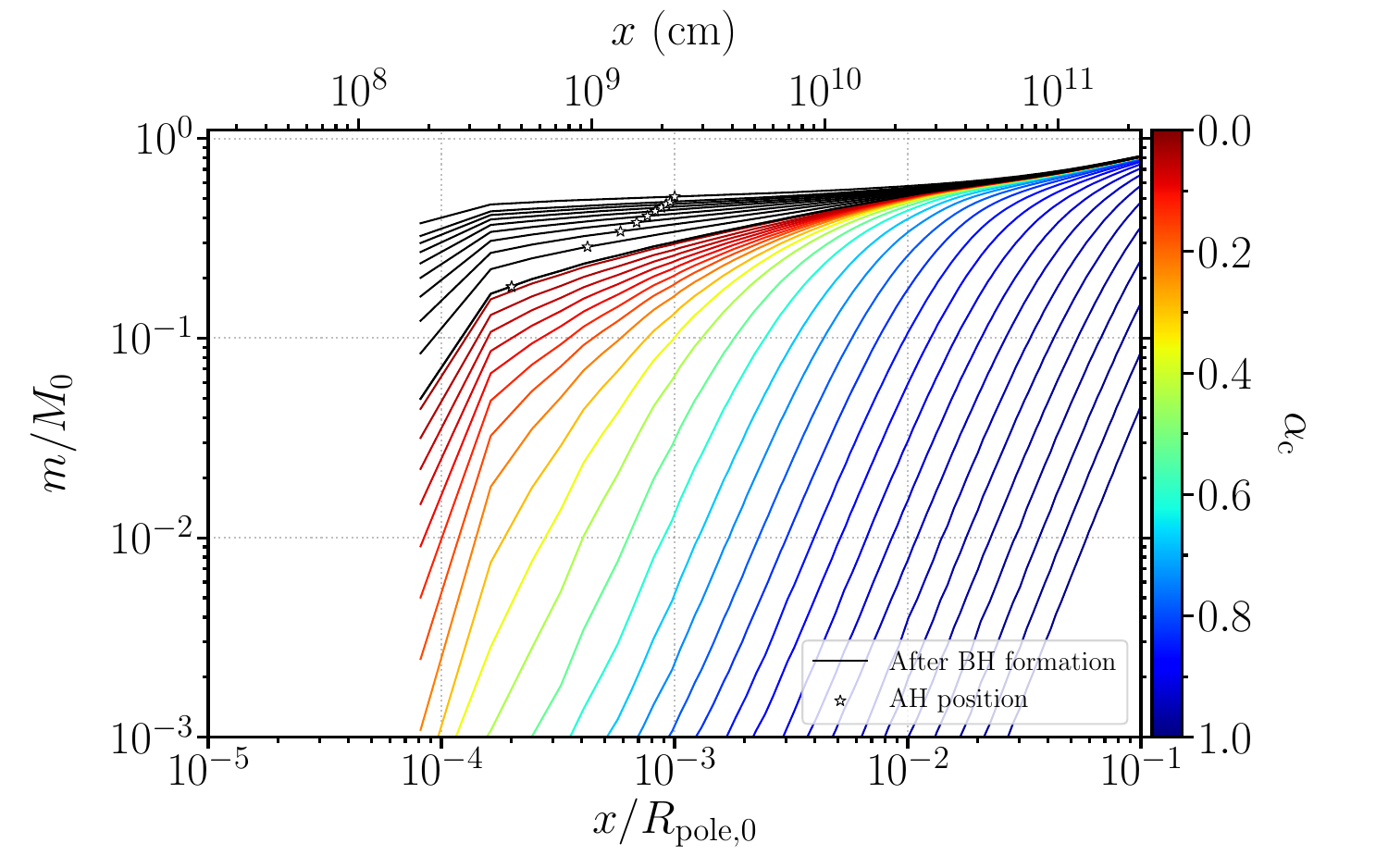}
    (b)\includegraphics[width=0.47\textwidth]{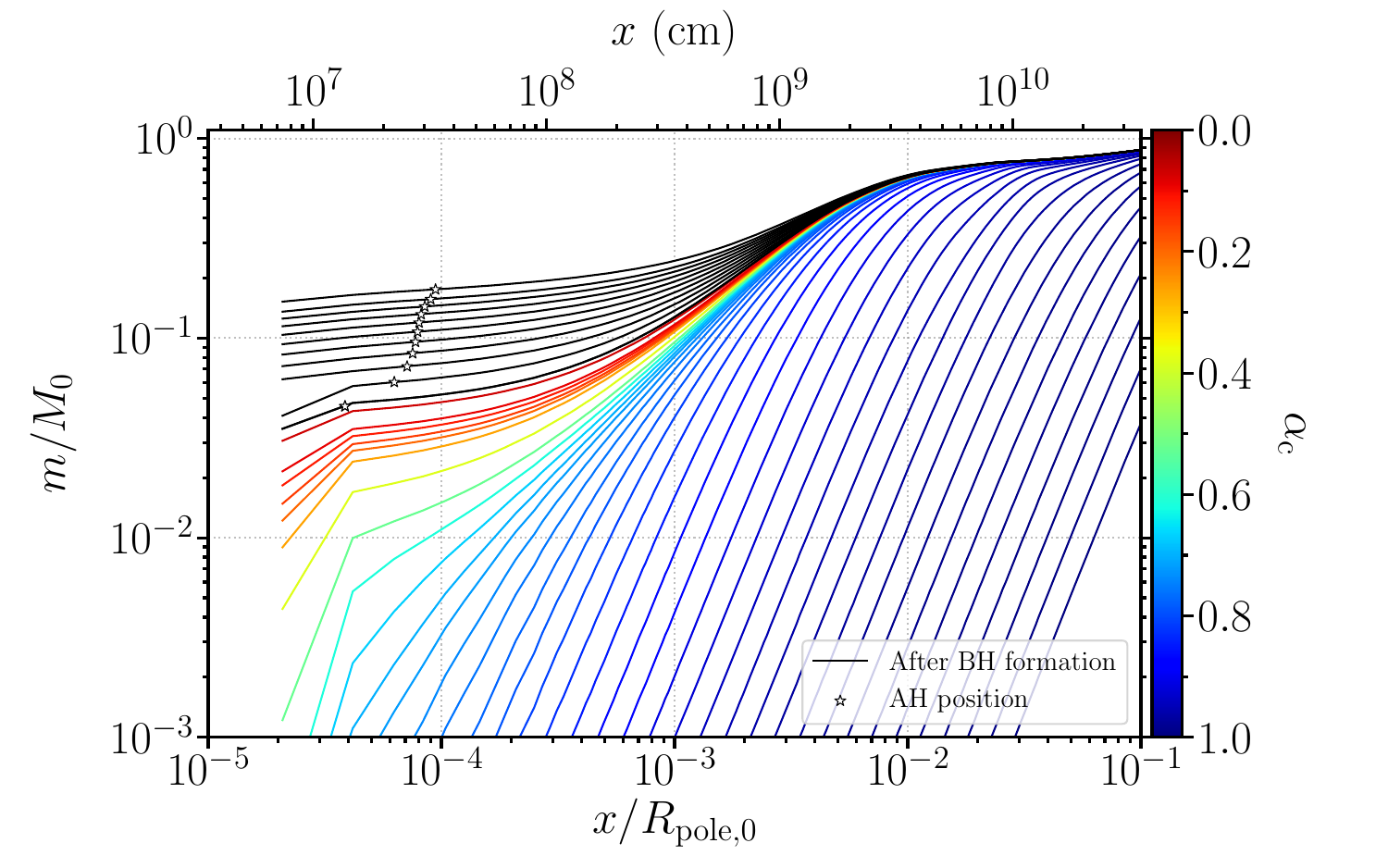}\\
    (c)\includegraphics[width=0.47\textwidth]{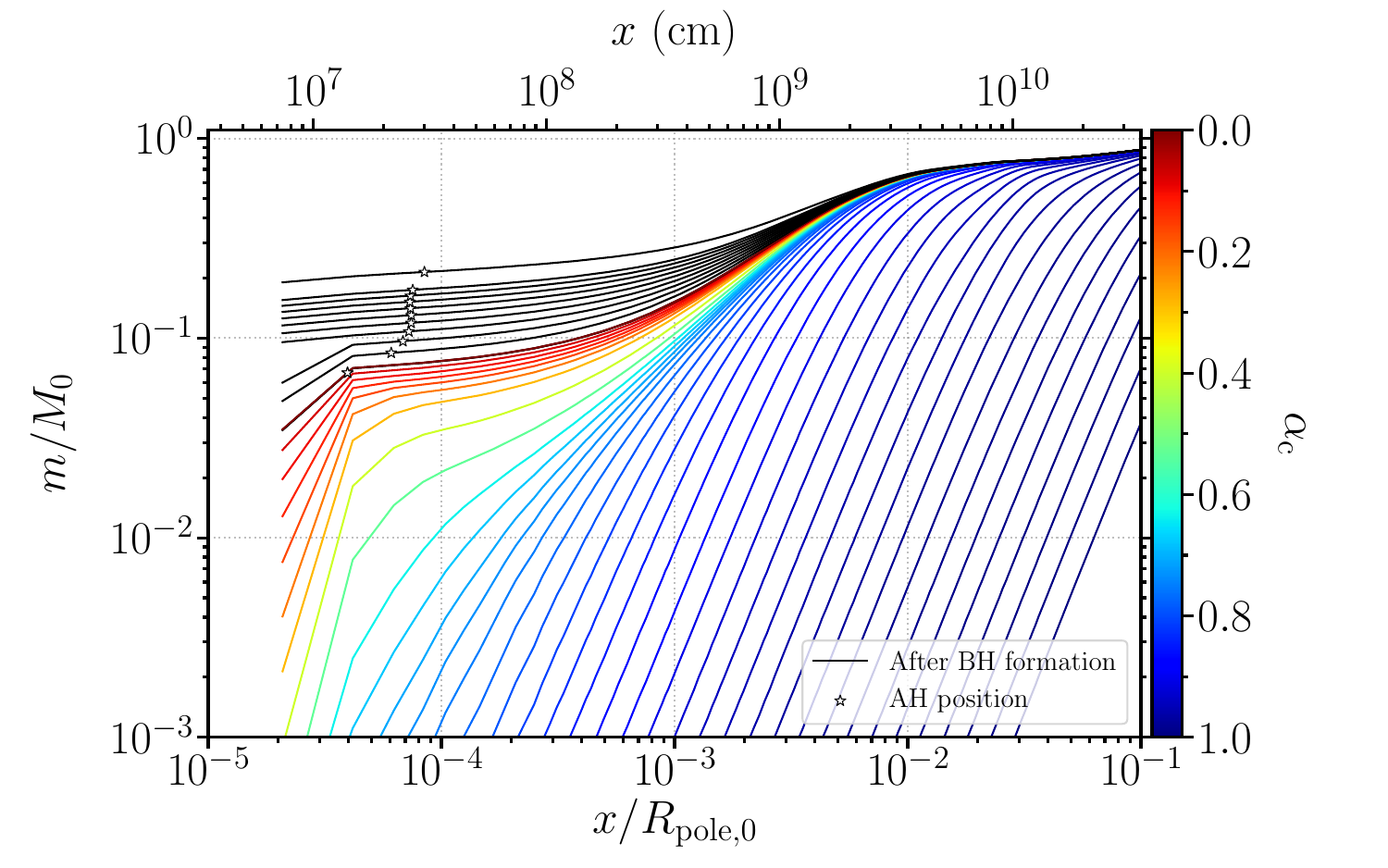}
    (d)\includegraphics[width=0.47\textwidth]{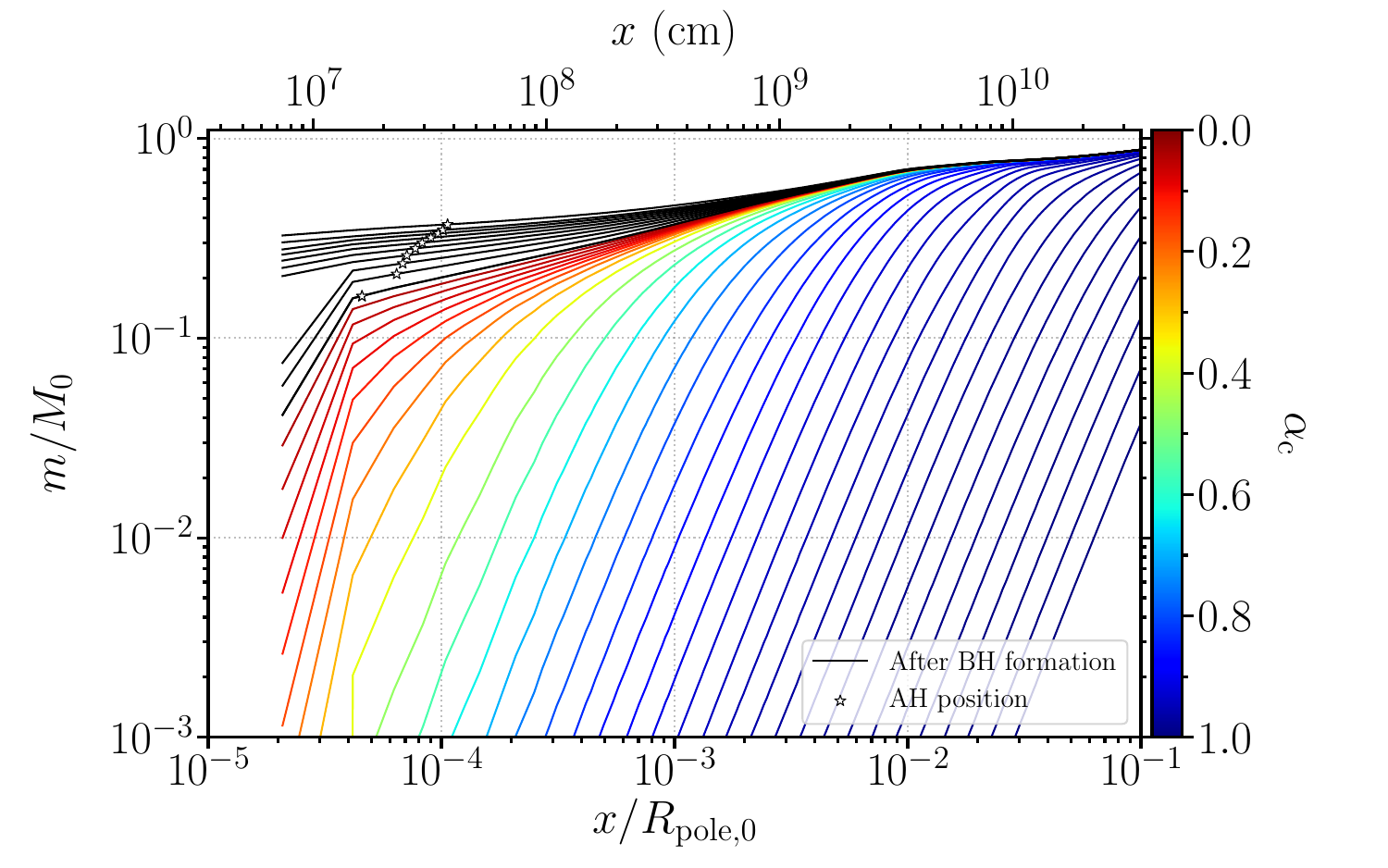}
    \caption{Enclosed mass normalized by the total core mass as a function of equatorial radius $x$. The models 5E4-09 (a) and 2E3-09 (b) are compared. The color of each curve indicates the central lapse, i.e., the stage of the collapse: $\alpha_\mathrm{c} \sim 1$ at the onset of the collapse and $\alpha_\mathrm{c} \rightarrow 0$ for the formation of a black hole. The black curves indicate the enclosed mass after black hole formation. The time interval of the black curves is approximately $t_\mathrm{g,0}:= GM_0/c^3$. The final time of the plot is $\approx 10\,t_\mathrm{g,0}$ after black hole formation. The star on each black curve indicates the location of the apparent horizon. Panels (c) and (d) show the 2E3-09 model with neutrino cooling artificially suppressed above $\rho_\mathrm{crit}=\SI{e10}{g.cm^{-3}}$ and \SI{e8}{g.cm^{-3}}, respectively.}
    \label{fig:mass-profile}
\end{figure*}

We first compare the collapse of two rapidly rotating stellar core models with different masses to illustrate the differences between the collapses driven by pair and general relativistic instabilities. Here, we take the 2E3-09 and 5E4-09 models, which begin the collapse due to pair and general-relativistic instabilities, respectively. The evolution tracks of their central densities and temperatures are shown in Fig.~\ref{fig:rho-T-cen}. The black diagonal line shows the relation between the central density and the central temperature for the marginally stable stellar cores against general-relativistic instability, and for the lower density (higher temperature) side, the spherical hydrostatic configuration is unstable~\citep{Shibata2025jan}. 

The 5E4-09 model begins its collapse from the region in which general-relativistic instability is expected to set in. On the other hand, the 2E3-09 model lies outside that region but is marginally stable to pair instability.

Figure~\ref{fig:mass-profile} compares the enclosed mass profiles of the models 5E4-09 (top left) and 2E3-09 (top right). The more massive model is closer to a monolithic collapse, in which the enclosed mass at a radius $r$ and a time $t$, $m(r,t)$, satisfies
\begin{align}
m(r,t) = m(r/a(t), t=0).
\end{align}
Here, $a(t)<1$ is a decreasing function of time (it equals the ratio of stellar radius at time $t$ to the initial radius; $a(0)=1$). Therefore, in this case, the enclosed-mass profile preserves its shape while shifting inward to smaller radii. For the 5E4-09 model, the shape of the enclosed mass is similar even at a late phase of the collapse ($\alpha_\mathrm{c}\sim0.1$). This characteristic of the collapse of supermassive-star core reflects the fact that the collapse is triggered by the f-mode radial instability~\citep{1964ApJ...140..417C} and proceeds approximately in a homogeneous manner.

By contrast, the lower-mass model shows a more runaway-like collapse at later phases with small values of the central lapse $\alpha_\mathrm{c}\lesssim 0.5$. In this model, only the central region, $x/R_\mathrm{pole,0} \lesssim 10^{-3}$, evolves significantly with time. This is because the lower-mass core has higher density and temperature, and as a result, neutrino emission cooling reduces pressure significantly during the collapse, accelerating the collapse of the central region.

\begin{figure}
    \centering
    \includegraphics[width=0.495\textwidth]{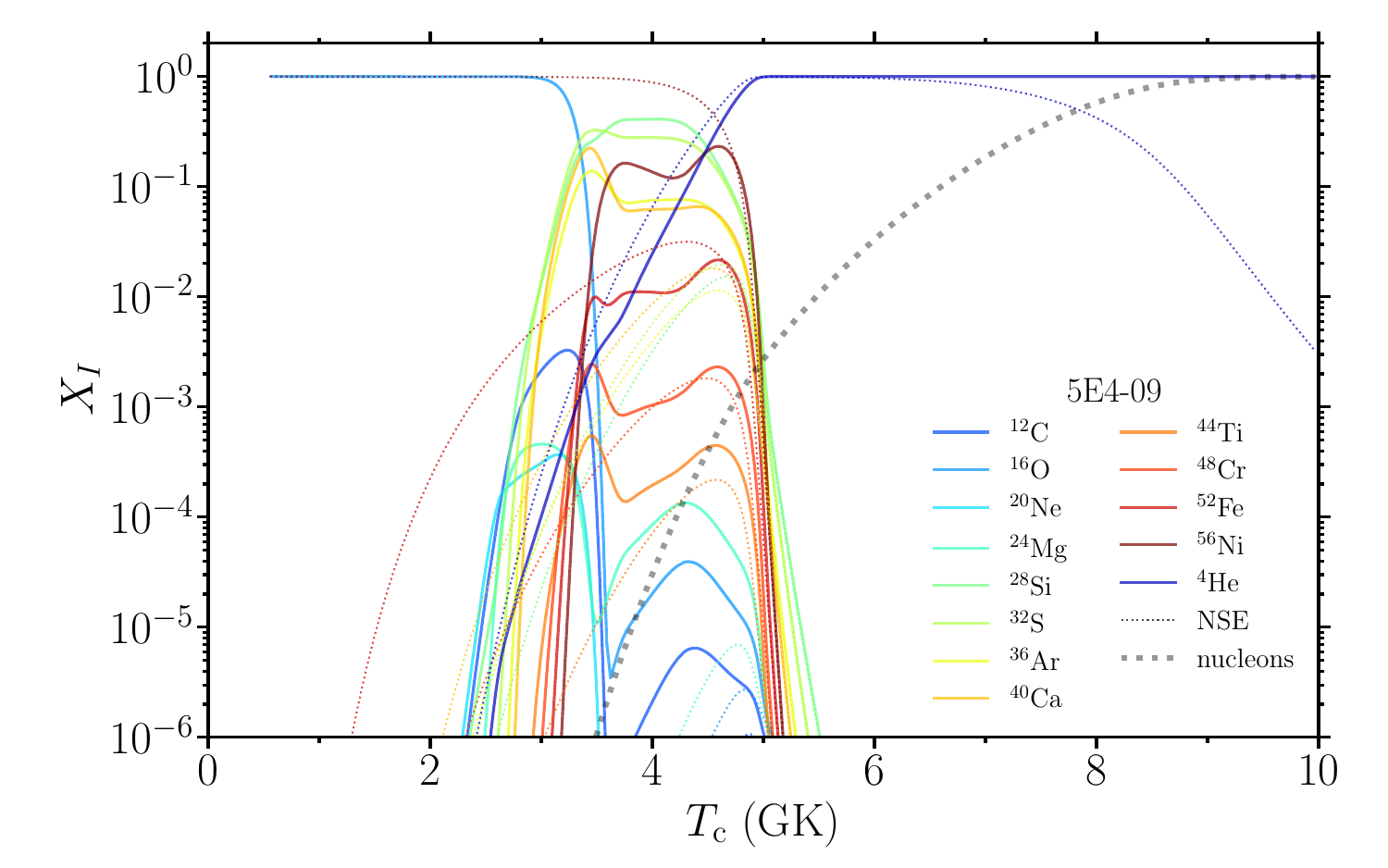}
    \includegraphics[width=0.495\textwidth]{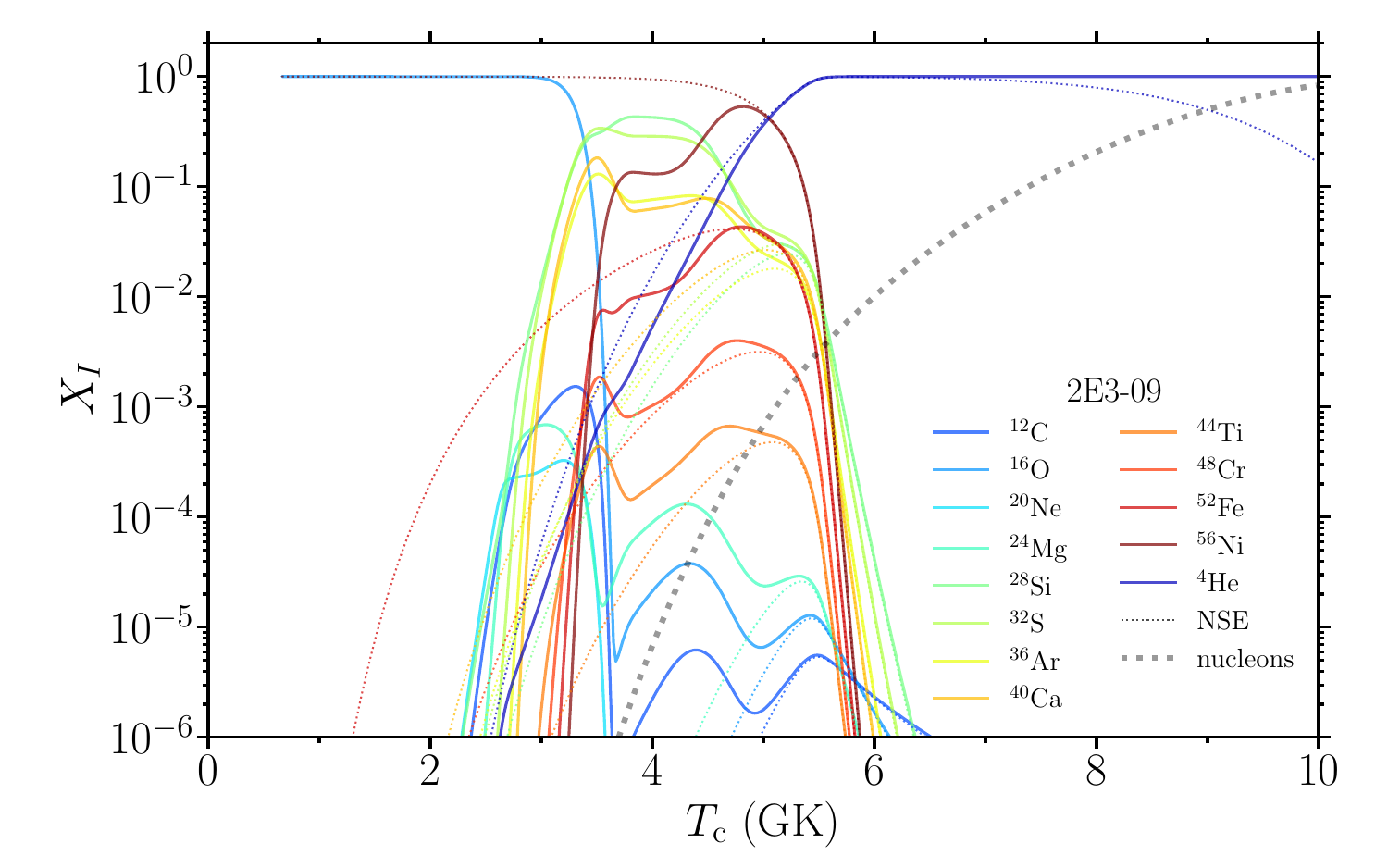}
    \includegraphics[width=0.495\textwidth]{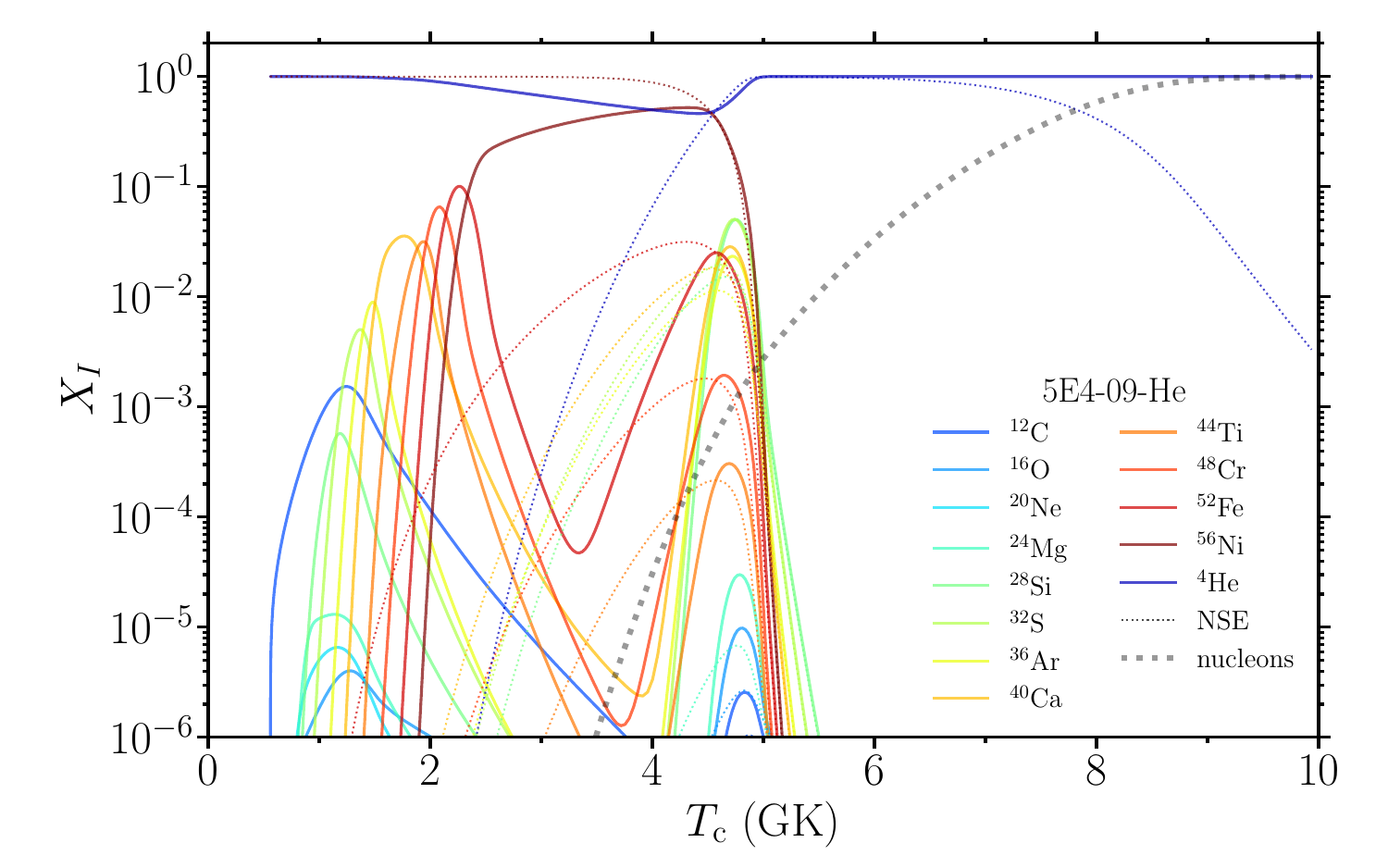}
    \caption{Central composition as a function of central temperature $T_\mathrm{c}$ during the collapse. The solid curves show the central mass fractions of nuclear species, while the dotted curves show the corresponding NSE mass fractions evaluated at the same central density and temperature. The thick gray dotted curve indicates the NSE mass fraction of free nucleons. From top to bottom, the panels show the 5E4-09 model, the 2E3-09 model, and a test model based on 5E4-09 but starting from a pure \isotope{4}{He} initial composition.}
    \label{fig:xcomp-cen}
\end{figure}

\begin{figure}
    \centering
    \includegraphics[width=0.495\textwidth]{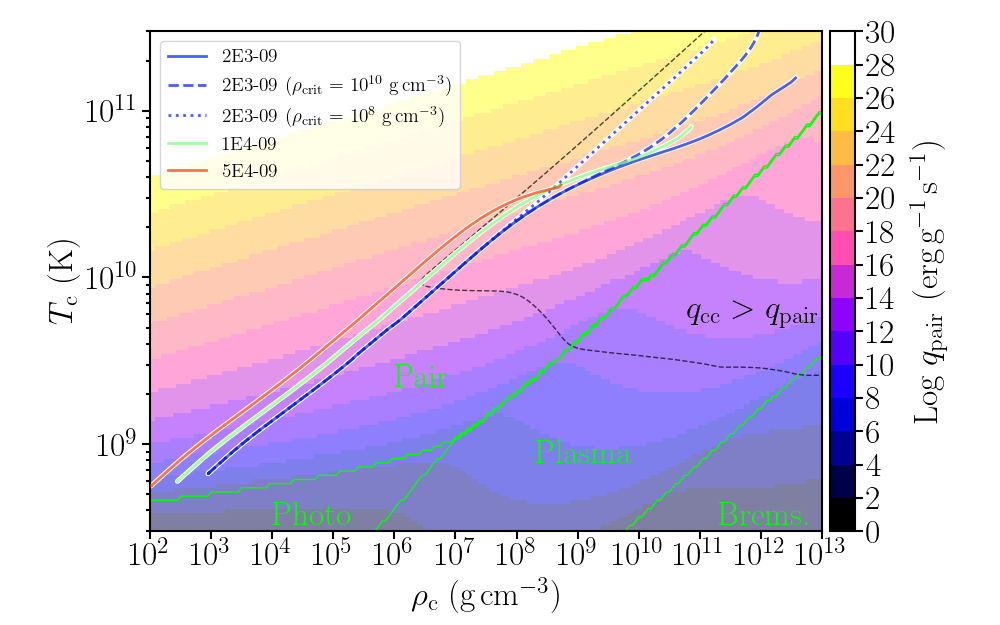}
    \caption{Same as Fig.~\ref{fig:rho-T-cen}, but the color denotes the specific cooling rate by pair processes ($q_\mathrm{pair}$). For the 2E3-09 model, additional trajectories are shown for simulations starting from the same initial data, in which cooling is artificially suppressed above $\rho_\mathrm{crit}=10^{10}$ and $10^{8}\ \mathrm{g\,cm^{-3}}$. The dashed curve encloses the upper-right region where charged-current weak interaction dominates the neutrino energy-loss rate. The 1E4-09 model is additionally plotted to illustrate that the core with $M_0 \lesssim 10^4M_\odot$ enters the region in which charged-current weak interaction dominates the cooling.}
    \label{fig:cooling}
\end{figure}

Differences in the collapse profile of the cores result in different black-hole masses at their formation. The mass of the black hole just after its formation is $\sim 20$\% of the total mass for the 5E4-09 case. It grows to more than 50\% of the total mass within a short time $10\,t_\mathrm{g,0}$, where $t_\mathrm{g,0}:= GM_0/c^3$ approximately denotes the light-crossing time of the black-hole horizon. By contrast, the black hole mass is only $\sim5$\% of the initial core mass at its formation for the 2E3-09 model. In addition, it is still less than 20\% after $10\,t_\mathrm{g,0}$. This is due to the runaway collapse behavior of this model; the central part collapses more rapidly.

The lower-mass core has to contract more before black hole formation. For the 2E3-09 model, the apparent-horizon radius at formation is $\sim 10^{-5}R_\mathrm{pole,0}$, about ten times smaller than that of the 5E4-09 model, $\sim 10^{-4}R_\mathrm{pole,0}$, where $R_\mathrm{pole,0}$ is the initial polar radius of the collapsing core. This is because stellar cores marginally stable to the pair instability are less compact at lower masses \citep{Shibata2025jan}; the 2E3-09 model has $R_\mathrm{pole,0}/r_\mathrm{g,0} = \SI{1.2e3}{}$, while it is $\SI{3e2}{}$ for the 5E4-09 model, where $r_\mathrm{g,0}=GM_0/c^2$ is the gravitational radius of the core. This naively implies that, in the case of the 2E3-09 model, the stellar radius would need to contract by an additional factor of $\approx 4$, relative to the 5E4-09 model, to reach $R/r_\mathrm{g,0}\sim 1$ for black hole formation. The required contraction is actually larger for the 2E3-09 model because the mass involved in the black hole formation is smaller due to its more runaway collapse. A less compact core can have a larger dimensionless angular momentum: the maximum value for this scales approximately as $C^{-1/2}$, where $C:=r_\mathrm{g,0}/R_\mathrm{pole,0}$ is an initial compactness. This favors the formation of a more rapidly rotating black hole and a more massive disk around the black hole.

This property of very massive stars leads to a numerical challenge in accurately solving the time evolution of the black-hole spacetime for lower-mass very-massive star models. With typical Einstein's equation solvers, it is necessary to resolve the apparent horizon with more than 20 grid points for an accurate time evolution of the black hole (depending on the timescale of the numerical simulation). A smaller initial core mass requires a finer grid spacing to capture its initial growth accurately. This numerical issue is discussed in \S~\ref{subsec:resolution}.

\subsection{Effects of nuclear burning and neutrino emission} \label{subsec:nuc-nu}

Figure~\ref{fig:xcomp-cen} shows the central nuclear composition as a function of the central temperature, $T_\mathrm{c}$, for models 5E4-09 and 2E3-09, which have the highest and lowest masses, respectively, among the main survey models. Since the central temperature increases monotonically with time in these models, this figure also represents the temporal evolution of the central composition. When the temperature reaches $\approx \SI{3}{GK}$, oxygen burning starts, producing heavier nuclei. At $\approx\SI{5}{GK}$, the composition approximately reaches NSE, as illustrated by the agreement between the solid and dotted curves in each panel. The dotted curves show the NSE abundances evaluated at the central density and temperature. For even higher temperatures, $T_\mathrm{c}\gtrsim \SI{6}{GK}$, heavy nuclei are photodisintegrated into \isotope{4}{He}. For the higher-mass model, because of its higher entropy, the NSE composition favors a larger abundance of \isotope{4}{He} relative to heavier nuclei.

For the 2E3-09 model, NSE is reached somewhat earlier than in the 5E4-09 model. This is because the lower-mass core has lower entropy and hence a higher density at a given temperature (note an approximate relation $s\propto T^3/\rho$), resulting in shorter nuclear-reaction timescales. The lower entropy also favors a larger abundance of \isotope{56}{Ni} at a given temperature than for the 5E4-09 model.

For the 5E4-09 model, the star may have already been gravitationally unstable during the helium-burning phase. Therefore, we performed a test simulation starting from pure \isotope{4}{He} instead of \isotope{16}{O} (see the bottom panel of Fig.~\ref{fig:xcomp-cen}). Because of the slow triple-alpha reaction, \isotope{4}{He} remains abundant throughout the evolution. NSE is reached somewhat earlier than in the model starting from pure \isotope{16}{O}. This is because the high-entropy NSE composition contains abundant \isotope{4}{He}; in the pure-\isotope{16}{O} model, \isotope{4}{He} has to be produced by the photodisintegration of heavier nuclei, which requires a high temperature.

In the present study, the dynamical nuclear reaction network includes only alpha nuclei. For sufficiently high temperature, however, \isotope{4}{He} can be further photodisintegrated into free nucleons. To estimate this effect, we include free neutrons and protons, in addition to the 13 alpha nuclei, when solving for the NSE abundances at a given density and temperature. The resulting free nucleon mass fraction is shown in Fig.~\ref{fig:xcomp-cen}. For both the 5E4-09 and 2E3-09 models, the free nucleons are expected to become more abundant than \isotope{4}{He} for $T\gtrsim8$--\SI{9}{GK}. This additional endothermic photodisintegration could further reduce the pressure in addition to the photodisintegration of \isotope{56}{Ni} into \isotope{4}{He}. Once photodisintegration is nearly complete, however, the increased number density of non-relativistic particles causes the pressure to increase more rapidly under further compression, effectively stiffening the equation of state (see, e.g., \citealt{Sekiguchi2011}).

In the models investigated in this work, however, most of the newly synthesized nuclei in the collapsing core are swallowed by the nascent black hole. Thus, the nuclear burning effects on the dynamics play only a minor role in the infalling phase. This is reasonable because the core masses in this study are much larger than the CO core mass that can make an explosion as pair-instability supernovae~(e.g., \citealt{2012A&A...542A.113Y, Takahashi2018}). We note, however, that the nuclear composition is non-trivial in the ejecta driven by the disk bounce, especially for lower-mass models (\S~\ref{subsec:ejecta}).

In the present models, neutrinos are assumed to escape without being absorbed or scattered after the emission. However, in the lower-mass models, the central density can reach sufficiently high for neutrinos to be trapped. Hence, the cooling efficiency would be reduced in reality \citep{Nagele2021nov}. To assess this effect, two additional simulations are performed for the model 2E3-09 with $\rho_\mathrm{crit}=10^{10}$ and $\SI{e8}{g.cm^{-3}}$.

Figure~\ref{fig:cooling} shows the behaviors of the model 2E3-09 with different critical densities $\rho_\mathrm{crit}$. It is found that the cooling rate due to electron-positron pair annihilation dominates over the other pair processes. The curves with $\rho_\mathrm{crit} = \SI{e10}{}$ and \SI{e8}{g.cm^{-3}} deviate from that of the fiducial 2E3-09 model when the density reaches the critical densities and the evolution becomes more adiabatic. The neutrino trapping effect modifies the dynamics of the collapse before the black hole formation. However, it has a limited impact on the subsequent dynamics after the black hole formation because the optically thick matter is swallowed into the black hole on the dynamical timescale.

The panel (c) of Fig.~\ref{fig:mass-profile} shows the 2E3-09 model with $\rho_\mathrm{crit}=\SI{e10}{g.cm^{-3}}$. As a result of adiabatic evolution for the last stage of the collapse ($\rho_\mathrm{c}\gtrsim \rho_\mathrm{crit}$), a runaway collapse is suppressed, and the mass of the black hole at formation is larger than that of the fiducial 2E3-09 model. For the same reason, the black hole formation time is slightly delayed for the model with suppressed cooling.

For the 2E3-09 model $\rho_\mathrm{crit}=\SI{e8}{g.cm^{-3}}$ case, the central density and temperature evolve in a nearly adiabatic manner (approximately $T\propto \rho^{1/3}$). As a result, the runaway collapse is suppressed, and the collapse profile becomes more monolithic. This is illustrated in panel (d) in Fig.~\ref{fig:mass-profile}.

\begin{figure*}
    \centering
    \includegraphics[width=0.49\textwidth]{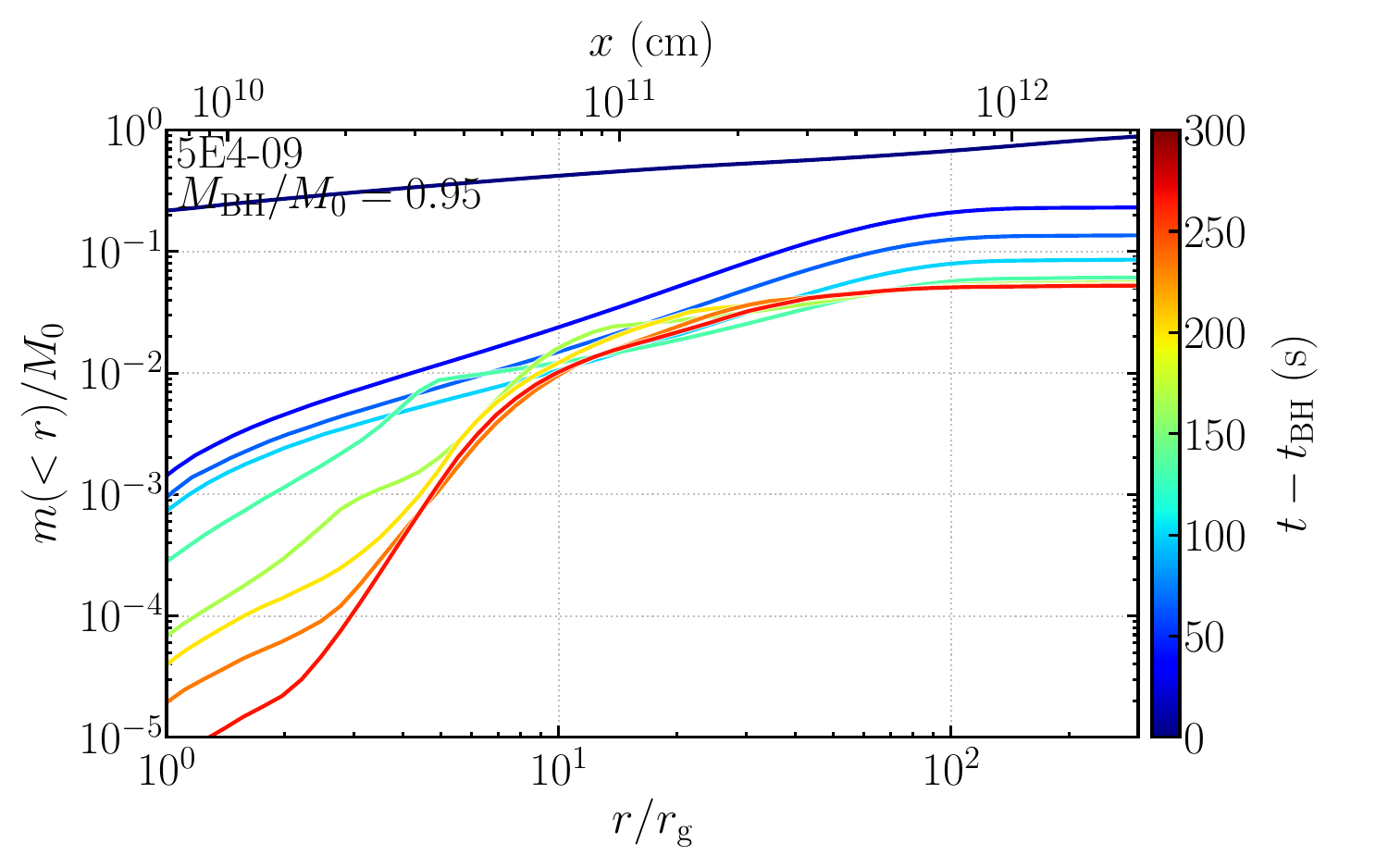}
    \includegraphics[width=0.49\textwidth]{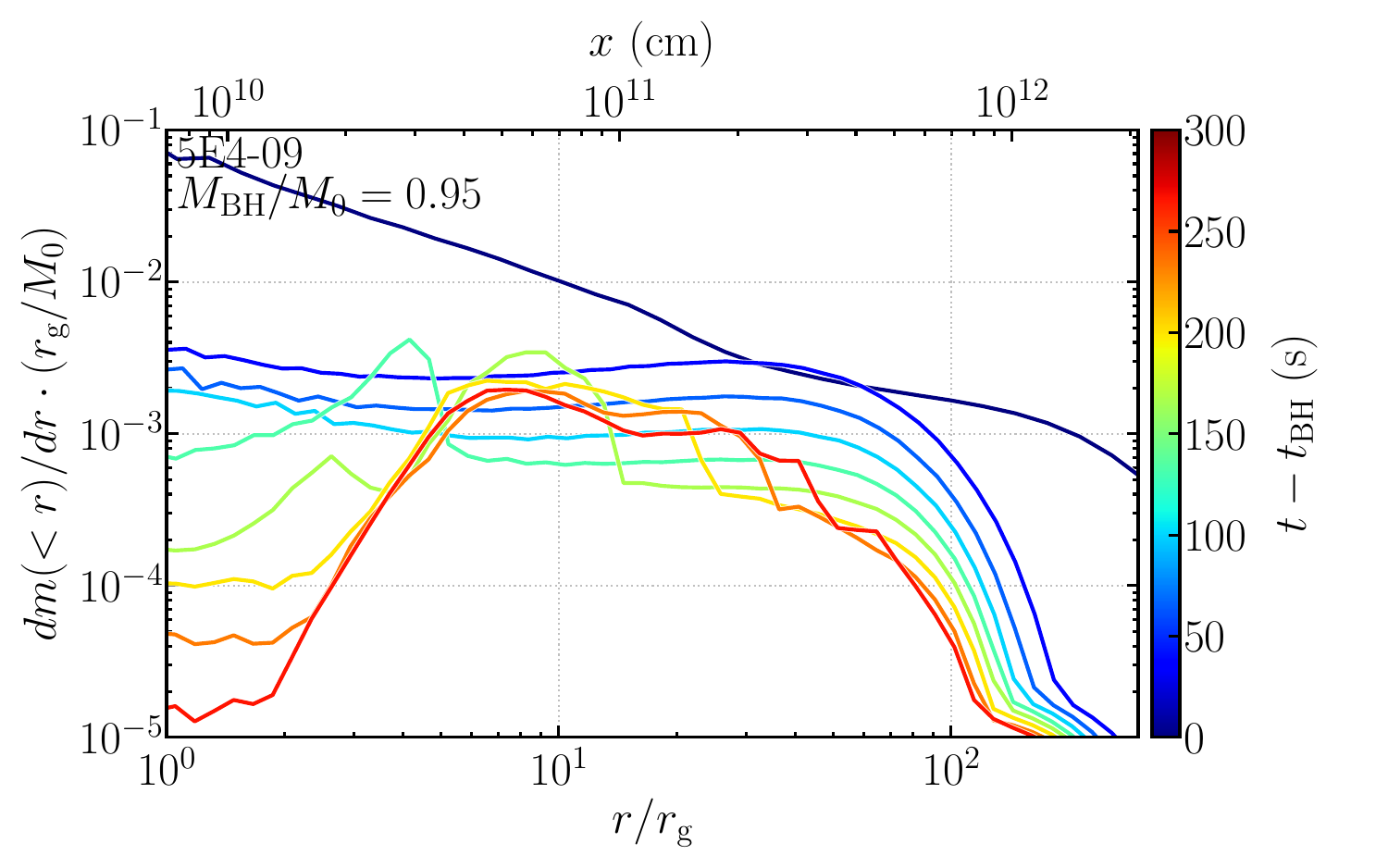}
    \includegraphics[width=0.49\textwidth]{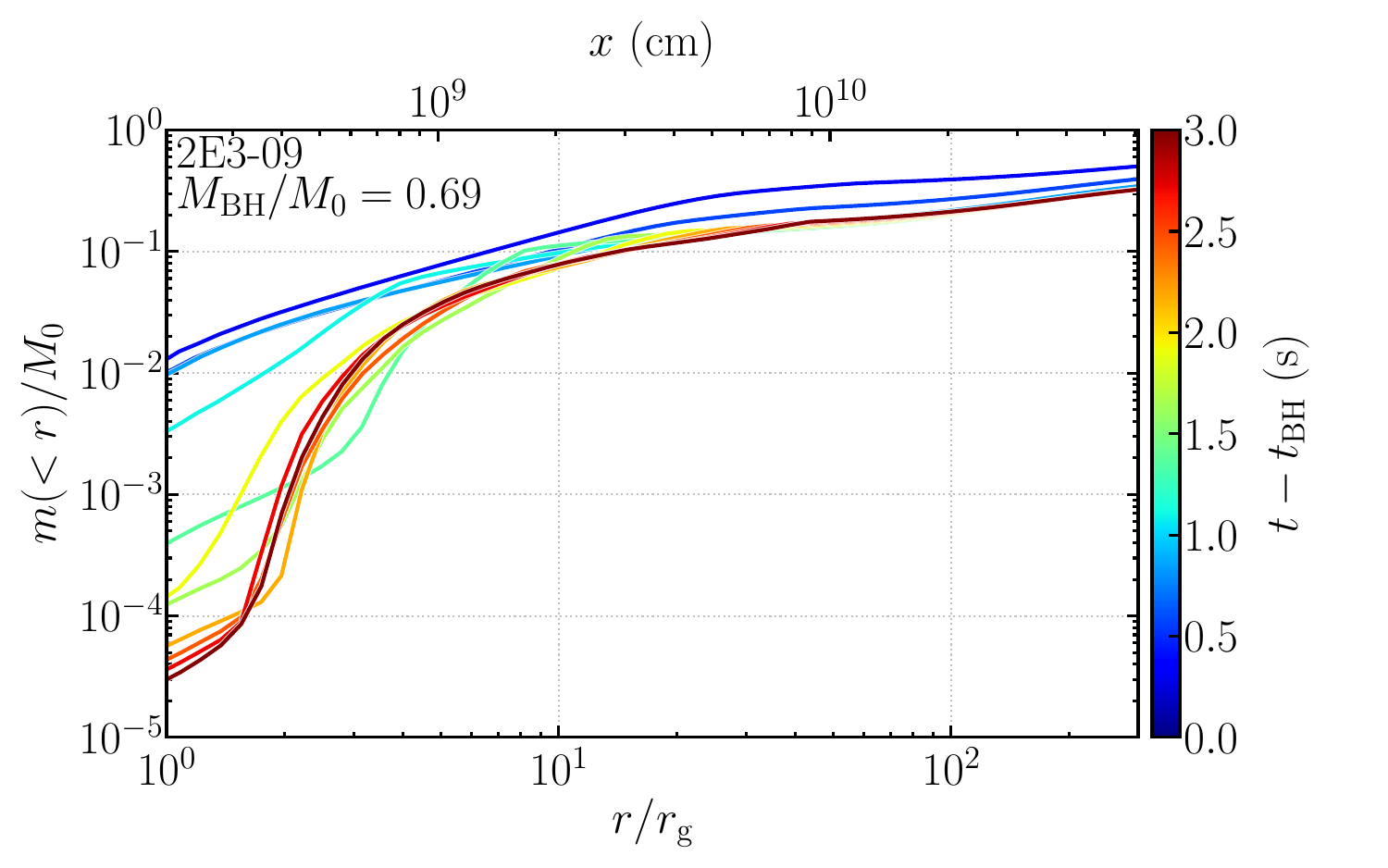}
    \includegraphics[width=0.49\textwidth]{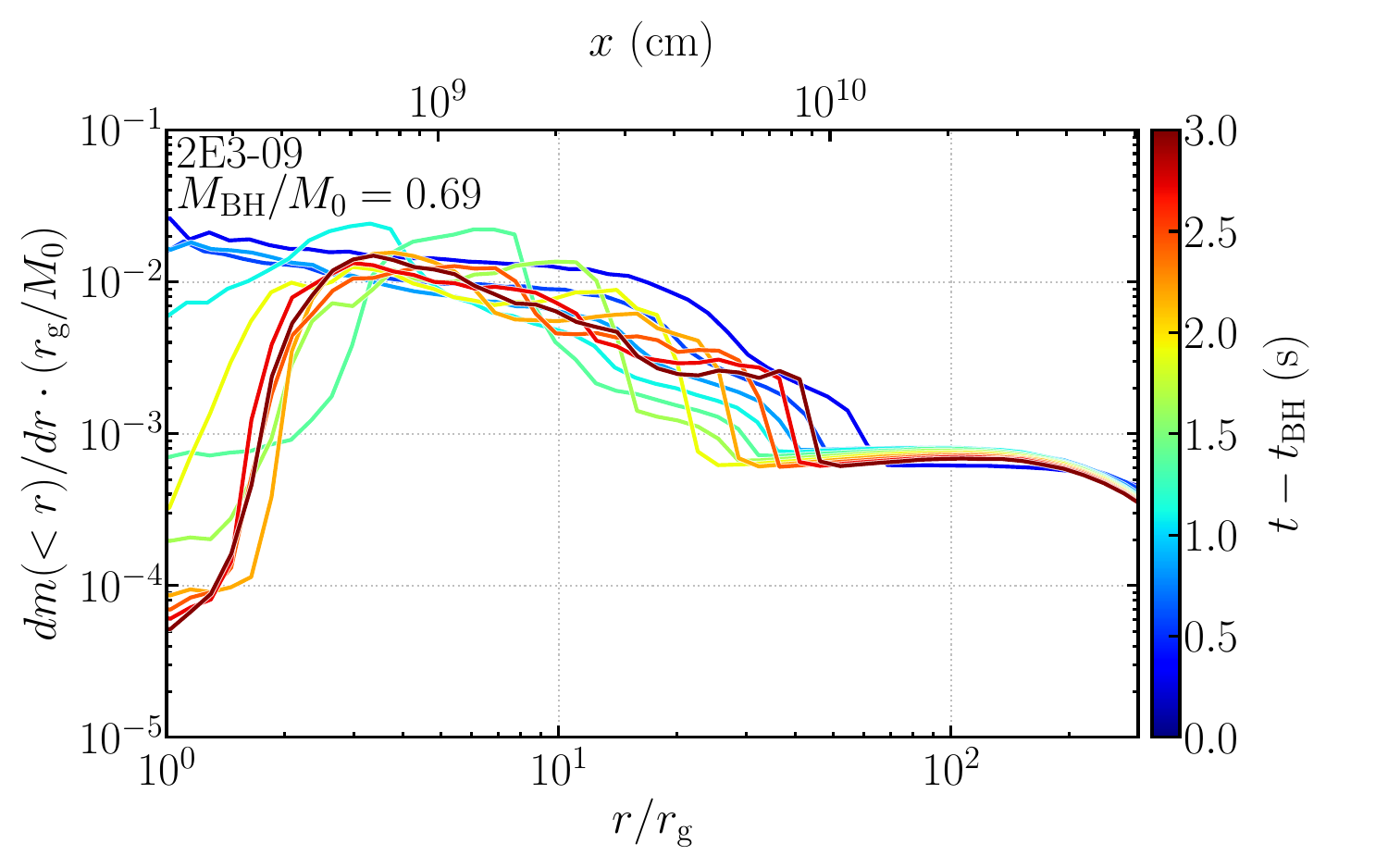}
    \caption{Mass distributions after the black hole formation for the 5E4-09 (top) and 2E3-09 (bottom) models. \textit{Left}: Cumulative mass distribution as a function of radius. The mass is normalized by the total core mass. \textit{Right}: Radial mass distribution obtained by differentiating the cumulative distribution. In all the panels, the radius is normalized by the gravitational radius of the black hole, $r_\mathrm{g}=GM_\mathrm{BH}/c^2$, evaluated using the instantaneous black hole mass. The color indicates the time after the black hole formation.} 
    \label{fig:mass-dist}
\end{figure*}

Charged-current weak interactions, which are ignored in this study, may be a more significant cooling mechanism during the collapse of lower-mass stars. In Fig.~\ref{fig:cooling}, the region enclosed by the dashed curve denotes the physical condition in which the energy emissivity due to the electron and positron captures on free nucleons, $q_\mathrm{cc}$, dominates over that of the pair processes (denoted by $q_\mathrm{pair}$, see Appendix~\ref{app:cc} for the details of the charged-current process). For the 5E4-09 model, the evolution path in the $\rho$--$T$ plane is outside the region of $q_\mathrm{cc}>q_\mathrm{pair}$ until the very last stage of the collapse, and thus, the collapse is not essentially affected by the charged-current weak interaction. For the 2E3-09 model, on the other hand, the path enters the $q_\mathrm{cc}>q_\mathrm{pair}$ region in the middle of the evolution. For the collapse of such a low-mass core, neutrino emission by the charged-current weak interaction significantly contributes to cooling. As a result, the runaway collapse is likely to be slightly accelerated before the neutrino trapping becomes important. We expect that this effect can play an important role in the collapse of relatively low-mass very massive stellar cores with mass less than $10^4M_\odot$ as illustrated by the 1E4-09 model in Fig.~\ref{fig:cooling}.

\subsection{Properties of formed disk and black hole} \label{subsec:disk-bh}
After the black hole formation, matter with a high angular momentum is supplied to the central region. This eventually leads to the disk formation around the black hole for the models considered in this paper. This process is illustrated in Fig.~\ref{fig:mass-dist}. The left panels show the cumulative mass as a function of radius for the 5E4-09 model (top) and the 2E3-09 models (bottom). It is defined by
\begin{align}
m(<r) = \int_{r_\mathrm{AH}<|\vec{x}|<r} \rho u^t\sqrt{-g} \,d^3x.
\end{align}
The integration includes only matter outside the apparent horizon of the black hole for which the radius is written as $r_\mathrm{AH}(\theta)$. 
For some time after black hole formation, the cumulative mass at a given radius decreases as matter is swallowed by the newly formed black hole. 
After $t-t_\mathrm{BH}\gtrsim$ \SI{130}{s} and \SI{1.1}{s} for the 5E4-09 and 2E3-09 models, respectively, the mass profile becomes nearly time-independent for $r/r_\mathrm{g}\gtrsim3$. 

To examine the radial distribution in more detail, the right panels show $dm(<r)/dr = 4\pi \bar \rho r^2$, where $\bar \rho$ denotes the angle-averaged density at each radius. The peaks of the distribution are located at $r/r_\mathrm{g}\approx6$ and 3 for the 5E4-09 and 2E3-09 models, respectively, indicating typical disk radii. The typical disk radius is smaller for the lower-mass model because the black hole is more rapidly rotating, and thus the radius of the innermost stable circular orbit (ISCO) in units of $r_\mathrm{g}$ is smaller~\citep{Bardeen:1972fi}.

\begin{figure}
    \centering
    \includegraphics[width=0.5\textwidth]{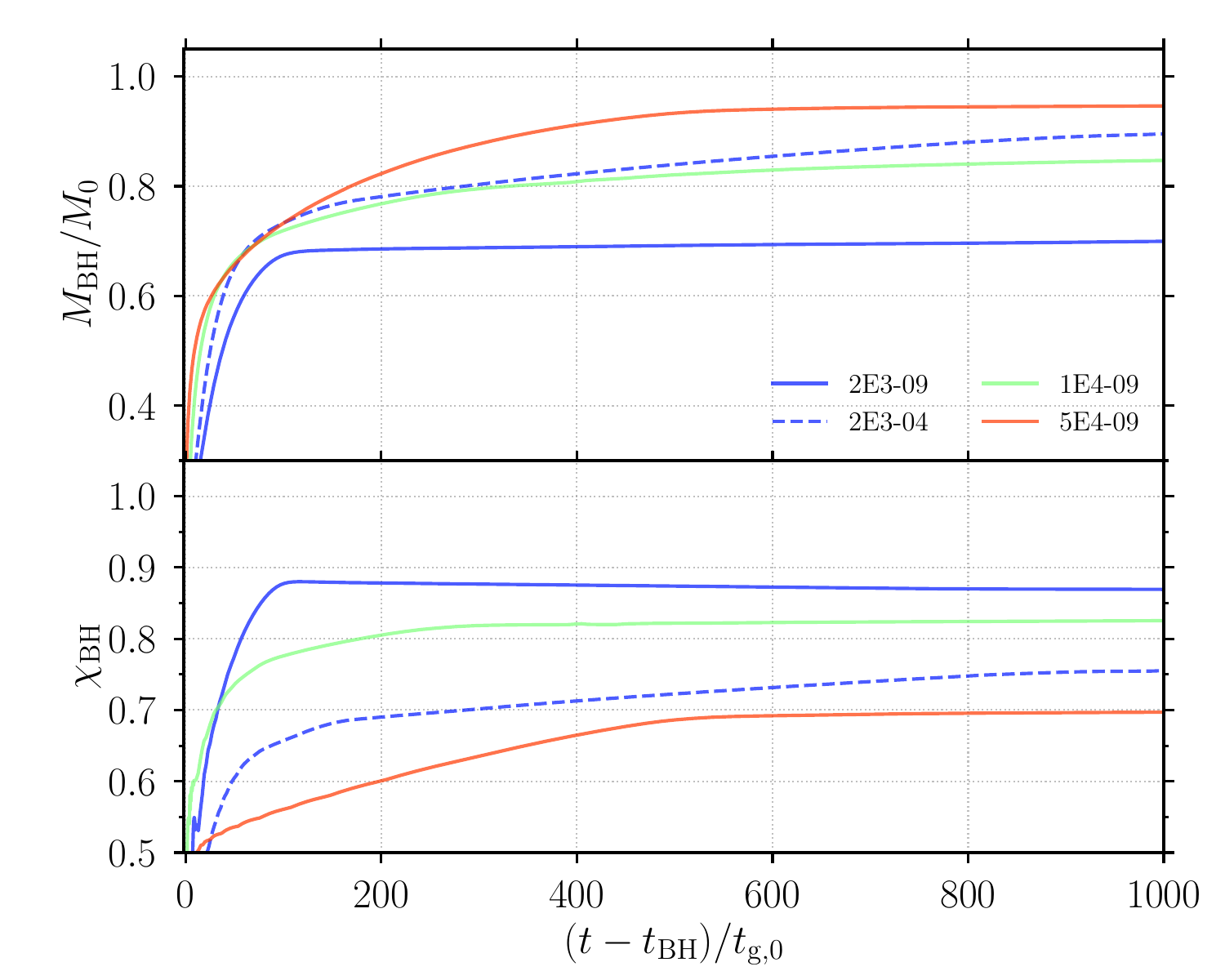}
    \caption{Time evolution of black hole mass (top) and dimensionless spin (bottom) for the 5E4-09, 1E4-09, 2E3-09, and 2E3-04 models. The black hole mass is normalized by the initial core mass, and the time is normalized by $t_\mathrm{g,0}=GM_0/c^3$.}
    \label{fig:BH-mass-spin}
\end{figure}

After the disk begins to be formed, the black hole mass and spin relax to saturated values because further infall of mass and angular momentum into the black hole is suppressed. This is illustrated in Fig.~\ref{fig:BH-mass-spin}, which shows the time evolution of the black hole mass and dimensionless spin. For a given time, the black hole mass and dimensionless spin, $M_\mathrm{BH}$ and $\chi_\mathrm{BH}$, are evaluated from the equatorial and polar circumferences of the apparent horizon in the same manner as in \cite{Shibata2016a, Fujibayashi2025mar}. To compare the evolution of black hole mass and spin for the models with quite different timescales, the time after the black hole formation is normalized by $t_\mathrm{g,0}$. The black hole mass and spin saturate at $(t-t_\mathrm{BH})/t_\mathrm{g,0} \approx 100$, 300, and 500 for the 2E3-09, 1E4-09, and 5E4-09 models, respectively. For those models, the physical times for the saturation are $\approx 1$, 15, and \SI{250}{s}, which correspond to the disk formation times (see in Fig.~\ref{fig:mass-dist} for the 2E3-09 and 5E4-09 models).

For the more slowly rotating 2E3-04 model, the disk formation and corresponding saturation of the black hole mass and spin occur at $t-t_\mathrm{BH} \approx 800t_\mathrm{g,0}$ (about \SI{8}{s}), which is later than the faster-rotating 2E3-09 model. This is because it takes more time for the matter that is located at a larger radius and has sufficiently high angular momentum to form the disk to fall into the central region.

We also find that the saturated black hole spin is higher in the lower-mass model. The dimensionless spin reaches $\approx 0.87$ at saturation for the 2E3-09 model, whereas it is $\approx 0.70$ and 0.83 for the 5E4-09 and 1E4-09 models. This reflects the larger dimensionless spin of less compact, lower-mass cores for a given degree of rotation $T_\mathrm{rot}/|W|$. This also leads to the smaller black hole mass at the saturation compared to the total core mass (see below). We note that the final value of the dimensionless spin for the 5E4-09 model is in good agreement with that found in \cite{Uchida2017oct} for the rapidly rotating model.

\begin{figure}
    \centering
    \includegraphics[width=0.5\textwidth]{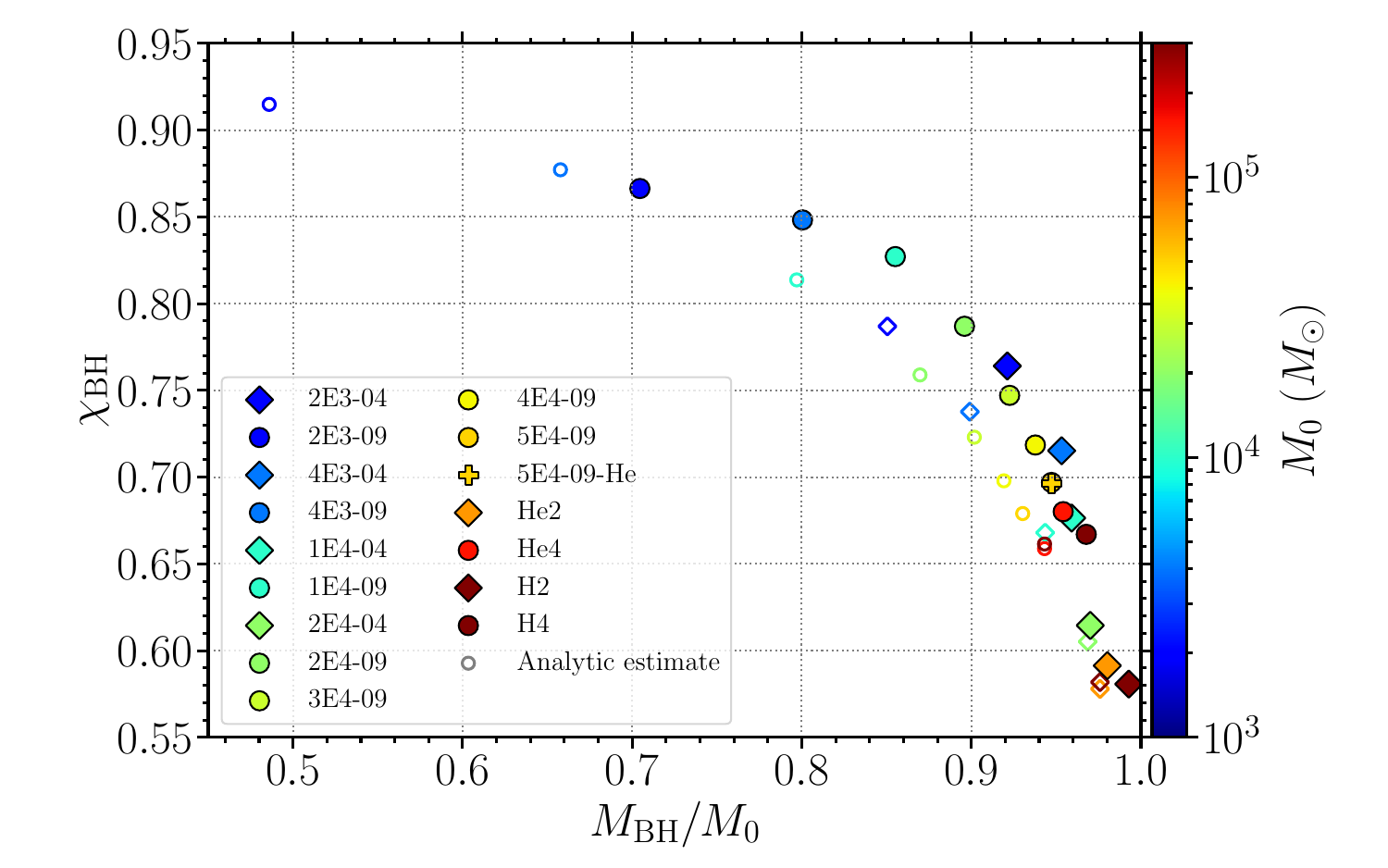}
    \caption{Relation between the black hole mass and dimensionless spin, evaluated at $1500\,t_\mathrm{g,0}$ after the black hole formation. The black hole mass is normalized by the initial core mass. The color indicates the initial core mass. The open markers denote analytical estimates based on the progenitor mass and angular momentum distributions.}
    \label{fig:fmass-spin}
\end{figure}

Figure~\ref{fig:fmass-spin} shows $M_\mathrm{BH}$ and $\chi_\mathrm{BH}$ for each model at $t-t_\mathrm{BH} = 1500\,t_\mathrm{g,0}$, by which time these quantities have nearly saturated. We compare them with the analytical estimates based on a method described in \cite{Stark1987sep, Shibata:2002br} (see also \citealt{Shibata2026jan} for procedure). We find that the black hole mass and dimensionless spin obtained in the numerical simulations follow trends similar to those predicted by the analytical estimates, with better agreement for the collapse of massive stars with $M_0 > \SI{1e4}{}M_\odot$.  However, the numerical results for the dimensionless spin are slightly higher than those estimates systematically.

This difference can be attributed to two effects. First, stellar matter located in the polar regions far from the center has small angular momentum but a long dynamical, or free-fall, timescale. As a result, such matter does not reach the black hole by the time that the disk forms and the black hole mass approximately saturates. This leads to a higher dimensionless spin of the black holes than predicted by the analytical estimate. Second, the infalling matter follows eccentric trajectories and can approach the center more closely than matter on circular orbits. Therefore, matter with specific angular momentum slightly higher than that at the ISCO can still fall onto the black hole. This increases both the black hole mass and spin.

For models with $M_0<\SI{1e4}{}M_\odot$, the black hole mass gradually increases and the dimensionless spin decreases spuriously because of finite-resolution effects (i.e., due to a numerical error for resolving the vicinity of the black holes, which continues to accumulate). This is particularly noticeable for the models with high values of $\chi_\mathrm{BH}$. We discuss this issue in \S~\ref{subsec:resolution}.

\subsection{Explosion properties} \label{subsec:explosion}
When the disk is formed around the black hole, the disk bounces and generates a shock wave, which propagates outward to cause a stellar core explosion. Indeed, the right panels of Fig.~\ref{fig:mass-dist} show an instantaneous increase in the radial mass distribution around $r/r_\mathrm{g}\approx 3$, which then propagates to larger radii as a shock. 

There are several differences  between the collapses driven by the general-relativistic and pair instabilities, which affect the details of the bounce process. One important difference is the amount of matter remaining outside the black hole at the time of the disk bounce. As shown in the right panels of Fig.~\ref{fig:mass-dist}, only $\sim5\%$ of the total core mass remains outside the black hole in the 5E4-09 model, whereas more than 30\% remains outside in the 2E3-09 model. This reflects the larger angular momentum of the lower-mass core. As a consequence of this, the disk is formed earlier in the lower-mass cores, while a larger amount of matter still remains outside the black hole.

\begin{figure*}
    \centering
    \includegraphics[width=0.32\textwidth]{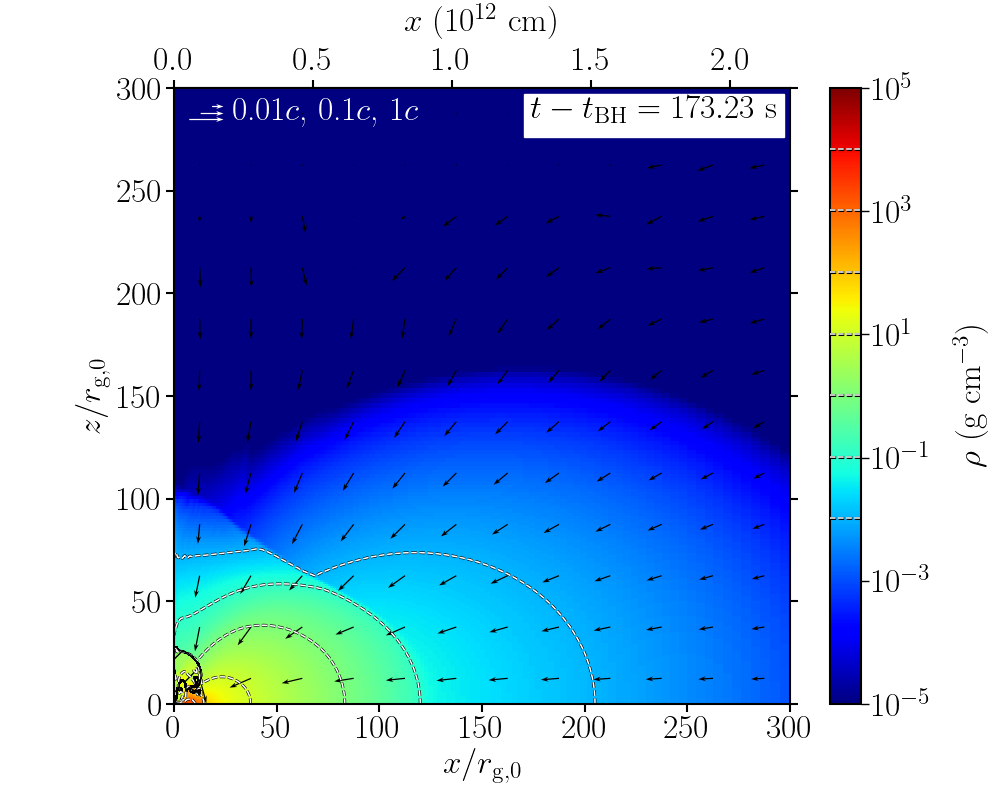}
    \includegraphics[width=0.32\textwidth]{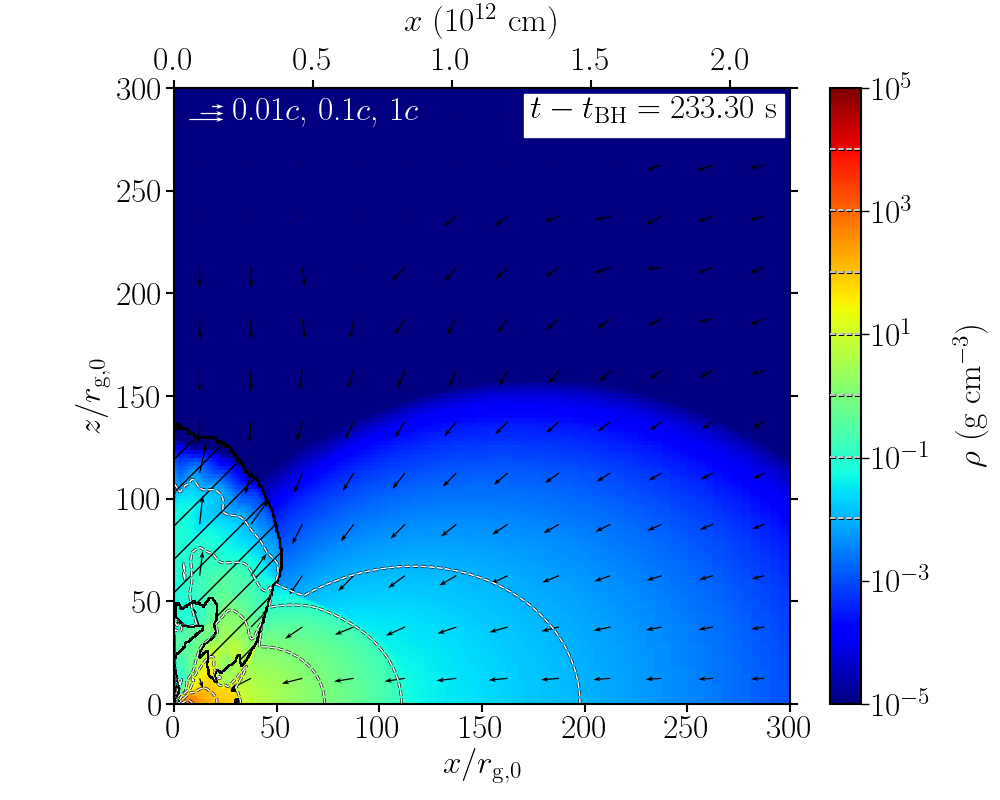}
    \includegraphics[width=0.32\textwidth]{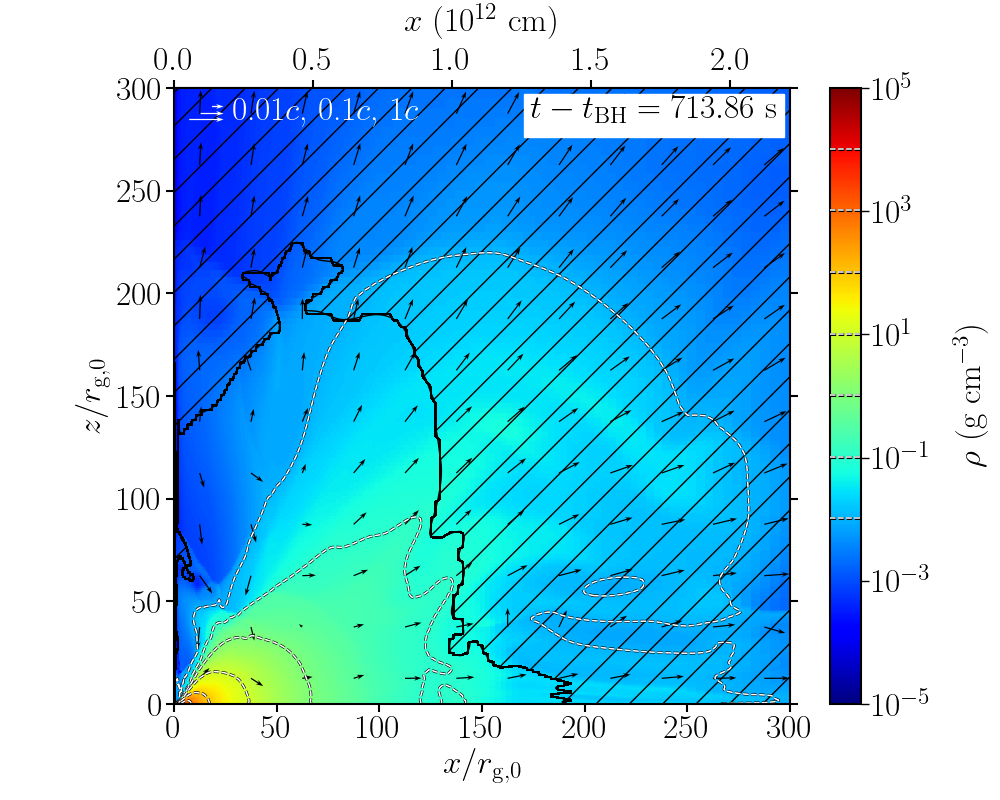}\\
    
    \includegraphics[width=0.32\textwidth]{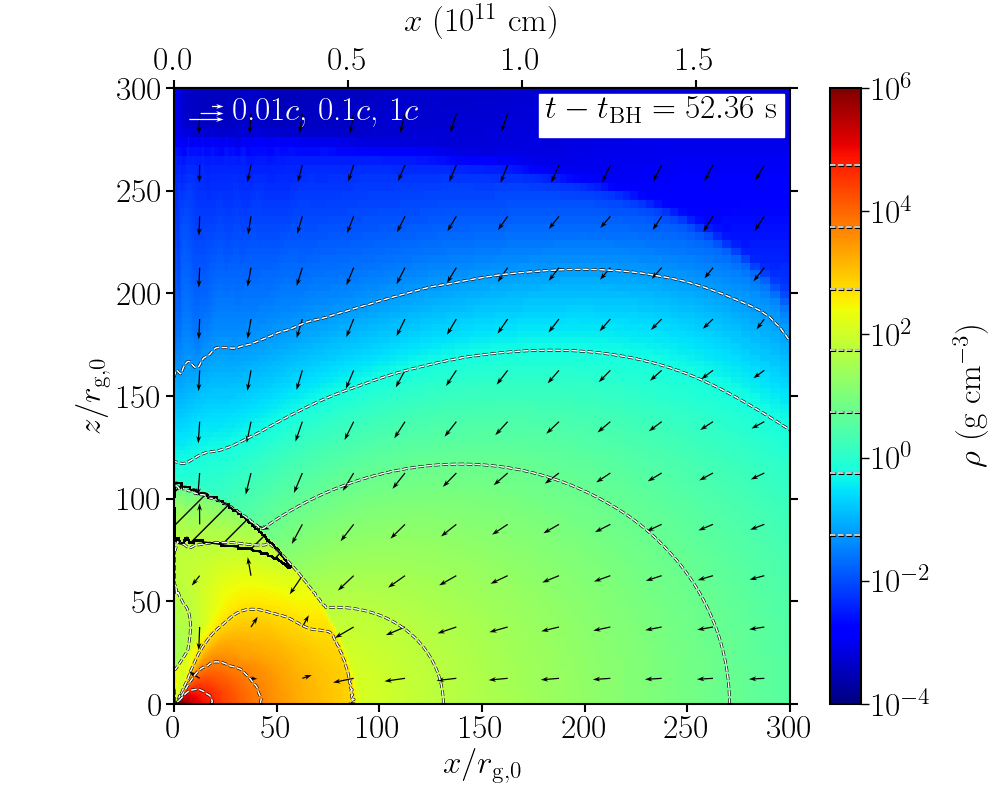}
    \includegraphics[width=0.32\textwidth]{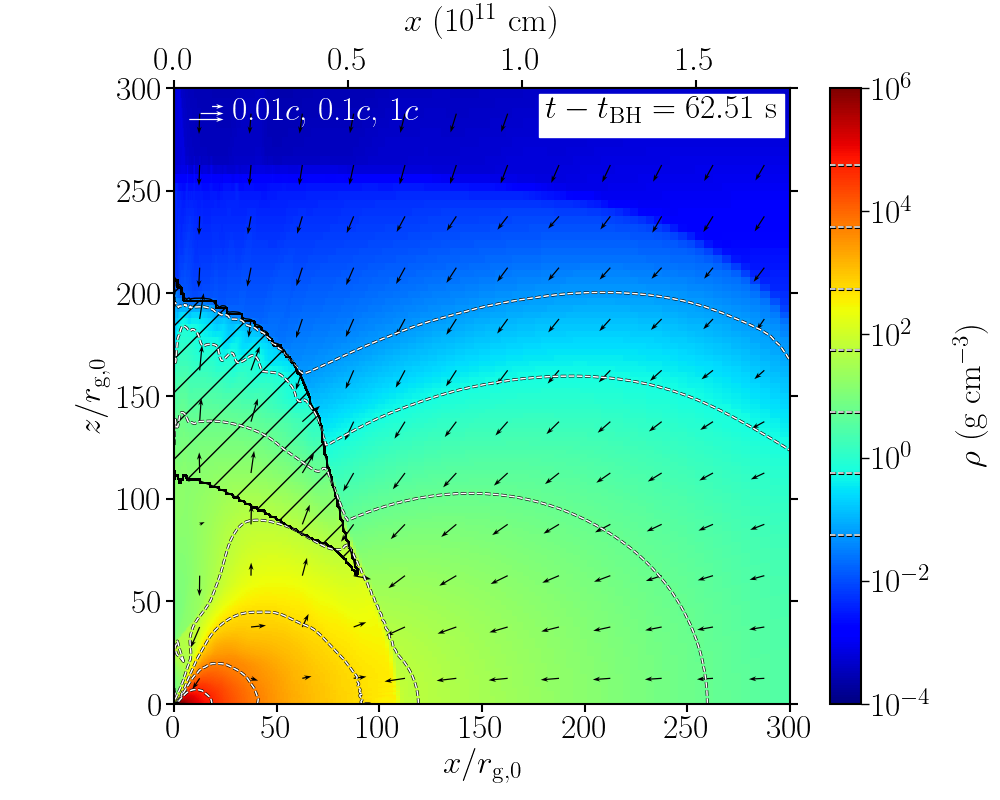}
    \includegraphics[width=0.32\textwidth]{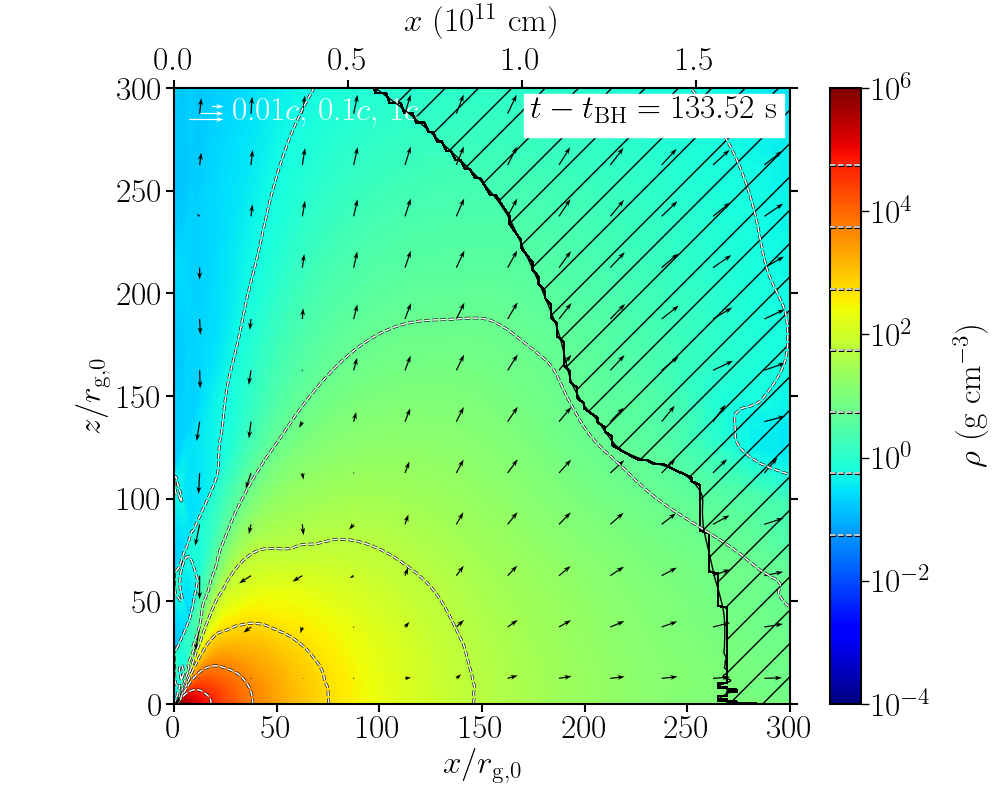}\\

    \includegraphics[width=0.32\textwidth]{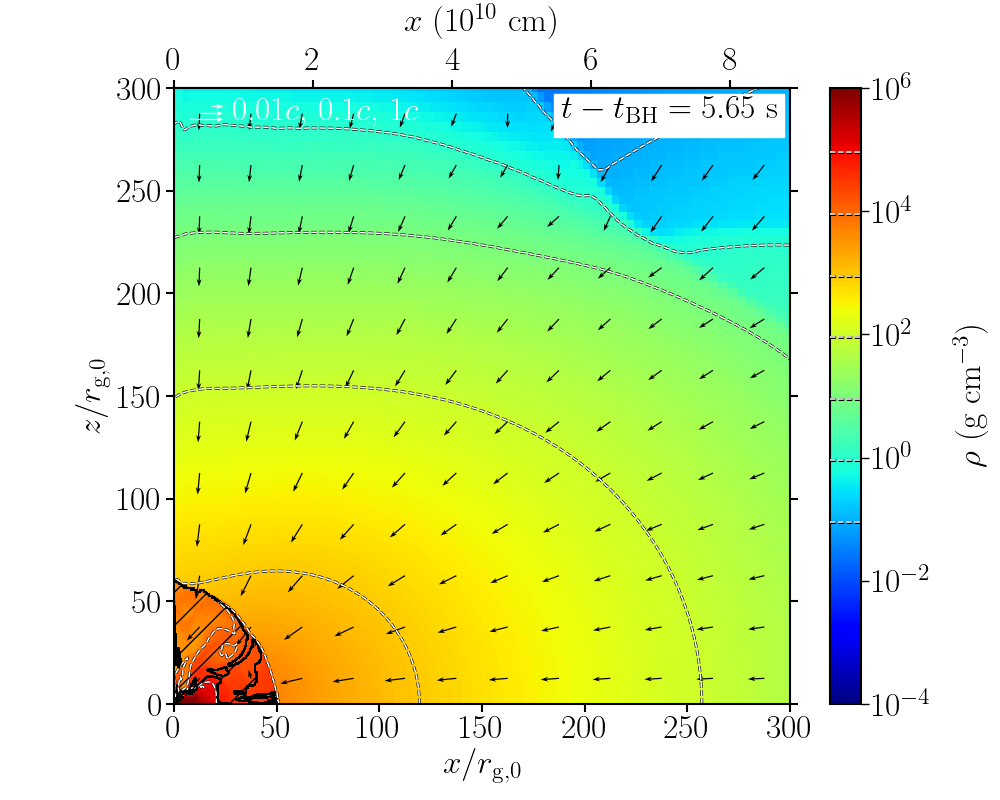}
    \includegraphics[width=0.32\textwidth]{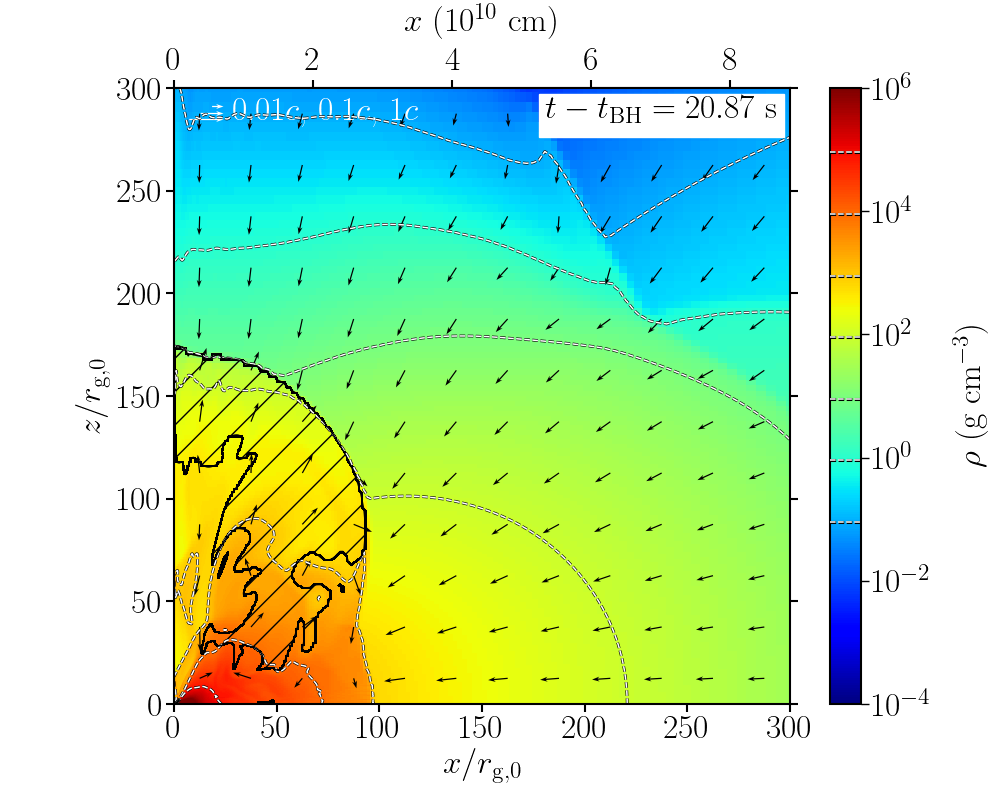}
    \includegraphics[width=0.32\textwidth]{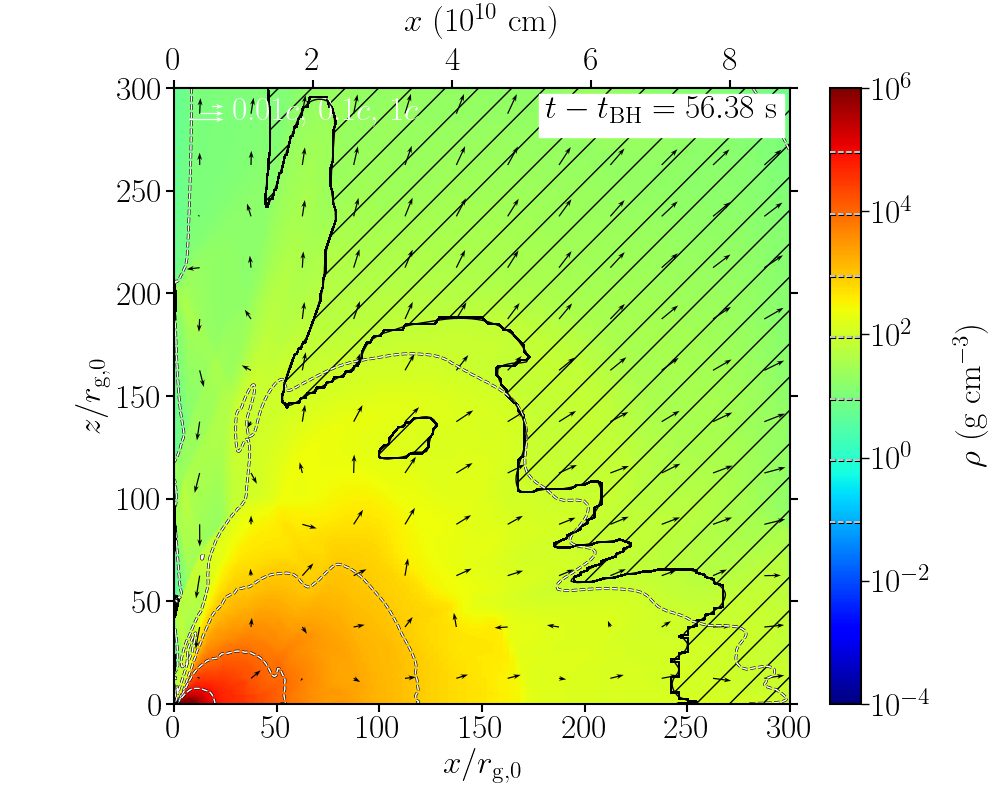}

    \caption{Time evolution of the shock propagation and mass ejection for the 5E4-09 model (upper row), the 4E3-09 model (middle row), and the 2E3-09 model (bottom row). The panels show the rest-mass density distribution in the meridional $x$--$z$ plane at selected times after black-hole formation, $t-t_{\rm BH}$, from left to right. The bottom and left axes show $x$ and $z$ normalized by the initial gravitational radius $r_{g,0}=GM_0/c^2$, while the top axis gives the corresponding physical scale. The color map denotes the rest-mass density $\rho$. The dashed curves denote density contours of $\rho/\rho_\mathrm{c,0}=10^{-4},10^{-3},\ldots,10^2$, where $\rho_\mathrm{c,0}$ is the initial central density. The arrows show the velocity field, with the reference arrows indicating $0.01c$, $0.1c$, and $1c$. The hatched regions denote unbound matter defined in the same way as \cite{Fujibayashi2025mar}.}
    \label{fig:density}
\end{figure*}

Figure~\ref{fig:density} shows the formation and propagation of the ejecta in the 5E4-09, 4E3-09, and 2E3-09 models. We primarily compare the 5E4-09 and 4E3-09 models to illustrate the systematic core-mass dependence, while the apparently exceptional behavior of the 2E3-09 model is discussed separately in \S~\ref{subsec:2e3-09}. To facilitate comparison between the two models with different initial core masses, all panels show the same spatial extent in units of the initial gravitational radius, $r_\mathrm{g,0}=GM_0/c^2$. We chose the 4E3-09 model as a representative low-mass model. The dashed curves denote contours of the density normalized by the initial central density, $\rho/\rho_\mathrm{c,0}=10^{-4},10^{-3},\ldots,10^2$, allowing us to compare the density structures independently of the characteristic density scale of each progenitor.

In the higher-mass 5E4-09 model, the disk bounce and the initial formation of unbound matter occur relatively close to the black hole. The shock subsequently propagates outward through the comparatively dilute material remaining outside the black hole \citep{Fujibayashi2025mar}. In the lower-mass 4E3-09 model, by contrast, the centrifugally supported structure extends to larger radii, and the shock encounters a larger amount of dense infalling matter. This difference reflects both the larger dimensionless angular momentum of the lower-mass progenitor and its more runaway-like collapse, which leaves a larger fraction of the stellar core outside the black hole when the disk is formed. Consequently, the ejecta-forming region is more extended in the lower-mass model, and a larger amount of kinetic energy is dissipated through interaction with the surrounding matter. This accounts for the lower characteristic ejecta velocity found toward lower initial core masses, as discussed below.

For still lower-mass cores with $M_0\lesssim10^3M_\odot$, the temperature in the downstream of the bounce shock would become sufficiently high, leading to much more efficient neutrino cooling and photodisintegration of heavy nuclei. These effects reduce the downstream pressure. When those effects are efficient enough, the disk bounce can no longer drive mass ejection as found in simulations of ordinary rotating massive star collapse (e.g., \citealt{Just2022aug,Dean2024apr,Fujibayashi2024jan}).

\begin{figure*}
    \centering
    \includegraphics[width=0.48\textwidth]{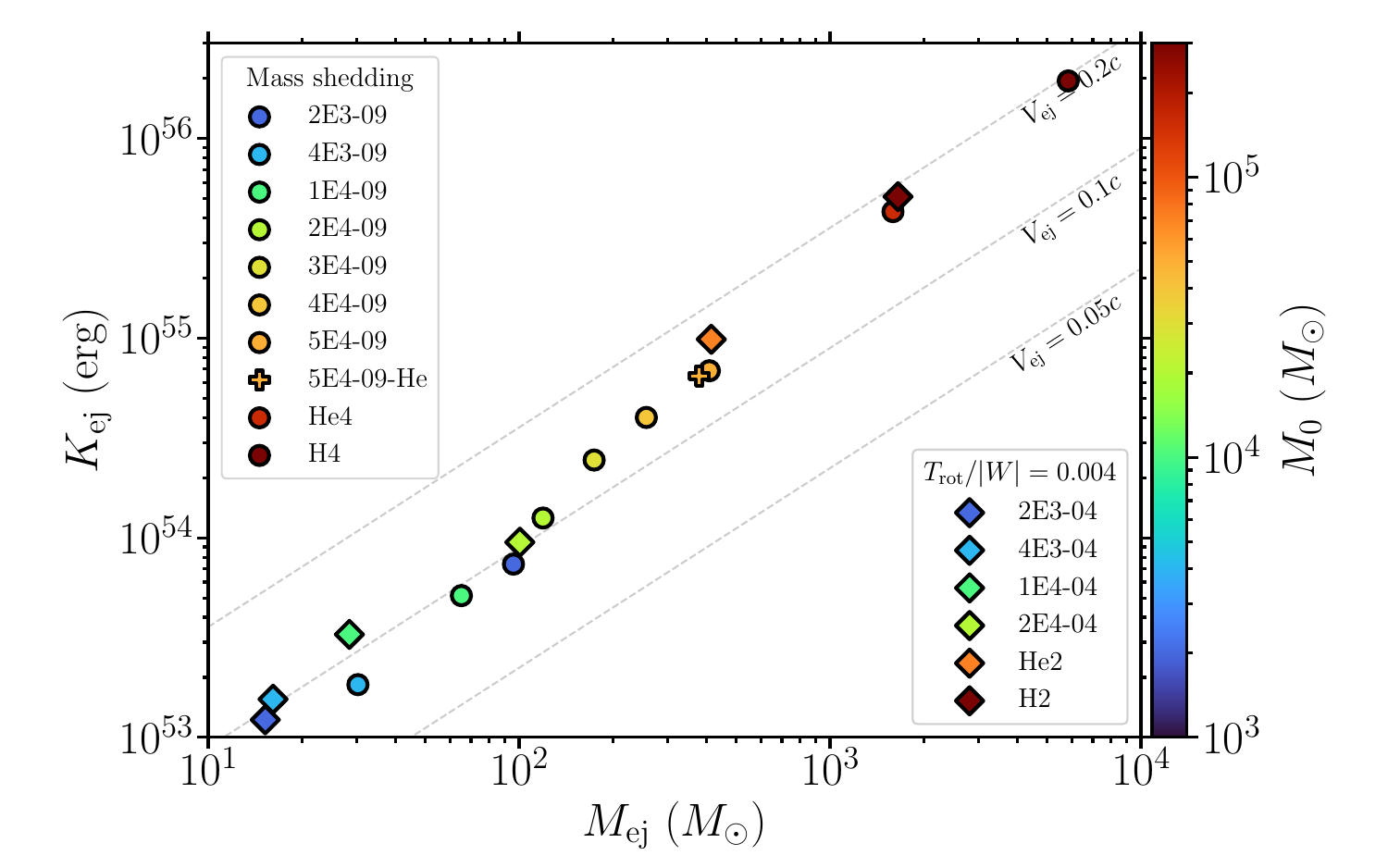}
    \includegraphics[width=0.48\textwidth]{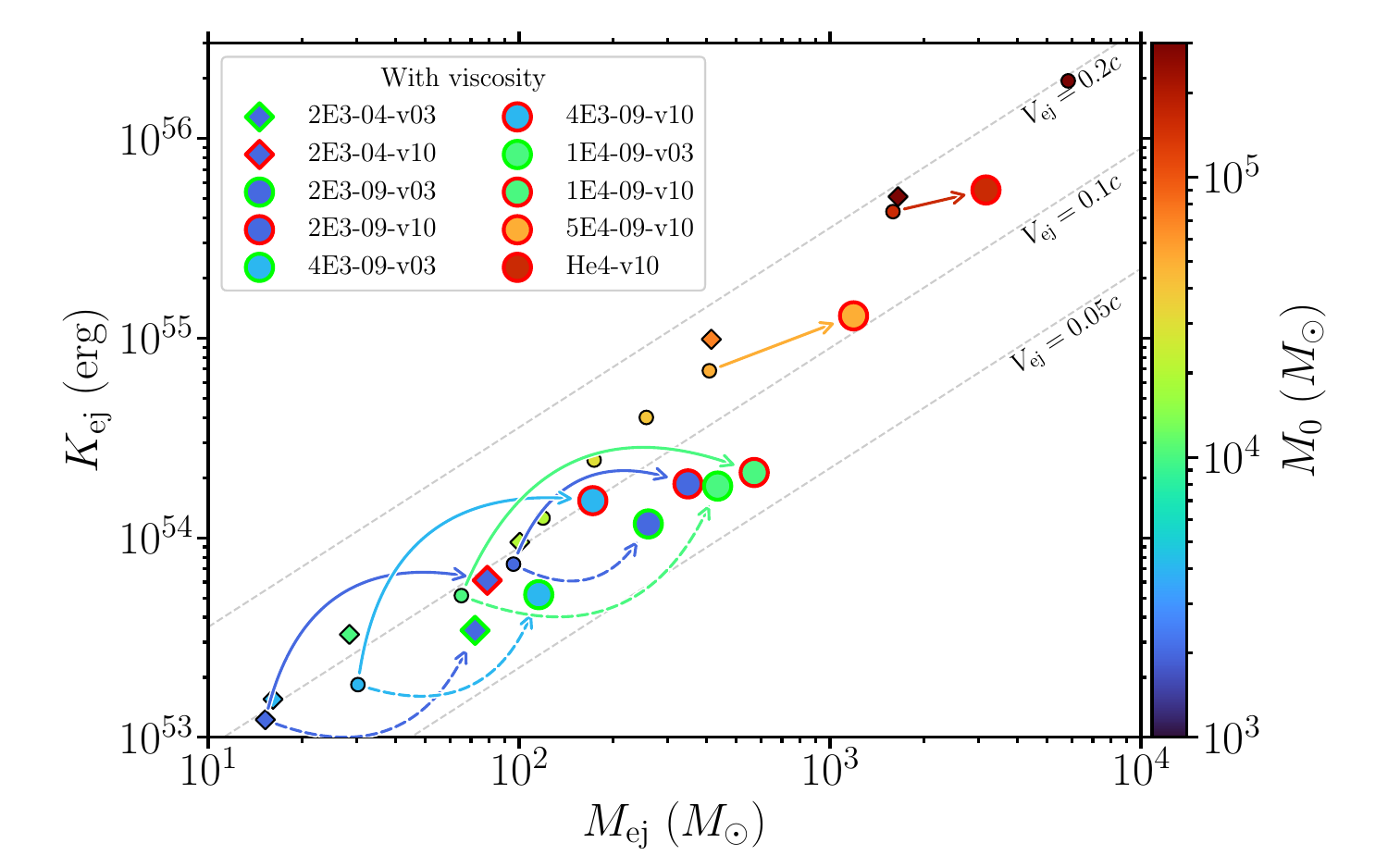}
    \includegraphics[width=0.48\textwidth]{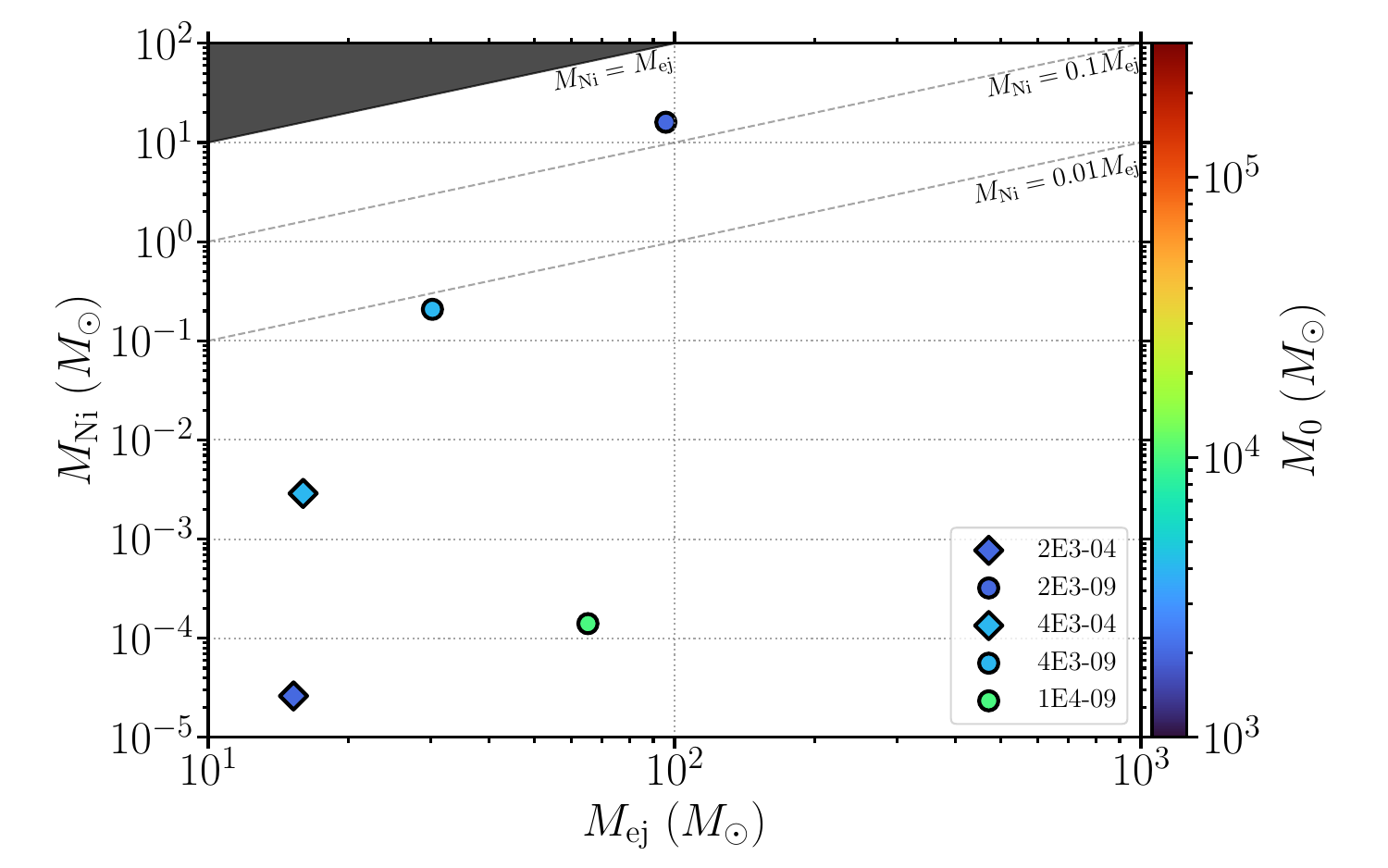}
    \includegraphics[width=0.48\textwidth]{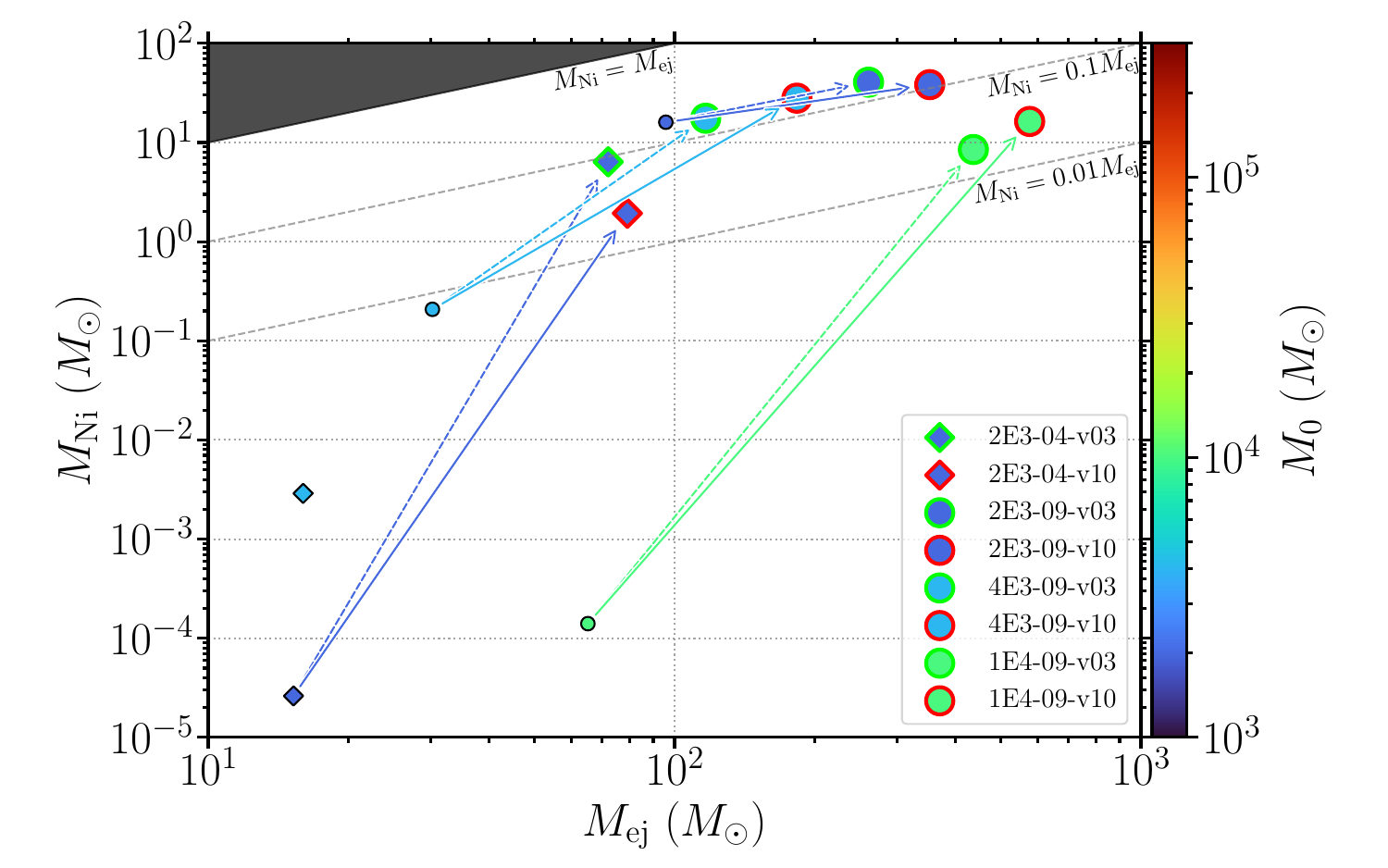}
    \includegraphics[width=0.48\textwidth]{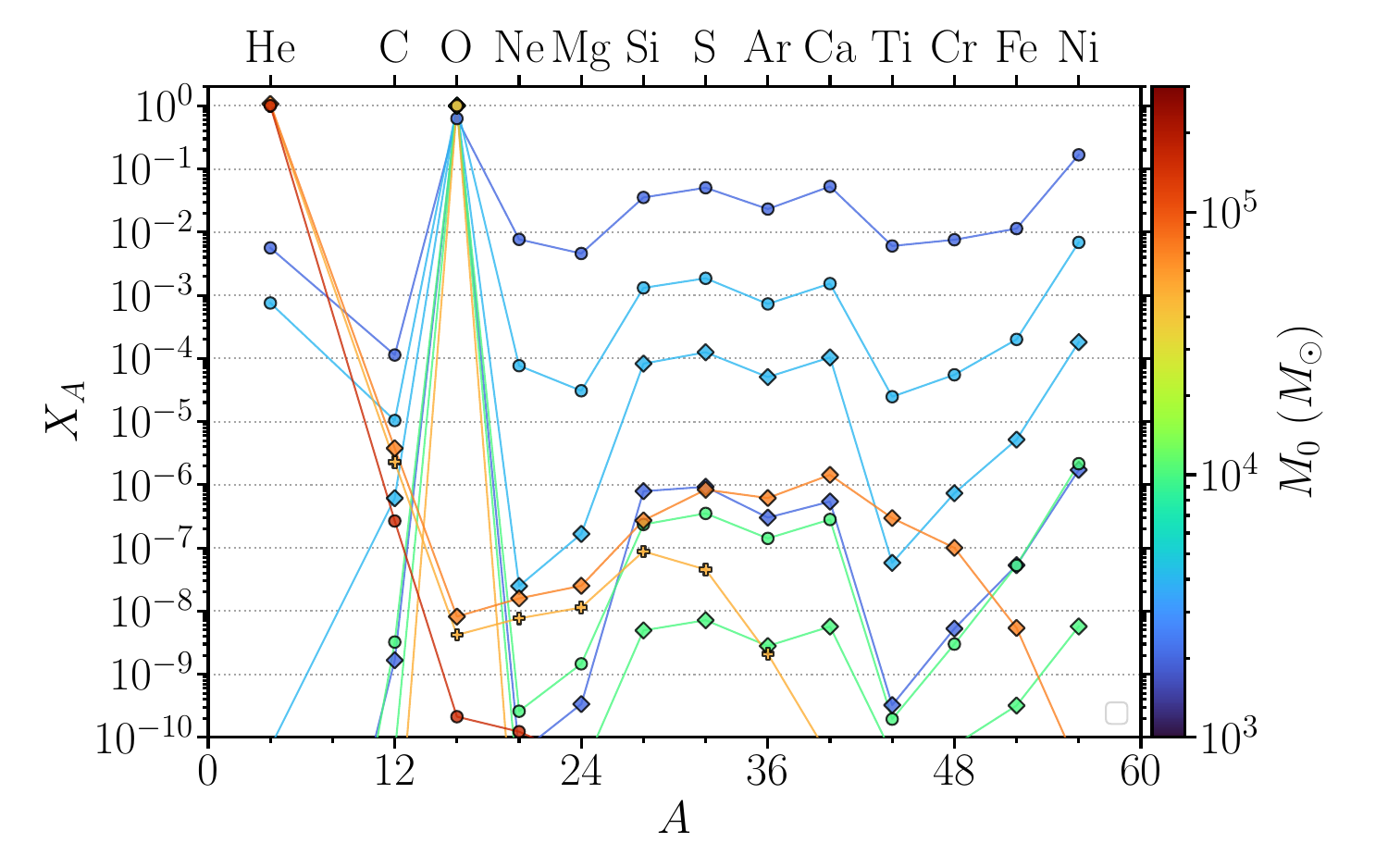}
    \includegraphics[width=0.48\textwidth]{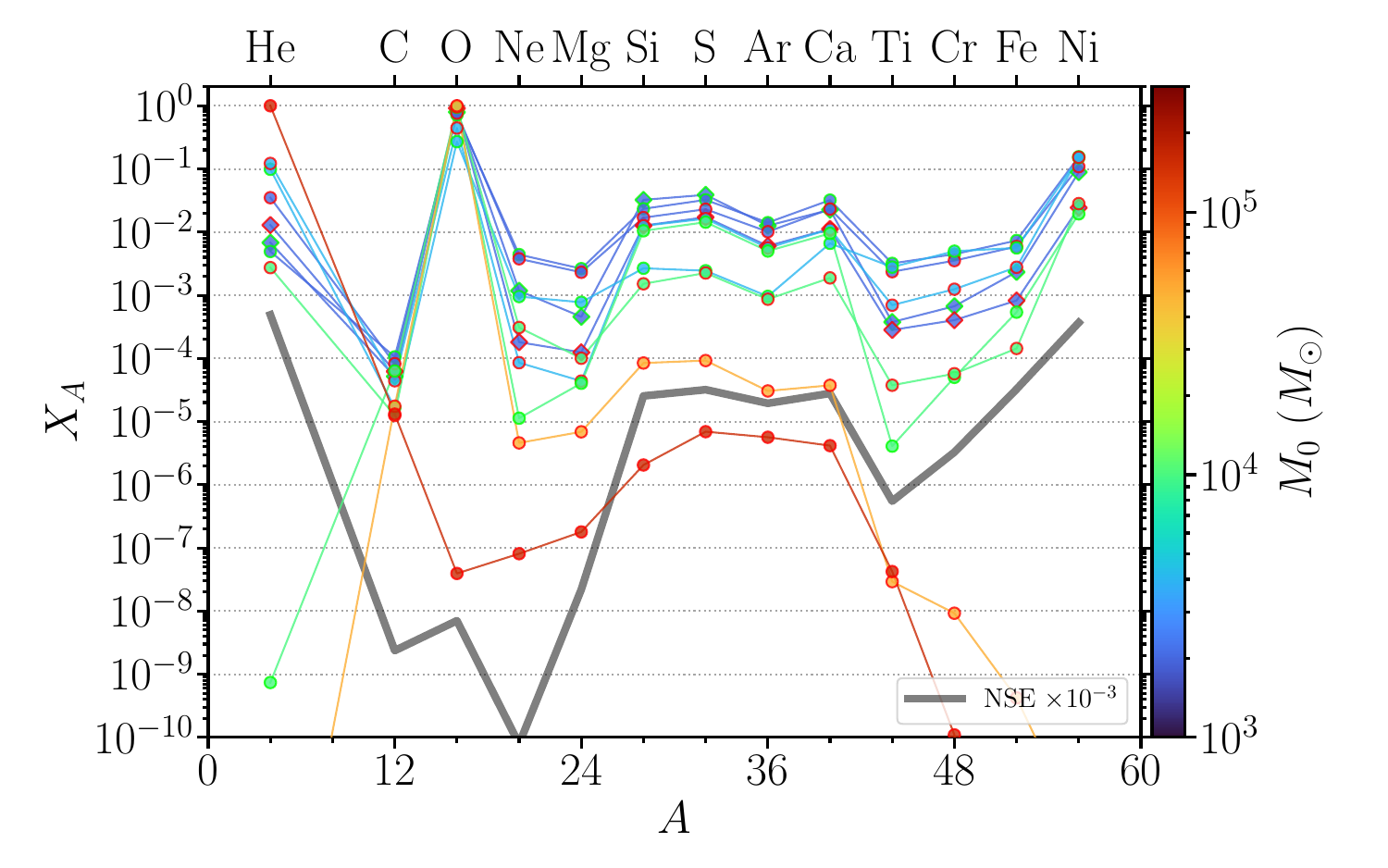}
    \caption{Ejecta properties for the models that produce mass ejection. Top: ejecta mass--kinetic energy relation. Middle: ejecta mass--\isotope{56}{Ni} mass relation. Bottom: ejecta mass fraction for each element as a function of mass number $A$. The left column shows only the disk-bounce ejecta for selected models, while the right column shows the total ejecta, including both the disk-bounce and viscosity-driven disk ejecta. For each model in the right column, the thin lines connect the disk-bounce component to the corresponding total ejecta. The color indicates the initial core mass and the identical symbols are self-consistently used to indicate each model (see the top panels to identify the model). The gray curve in the bottom-right panel shows the NSE abundance distribution calculated at $\rho=\SI{1e6}{\gram\per\centi\meter\cubed}$ and $T=\SI{5}{GK}$, multiplied by $10^{-3}$.}
    \label{fig:ejecta-E-M}
\end{figure*}

We calculate the mass and asymptotic kinetic energy of the ejecta, $M_\mathrm{ej}$ and $K_\mathrm{ej}$, in the same manner as \cite{Fujibayashi2025mar}. Figure~\ref{fig:ejecta-E-M} shows those driven by the bounce of the disk. We find a clear correlation between $M_\mathrm{ej}$ and $K_\mathrm{ej}$. In the high-mass end (i.e., for the collapse of supermassive stars with high $T_\mathrm{rot}/|W|$), the correlation is found to be the one with a constant average velocity of $V_\mathrm{ej}=\sqrt{2K_\mathrm{ej}/M_\mathrm{ej}}\approx 0.2 c$ as found in \cite{Fujibayashi2025mar}. For lower core masses, the average velocity of the ejecta decreases. This stems from the larger mass that remains outside the central region at the disk bounce, which dissipates more kinetic energy and reduces the ejecta velocity.

\subsection{Ejecta composition}\label{subsec:ejecta}
Lower-mass cores tend to have higher central temperatures. As a result, nuclear burning proceeds more efficiently, and a larger amount of newly synthesized material is ejected. The middle-left panel of Fig.~\ref{fig:ejecta-E-M} shows the relation between the ejecta mass and the mass of synthesized \isotope{56}{Ni} in the disk-bounce ejecta. We find that the fraction of \isotope{56}{Ni} in this ejecta is higher for lower-mass cores, and that significant production of \isotope{56}{Ni} occurs only for models with $M_0\lesssim 10^4M_\odot$. 

The bottom-left panel shows the mass fraction as a function of mass number. Except for the He2 and He4 models, in which the initial composition is pure \isotope{4}{He}, the most abundant species in the ejecta is \isotope{16}{O}, reflecting the initial composition. The patterns of the heavier, newly synthesized species ($A\geq20$) are similar among stars initialized with \isotope{16}{O}.

We briefly discuss the result for the 5E4-09 model with a pure \isotope{4}{He} initial composition. As shown in the upper-left panel of Fig.~\ref{fig:ejecta-E-M}, this model produces a disk-bounce ejecta mass and kinetic energy similar to those of the fiducial 5E4-09 model initialized with \isotope{16}{O} (compare the plots of 5E4-09 and 5E4-09-He). The ejecta composition, however, is markedly different, reflecting the different initial composition: most of the ejecta remains in the form of \isotope{4}{He}. In addition, the mass of nuclei with $A\geq 20$ is larger than in the fiducial 5E4-09 model. This difference can be understood as follows. In the \isotope{16}{O}-initial model, a fraction of the oxygen nuclei first has to be photodisintegrated to supply \isotope{4}{He}, which is then used to build heavier nuclei. In contrast, the \isotope{4}{He}-initial model already contains abundant alpha particles. Therefore, once carbon seeds are produced through the triple-alpha reaction, subsequent alpha-capture reactions can proceed more readily.

The He2 and He4 models show a similar qualitative trend. Their disk-bounce ejecta are dominated by \isotope{4}{He}, reflecting their pure-\isotope{4}{He} initial composition. Nevertheless, despite their large initial core masses and relatively low temperatures, these models contain a non-negligible fraction of nuclei with $A\geq 20$ in the ejecta. This is consistent with the idea that the initially abundant alpha particles allow alpha-capture reactions to proceed once carbon seeds are produced through the triple-alpha reaction.

\subsection{Effects of viscosity}
For selected models, we turn on the viscosity after the disk settles into a quasi-steady state and the disk-bounce ejecta mass approximately saturates. The shear viscous hydrodynamics is modeled with a formalism used in \cite{shibata2017apr}. The prescription for the viscous coefficient is the same as the ``SS" prescription presented in \cite{Fujibayashi2025mar}, which is essentially the standard ``alpha"-prescription for the coefficient~\citep{Shakura1973a}. We choose the dimensionless viscous parameter as $\alpha_\mathrm{vis}=0.03$ or 0.1.

The inclusion of viscosity increases both the ejecta mass and the kinetic energy by driving additional mass ejection from the disk, as shown in the upper-right panel of Fig.~\ref{fig:ejecta-E-M}. On the other hand, the average ejecta velocity decreases in the viscous models because the viscosity-driven component is slower than the disk-bounce component.

The enhancement of the ejecta mass and kinetic energy is more significant for lower-mass models, reflecting their larger disk masses. For example, for the models with $\alpha_{\rm vis}=0.10$, the ejecta mass and kinetic energy in the 2E3-09 model increase by factors of approximately 4 and 3, respectively, whereas in the He4 model they increase only by factors of approximately 2.0 and 1.3.

\begin{figure}
    \centering
    \includegraphics[width=0.5\textwidth]{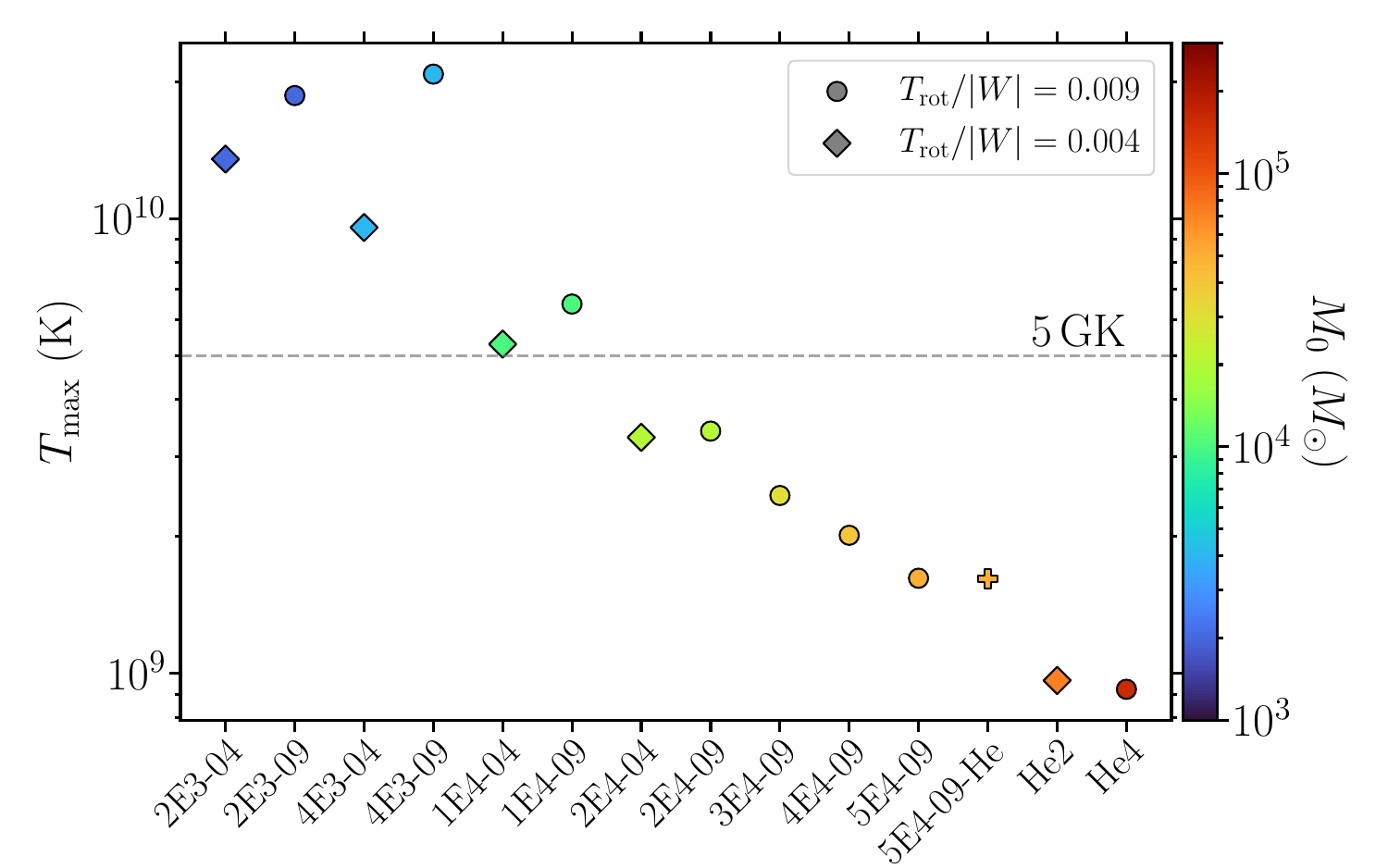}
    \caption{Maximum temperature achieved inside the disk for each model. The horizontal line denotes $T=\SI{5}{GK}$, the temperature above which the composition is expected to approach NSE.}
    \label{fig:Tmax}
\end{figure}

The viscosity-driven ejecta contains a significant fraction of nucleosynthesis products because it is launched from the disk, where the temperature can become sufficiently high for efficient nuclear burning. In particular, when the maximum disk temperature exceeds $\approx \SI{5}{GK}$, the disk matter approaches NSE, and the viscosity-driven ejecta becomes rich in \isotope{56}{Ni}. Figure~\ref{fig:Tmax} shows that this condition is satisfied for models with $M_0\lesssim 10^4M_\odot$. Consistently, the \isotope{56}{Ni} mass increases substantially in the viscous models 1E4-09, 4E3-09, and 2E3-04, as shown in the middle-right panel of Fig.~\ref{fig:ejecta-E-M}. The bottom-right panel of Fig.~\ref{fig:ejecta-E-M} also shows that the mass fractions of nuclei with $A\geq 28$, as well as \isotope{4}{He}, are similar to the NSE abundance pattern. These results indicate that, in lower-mass models, disk matter that has experienced NSE is efficiently ejected by viscosity.

\subsection{Slower-rotation models}
For the models with $M_0=\SI{2e3}{}$, \SI{4e3}{}, \SI{1e4}{}, and $\SI{2e4}{}M_\odot$, we performed additional simulations with reduced rotation, adopting $T_\mathrm{rot}/|W|=0.004$. These models are referred to as 2E3-04, 4E3-04, 1E4-04, and 2E4-04, respectively. As expected, they leave behind more slowly rotating black holes, with black-hole masses that are larger relative to the initial core mass. The disk and ejecta masses are also smaller than those in the corresponding models with the same initial core mass but more rapid rotation.

As the top panel of Fig.~\ref{fig:ejecta-E-M} shows, the disk-bounce mass ejection is suppressed for these slower rotation models. This implies that the ratio of the disk mass to the disk-bounce ejecta mass can be larger for the slower rotation model. As a result, the inclusion of viscosity can substantially change the ejecta properties in these slower-rotation models. In particular, the viscous 2E3-04 model produces a much larger ejecta mass and kinetic energy than the model with only disk-bounce ejecta, as shown in the upper-right panel of Fig.~\ref{fig:ejecta-E-M}. In this model, both the ejecta mass and kinetic energy become about five times larger than in the corresponding model without viscosity. The viscosity-driven ejecta is also rich in \isotope{56}{Ni}, as seen in the middle-right and bottom-right panels. The viscous component becomes important because of the weaker shock generated at disk formation in the slower-rotation model. The disk-bounce ejecta mass is reduced much more strongly than the mass remaining outside the black hole, leaving a comparatively large disk reservoir relative to the prompt ejecta and allowing the subsequent viscosity-driven outflow to dominate.

\subsection{Possible exceptional behavior of the 2E3-09 model} \label{subsec:2e3-09}
The 2E3-09 model shows exceptional behavior in the present simulations. Among all the models, it has the largest disk-bounce ejecta-to-core mass ratio, $M_\mathrm{ej}/M_0$. As shown in the upper-left panel of Fig.~\ref{fig:ejecta-E-M}, this model deviates from the overall trend that lower-mass models produce smaller masses and kinetic energies of disk-bounce ejecta.

The origin of this behavior is likely to be the early formation of a compact disk after black hole formation. For the 2E3-09 model, the collapsing matter has relatively large specific angular momentum, and the resulting black hole has a high dimensionless spin at the onset of massive disk formation, $\chi_{\rm BH}\gtrsim 0.85$, the largest among our models. Because of the high black hole spin, the disk bounce can occur deeper in the gravitational potential, potentially producing a stronger shock. In addition, a large amount of matter remains outside the black hole during disk formation. As a result, the stronger shock can sweep up and unbind a large amount of matter (see bottom row of Fig.~\ref{fig:density}). This is the main difference from both the higher-mass models and the more slowly rotating models. The comparison with the 2E3-04 model further suggests that the exceptional behavior is not a generic consequence of the low initial core mass alone, but may require sufficiently rapid initial rotation. However, this interpretation remains tentative because the initial black hole and disk properties in the 2E3-09 model are sensitive to finite-resolution effects (see \S~\ref{subsec:resolution}). Therefore, the robustness of the exceptionally large disk-bounce ejecta needs to be tested with future higher-resolution simulations.

We note that this exceptional mass ejection is not caused primarily by the microphysical effects included in this study. Simulations performed without nuclear energy generation and neutrino cooling show qualitatively similar mass ejection. These tests suggest that the exceptional mass ejection is primarily hydrodynamic rather than being directly caused by the microphysical effects included in this study. Its magnitude and detailed mechanism, however, remain sensitive to the initial black-hole and disk properties.

With the current numerical code, it is difficult to perform stable simulations of even lower-mass stellar cores, $M_0\lesssim10^3M_\odot$, because such models are expected to form rapidly spinning black holes, with initially small mass at the formation (Lam et al., 2026, in preparation). The simulations for such cases require very high spatial resolution to resolve the formation and evolution of the black hole accurately. A systematic investigation of rapidly rotating lower-mass stellar core collapses is therefore left for future work.

\begin{figure}
    \centering
    \includegraphics[width=0.5\textwidth]{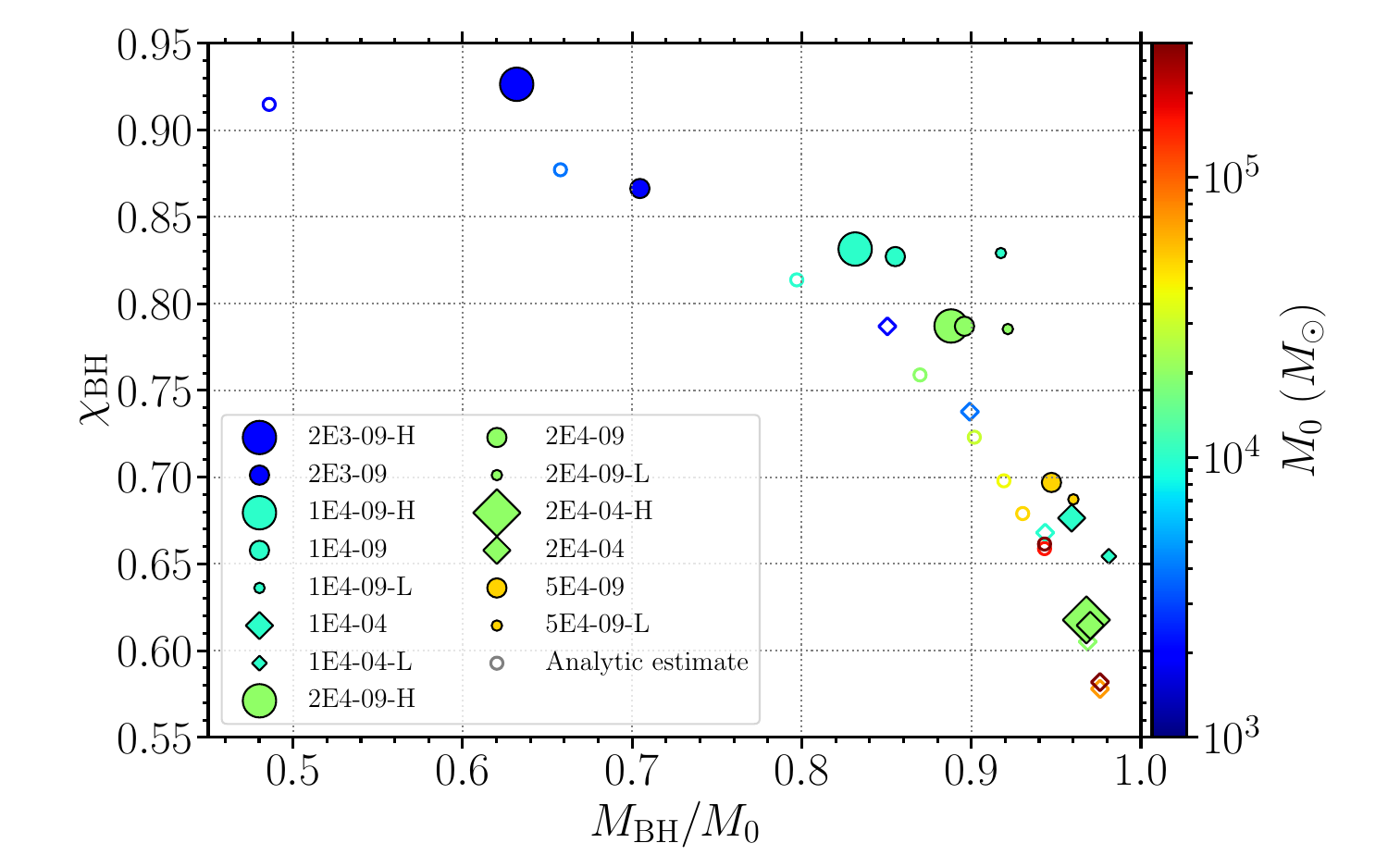}
    \includegraphics[width=0.5\textwidth]{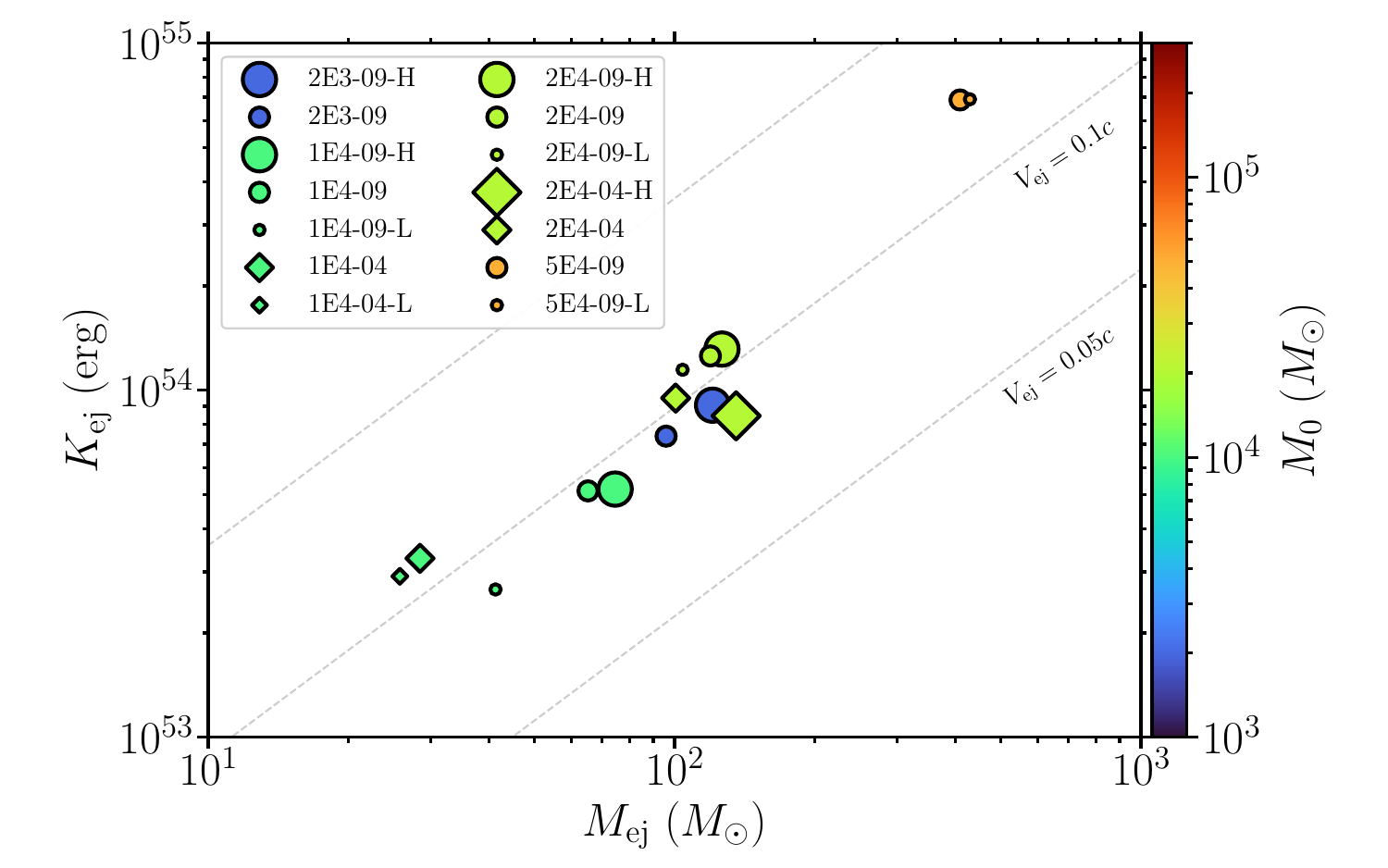}
    \caption{Resolution dependence of the black-hole properties and disk-bounce ejecta. The upper panel shows the relation between black-hole mass and dimensionless spin, and the lower panel shows the ejecta mass–kinetic-energy relation. The larger markers indicate the higher resolution runs. The same progenitor models are denoted by identical symbols; e.g., the results for the 2E4-04 model are denoted by the green diamond. The open markers in the upper panel denote the analytical prediction based on the mass and angular momentum distributions of the progenitor cores. }
    \label{fig:resolution}
\end{figure}

\subsection{Resolution dependence} \label{subsec:resolution}
We here describe the dependence of the results on numerical resolution. This is particularly important for lower-mass models because they form smaller-mass and more rapidly rotating black holes. The black hole mass at the formation is in particular small, e.g., $\sim 10^{-2}M_0$. Resolving such a black hole requires a high resolution and is, therefore, more challenging in these models, and as a result, the black hole mass and spin can be more affected by finite-resolution effects.

The upper panel of Fig.~\ref{fig:resolution} shows the relation between the black hole mass and dimensionless spin for runs with different resolutions. The larger markers denote the results with higher-resolution runs for the same progenitor model. The resolution dependence is most clearly seen in the lower-mass models. With an insufficient resolution, the black hole mass tends to increase spuriously after its formation, while the dimensionless spin decreases. This behavior reflects the difficulty in accurately resolving the black hole spacetime in these models. For this reason, we do not show the late-time black hole mass and spin in the viscous models in Table~\ref{tab:key-result} as they should not be taken as quantitative results.

We also compare the ejecta properties obtained at different resolutions in the lower panel of Fig.~\ref{fig:resolution}. In several models, the higher-resolution runs give somewhat larger ejecta masses and kinetic energies. This is consistent with the smaller black hole mass and higher black hole spin obtained at higher resolution, which leave a more massive disk outside the black hole. However, the differences in the results between runs with different resolutions are modest. For each model, the ejecta mass and kinetic energy obtained in the highest-resolution run differ from those in the second-highest-resolution run by less than $\sim 30\%$. The qualitative trends discussed above remain unchanged: lower-mass, rapidly rotating models tend to produce larger disk masses, viscosity enhances the ejecta mass and kinetic energy, and the viscosity-driven ejecta from models with sufficiently high disk temperatures becomes rich in \isotope{56}{Ni}.

The 2E3-09 model requires particular caution because it shows the strongest resolution dependence in the black hole properties among our models. Within the range of resolutions explored here, the early disk formation and relatively large disk-bounce ejecta persist. Nevertheless, even the highest-resolution run in the present study may not sufficiently resolve the initial black hole to establish convergence. At even higher resolution, the initial black-hole and disk masses could change, potentially altering the formation and propagation of the bounce shock. We therefore regard both the magnitude and the detailed mechanism of the exceptional mass ejection in the 2E3-09 model as tentative.

Overall, the broad trends found in this study are not altered within the range of resolutions explored here. However, accurate long-term evolution of the black hole properties in the lowest-mass, rapidly rotating models requires higher spatial resolution than employed in the present study. Therefore, a systematic survey of collapses with $M_0\lesssim1000M_\odot$ is beyond the scope of the present numerical setup. We leave such an extension to future work using an improved numerical implementation with adaptive mesh refinement based on {\tt SACRA-2D} (\citealt{Lam:2025pmz}, Lam et al., 2026, in preparation).

\begin{table*}[]
    \centering
    \caption{Key results of the simulations. From left to right, the columns show the model name, initial core mass, ejecta mass, asymptotic kinetic energy, average ejecta velocity, ejected \isotope{56}{Ni} mass, and the mass and dimensionless spin of the formed black hole. For non-viscous simulations, the black hole mass and dimensionless spin are measured at $t-t_\mathrm{BH}=1500GM_0/c^3$. For viscous simulations, the black hole mass and dimensionless spin are not listed, because their late-time values are affected by finite-resolution errors in the evolution of the black hole spacetime and are not used in the present analysis. The \isotope{56}{Ni} mass is indicated by a dash, --, for models with $M_\mathrm{Ni}/M_\mathrm{ej}<10^{-6}$.}
    \begin{center}
    \begin{tabular}{lccccccccc}
    \hline
    \hline
    Model & $M_0$ & $M_\mathrm{ej}$ & $K_\mathrm{ej}$ & $V_\mathrm{ej}$ & $M_\mathrm{Ni}$ & $M_\mathrm{BH}$ & $\chi_\mathrm{BH}$\\
        & ($M_\odot$)  & ($M_\odot$) & (erg) & $(c)$ & $(M_\odot)$ & $(M_\odot)$ &\\
    \hline
    Standard resolution runs\\
    2E3-04  & \SI{2.0e+03}{} & \SI{1.5e+01}{} & \SI{1.2e+53}{} & 0.09 & \SI{2.6e-05}{} & \SI{1.8e+03}{} & 0.76\\
    2E3-09  & \SI{2.0e+03}{} & \SI{9.6e+01}{} & \SI{7.4e+53}{} & 0.09 & \SI{1.6e+01}{} & \SI{1.4e+03}{} & 0.87\\
    4E3-04  & \SI{4.0e+03}{} & \SI{1.6e+01}{} & \SI{1.6e+53}{} & 0.10 & \SI{2.9e-03}{} & \SI{3.8e+03}{} & 0.72\\
    4E3-09  & \SI{4.0e+03}{} & \SI{3.0e+01}{} & \SI{1.8e+53}{} & 0.08 & \SI{2.1e-01}{} & \SI{3.2e+03}{} & 0.85\\
    1E4-04  & \SI{1.0e+04}{} & \SI{2.8e+01}{} & \SI{3.3e+53}{} & 0.11 &      -- & \SI{9.6e+03}{} & 0.68\\
    1E4-09  & \SI{1.0e+04}{} & \SI{6.5e+01}{} & \SI{5.1e+53}{} & 0.09 & \SI{1.4e-04}{} & \SI{8.6e+03}{} & 0.83\\
    2E4-04  & \SI{2.0e+04}{} & \SI{1.0e+02}{} & \SI{9.5e+53}{} & 0.10 &      -- & \SI{1.9e+04}{} & 0.61\\
    2E4-09  & \SI{2.0e+04}{} & \SI{1.2e+02}{} & \SI{1.3e+54}{} & 0.11 &      -- & \SI{1.8e+04}{} & 0.79\\
    3E4-09  & \SI{3.0e+04}{} & \SI{1.7e+02}{} & \SI{2.5e+54}{} & 0.12 &      -- & \SI{2.8e+04}{} & 0.75\\
    4E4-09  & \SI{4.0e+04}{} & \SI{2.6e+02}{} & \SI{4.0e+54}{} & 0.13 &      -- & \SI{3.8e+04}{} & 0.72\\
    5E4-09  & \SI{5.0e+04}{} & \SI{4.1e+02}{} & \SI{6.9e+54}{} & 0.14 &      -- & \SI{4.7e+04}{} & 0.70\\
 5E4-09-He  & \SI{5.0e+04}{} & \SI{3.8e+02}{} & \SI{6.5e+54}{} & 0.14 &      -- & \SI{4.7e+04}{} & 0.70\\
       He2$^\mathrm{a}$  & \SI{7.1e+04}{} & \SI{4.2e+02}{} & \SI{9.9e+54}{} & 0.16 &      -- & \SI{6.9e+04}{} & 0.59\\
       He4$^\mathrm{a}$  & \SI{1.6e+05}{} & \SI{1.6e+03}{} & \SI{4.3e+55}{} & 0.17 &      -- & \SI{1.5e+05}{} & 0.68\\
        H2$^\mathrm{b}$  & \SI{3.2e+05}{} & \SI{1.7e+03}{} & \SI{5.1e+55}{} & 0.18 &      -- & \SI{3.2e+05}{} & 0.58\\
        H4$^\mathrm{b}$  & \SI{6.9e+05}{} & \SI{5.8e+03}{} & \SI{1.9e+56}{} & 0.19 &      -- & \SI{6.7e+05}{} & 0.67\\
\hline
 Different resolution runs\\
  2E3-09-H  & \SI{2.0e+03}{} & \SI{1.2e+02}{} & \SI{9.1e+53}{} & 0.09 & \SI{1.7e+01}{} & \SI{1.3e+03}{} &  0.93\\
  1E4-09-H  & \SI{1.0e+04}{} & \SI{7.5e+01}{} & \SI{5.2e+53}{} & 0.09 &      -- & \SI{8.3e+03}{} &  0.83\\
  1E4-09-L  & \SI{1.0e+04}{} & \SI{4.1e+01}{} & \SI{2.7e+53}{} & 0.08 & \SI{5.0e-05}{} & \SI{9.2e+03}{} &  0.83\\
  1E4-04-L  & \SI{1.0e+04}{} & \SI{2.6e+01}{} & \SI{2.9e+53}{} & 0.11 &      -- & \SI{9.8e+03}{} &  0.65\\
  2E4-09-H  & \SI{2.0e+04}{} & \SI{1.3e+02}{} & \SI{1.3e+54}{} & 0.11 &      -- & \SI{1.8e+04}{} &  0.79\\
  2E4-09-L  & \SI{2.0e+04}{} & \SI{1.0e+02}{} & \SI{1.1e+54}{} & 0.11 &      -- & \SI{1.9e+04}{} &  0.79\\
  2E4-04-H  & \SI{2.0e+04}{} & \SI{1.4e+02}{} & \SI{8.4e+53}{} & 0.08 &      -- & \SI{1.9e+04}{} &  0.62\\
  5E4-09-L  & \SI{5.0e+04}{} & \SI{4.3e+02}{} & \SI{6.9e+54}{} & 0.13 &      -- & \SI{4.8e+04}{} &  0.69\\
\hline
 Viscous runs\\
2E3-04-v03  & \SI{2.0e+03}{} & \SI{7.2e+01}{} & \SI{3.4e+53}{} & 0.07 & \SI{6.4e+00}{} &         -- &    --\\
2E3-04-v10  & \SI{2.0e+03}{} & \SI{7.9e+01}{} & \SI{6.1e+53}{} & 0.09 & \SI{1.9e+00}{} &         -- &    --\\
2E3-09-v03  & \SI{2.0e+03}{} & \SI{2.6e+02}{} & \SI{1.2e+54}{} & 0.07 & \SI{4.1e+01}{} &         -- &    --\\
2E3-09-v10  & \SI{2.0e+03}{} & \SI{3.5e+02}{} & \SI{1.9e+54}{} & 0.08 & \SI{3.8e+01}{} &         -- &    --\\
4E3-09-v03  & \SI{4.0e+03}{} & \SI{1.2e+02}{} & \SI{5.2e+53}{} & 0.07 & \SI{1.8e+01}{} &         -- &    --\\
4E3-09-v10  & \SI{4.0e+03}{} & \SI{1.7e+02}{} & \SI{1.5e+54}{} & 0.10 & \SI{2.8e+01}{} &         -- &    --\\
1E4-09-v03  & \SI{1.0e+04}{} & \SI{4.4e+02}{} & \SI{1.8e+54}{} & 0.07 & \SI{8.5e+00}{} &         -- &    --\\
1E4-09-v10  & \SI{1.0e+04}{} & \SI{5.7e+02}{} & \SI{2.1e+54}{} & 0.06 & \SI{1.6e+01}{} &         -- &    --\\
5E4-09-v10  & \SI{5.0e+04}{} & \SI{1.2e+03}{} & \SI{1.3e+55}{} & 0.11 &      -- &         -- &    --\\
   He4-v10  & \SI{1.6e+05}{} & \SI{3.2e+03}{} & \SI{5.5e+55}{} & 0.14 &      -- &         -- &    --\\
\hline
    \end{tabular}\\
    $^\mathrm{a}$ Updated models from \cite{Fujibayashi2025mar}, which includes neutrino cooling and alpha-network.\\
    $^\mathrm{b}$ Models reported in \cite{Fujibayashi2025mar}
    \label{tab:key-result}
    \end{center}
\end{table*}

\section{Discussion} \label{sec:discussion}
\subsection{Electromagnetic signals}
The circumstellar environment of the progenitor stars considered in this work is uncertain. Very massive stars may form by rapid accretion in primordial gas clouds as the latest simulation study indicates~\citep{2026arXiv260627427P}. In the accretion-driven channel, radiative feedback from the growing protostar can suppress the surrounding accretion flow in some cases, while sufficiently rapid accretion may allow the star to continue growing without strong feedback  (e.g., \citealt{Hosokawa2013dec}, \citealt{Sakurai2016jun}, and \citealt{Toyouchi2023jan}). Thus, unlike the case of rapidly accreting supermassive stars, a dense accretion flow around the progenitor is not guaranteed, but it also cannot be excluded. 

Very massive stars may also form through stellar mergers in dense stellar systems \citep{Portegies_Zwart2004_Nature,Sakurai2017dec}. In this channel as well, the gas distribution around the star at the time of collapse is uncertain. Stellar mergers may eject material into the surroundings, but whether such material remains sufficiently dense and close to the star until collapse depends on the time interval between the last merger and the final collapse.

This uncertainty affects the expected electromagnetic signal. If dense gas remains around the progenitor at the time of explosion, interaction between the ejecta and the surrounding material may power a luminous long-duration transient \citep{Jockel2026jan}. If, on the other hand, the surrounding gas has been substantially evacuated, the emission is expected to be dominated by shock-cooling emission from the ejecta itself. Its characteristic luminosity and timescale may be estimated as \citep{Arnett1980apr}
\begin{align}
L&\sim \frac{K_\mathrm{ej}/2}{t_0} \notag\\
&\approx \SI{3e42}{erg.s^{-1}}\, \bigg(\frac{\kappa}{\SI{0.35}{cm^2.g^{-1}}}\bigg)^{-1}\notag \\
&\hspace{0.5cm}\times 
\bigg(\frac{M_\mathrm{ej,tot}}{10^4M_\odot}\bigg)^{-1}
\bigg(\frac{K_\mathrm{ej}}{\SI{e54}{erg}}\bigg)
\bigg(\frac{R_0}{\SI{e15}{cm}}\bigg),\\
t_\mathrm{ch}& \approx \sqrt{2t_0\frac{R_0}{v_\mathrm{sc}}}\notag\\
&\approx \SI{7.2}{yr}\,\bigg(\frac{\kappa}{\SI{0.35}{cm^2.g^{-1}}}\bigg)^{1/2}\notag\\
&\hspace{0.5cm}\times\bigg(\frac{M_\mathrm{ej,tot}}{10^4M_\odot}\bigg)^{3/4}\bigg(\frac{K_\mathrm{ej}}{\SI{e54}{erg}}\bigg)^{-1/4},
\end{align}
where $t_0:=9\kappa M_\mathrm{ej,tot}/(4\pi^3 cR_0)$, $v_\mathrm{sc}=\sqrt{10K_\mathrm{ej}/3M_\mathrm{ej,tot}}$, $R_0$ is the stellar radius at explosion, $\kappa$ is the opacity, which is $\approx 0.35\,\mathrm{cm^2\,g^{-1}}$ for the primordial gas, and the internal energy available after shock breakout is assumed to be approximately one half of the ejecta kinetic energy. For the assumed parameter values, the characteristic diffusion timescale $t_\mathrm{ch}$ is much longer than the decay timescales of the \isotope{56}{Ni}--\isotope{56}{Co} decay chain, so that a large fraction of the radioactive energy is deposited at early times and can be degraded by adiabatic expansion before escaping as radiation (see below for discussion of the energetics). Here, $M_\mathrm{ej,tot}$ denotes the total ejecta mass relevant for photon diffusion, including both the ejecta obtained in the present simulations and, if expelled, the stellar envelope. This quantity can therefore be larger than the ejecta mass listed in Table~\ref{tab:key-result}. For very massive stars that are not strongly affected by wind mass loss, the CO core mass is typically about one-half of the zero-age-main-sequence mass (e.g., \citealt{Heger2002mar}; \citealt{Takahashi2016feb}). Thus, in the above estimate, we adopt a total ejecta mass comparable to the mass of the collapsing core, allowing for a contribution from the stellar envelope. We note that the binding energy of the outer envelope is expected to be minor compared to the kinetic energy of the ejecta obtained in the present simulation.

This estimate illustrates the strong dependence of the shock-cooling luminosity on the progenitor radius. The final radius of a progenitor star depends sensitively on the metallicity and uncertain mixing processes in the star, and it can be either a compact blue supergiant or an extended red supergiant (e.g., \citealt{Kasen2011jun, 2012A&A...542A.113Y, Takahashi2018}). In the above estimation, we assume that the exploded star is inflated to a radius of $R_0=\SI{e15}{cm}$. However, if it exploded as a more compact star, the transient can be much fainter.

If the progenitor retains a hydrogen-rich envelope, as may be expected for metal-poor very massive stars with weak wind mass loss, a recombination front recedes inward through the ejecta (e.g., \citealt{Popov1993sep,Kasen2009oct}). The time at which the recombination starts is estimated as (\citealt{Popov1993sep})
\begin{align}
t_\mathrm{i} \approx \SI{0.44}{yr}\ \bigg(\frac{\kappa}{\SI{0.35}{cm^2.g^{-1}}}\bigg)^{-1/2}\bigg(\frac{R_0}{\SI{e15}{cm}}\bigg)^{1/2}.
\end{align}
Hereafter, we do not show the dependency on the recombination temperature. This indicates that the recombination starts much shorter than the expected duration of the shock-cooling emission. The luminosity and duration of the subsequent ``plateau" phase are estimated as
\begin{align}
L_\mathrm{pl} &\approx \SI{1.5e44}{erg.s^{-1}}\,\bigg(\frac{\kappa}{\SI{0.35}{cm^2.g^{-1}}}\bigg)^{-1/3}\notag\\ &\hspace{0.5cm}\times\bigg(\frac{M_\mathrm{ej,tot}}{10^4M_\odot}\bigg)^{-1/2}
\bigg(\frac{K_\mathrm{ej}}{\SI{e54}{erg}}\bigg)^{5/6}\bigg(\frac{R_0}{\SI{e15}{cm}}\bigg)^{2/3}.\\
t_\mathrm{pl} &\approx \SI{4.8}{yr}\,\bigg(\frac{\kappa}{\SI{0.35}{cm^2.g^{-1}}}\bigg)^{1/6}\notag\\
&\hspace{0.5cm}\times\bigg(\frac{M_\mathrm{ej,tot}}{10^4M_\odot}\bigg)^{1/2}
\bigg(\frac{K_\mathrm{ej}}{\SI{e54}{erg}}\bigg)^{-1/6}\bigg(\frac{R_0}{\SI{e15}{cm}}\bigg)^{1/6}.
\end{align}
Therefore, the transient from the explosion of very massive stars is likely characterized by a long-lasting plateau phase. 

A radiation-hydrodynamics calculation of a supernova from a $\sim 5\times 10^4M_\odot$ primordial star indeed shows a long-lasting plateau phase \citep{Moriya2021may}. However, the plateau phase ends about \SI{2.2}{yr} after the explosion in the rest frame, approximately a factor of two earlier than predicted by the above estimate. Using their model parameters, $M_\mathrm{ej,tot}=5.5\times10^4M_\odot$, $R_0=256R_\odot\approx\SI{1.8e13}{cm}$, and their reported explosion energy of $\SI{9e54}{erg}$ as $K_\mathrm{ej}$, the above expression gives $t_\mathrm{pl}\approx \SI{4}{yr}$. In their model, however, only the outer $\approx 2.5\times10^4M_\odot$ of the ejecta contains hydrogen. Using this envelope mass in the scaling $t_\mathrm{pl}\propto M_\mathrm{ej}^{1/2}$, the estimate is reduced to $\sim \SI{2.7}{yr}$, which is closer to the radiation-hydrodynamic result.

Radioactively powered emission is unlikely to be the dominant component for the present models unless the progenitor radius is much smaller than assumed above or an exceptionally large amount of \isotope{56}{Ni} is synthesized. The radiated energy during the plateau phase is estimated as
\begin{align}
E_\mathrm{pl} &= L_\mathrm{pl}t_\mathrm{pl} \approx \SI{2.3e52}{erg}\,\bigg(\frac{\kappa}{\SI{0.35}{cm^2.g^{-1}}}\bigg)^{-1/6}\notag\\
&\hspace{1.0cm}\bigg(\frac{K_\mathrm{ej}}{\SI{e54}{erg}}\bigg)^{2/3}\bigg(\frac{R_0}{\SI{e15}{cm}}\bigg)^{5/6}.
\end{align}
This energy is comparable to the total radioactive energy released by the \isotope{56}{Ni} decay chain, $E_\mathrm{Ni} \approx \SI{2e50}{erg}\, (M_\mathrm{Ni}/M_\odot)$, only for $M_\mathrm{Ni}$ more than $100M_\odot$. Moreover, this comparison is optimistic for radioactive heating, because a substantial fraction of the decay energy is deposited while the ejecta is still optically thick and can be degraded by adiabatic expansion before escaping. We therefore expect shock-cooling emission, rather than \isotope{56}{Ni}-powered emission, to be the more plausible dominant electromagnetic signal of the explosions studied here. Note that the radioactive heating can keep the temperature of the hydrogen envelope high and extend the plateau phase \citep{Kasen2009oct,Sukhbold2016apr,Goldberg2019jul,Matsumoto2025jan}.

The very massive stars considered in the present study are likely to be formed in the early universe, where low-metallicity environments are available. Cosmological time dilation further lengthens the observed duration, and the emission is redshifted to infrared wavelengths. Therefore, such transients may be promising targets for deep infrared observations with facilities such as \textit{JWST}, \textit{Euclid}, and \textit{Nancy Grace Roman Space Telescope}, although quantitative detectability estimates require detailed modeling of the light curves and the spectra of the transient, together with cosmological redshifting and survey sensitivities.

\subsection{Effects of charged-current weak interactions}\label{subsec:cc}

\begin{figure}
    \centering
    \includegraphics[width=0.5\textwidth]{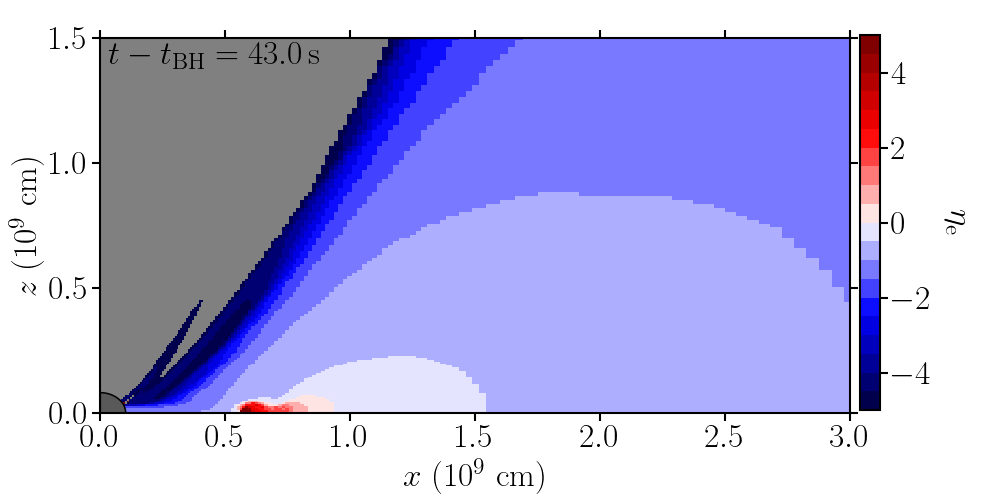}
    \includegraphics[width=0.5\textwidth]{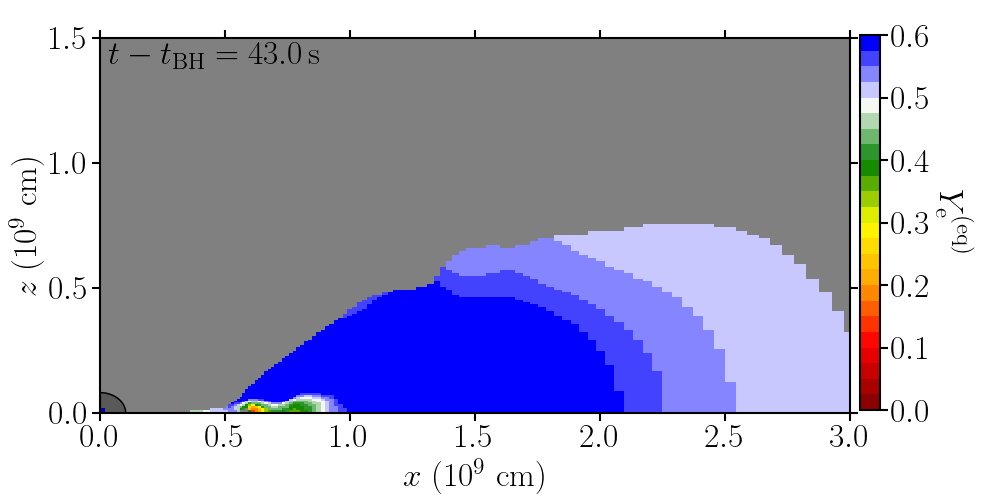} \includegraphics[width=0.5\textwidth]{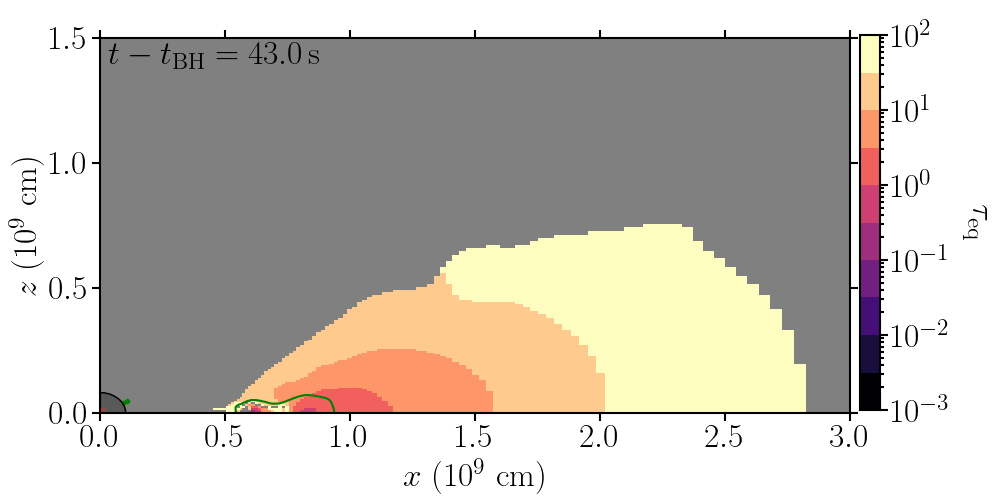}
    \caption{Electron degeneracy parameter (top), the equilibrium electron fraction (middle), and the equilibration timescale (bottom) at 43\,s after the black hole formation for the 2E3-09 model without viscosity. The green contour in the bottom panel encloses the region where $\eta_\mathrm{e}>0$.}
    \label{fig:electron-fraction-2E3-09}
\end{figure}

\begin{figure}
    \centering
    \includegraphics[width=0.5\textwidth]{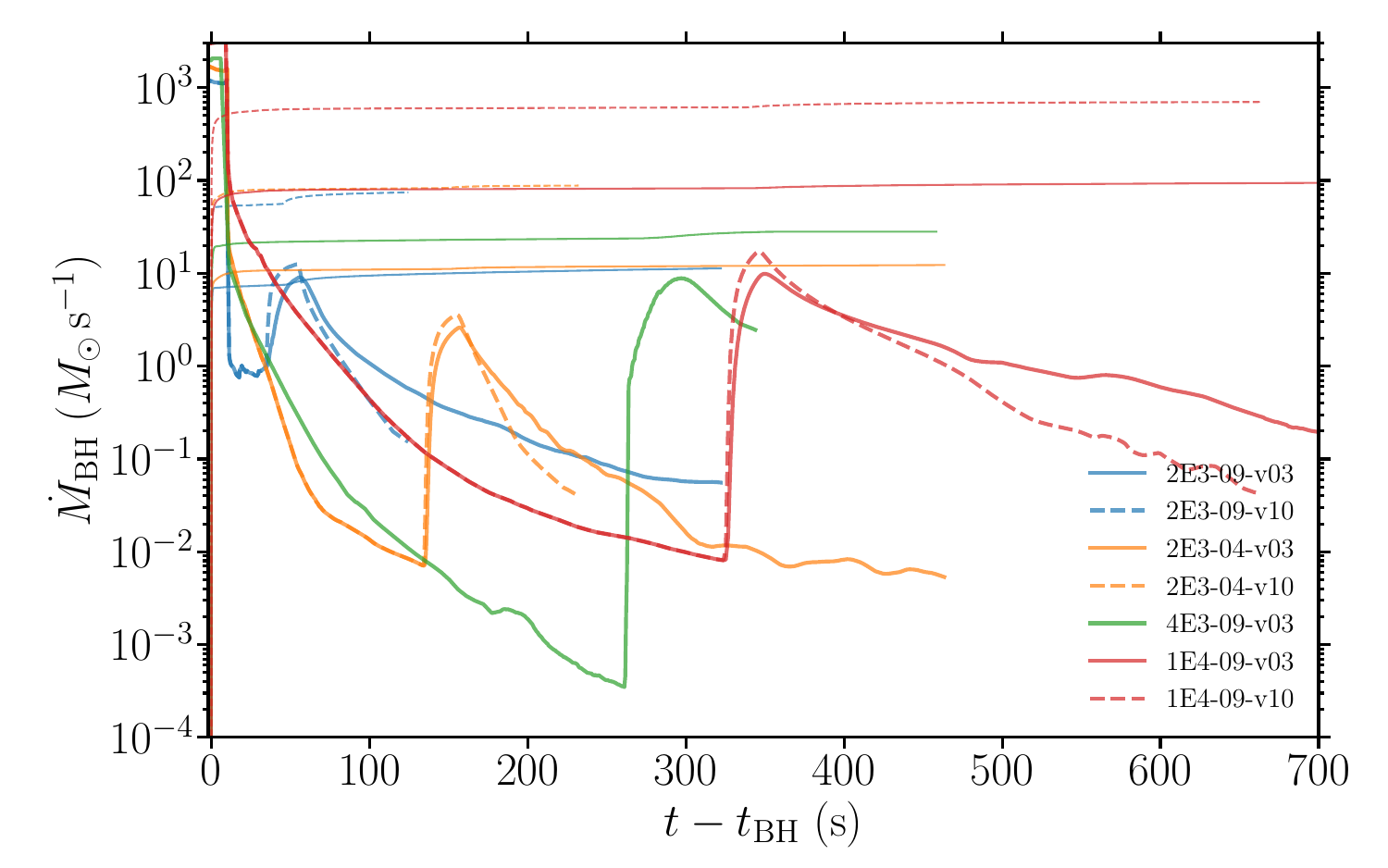} \includegraphics[width=0.5\textwidth]{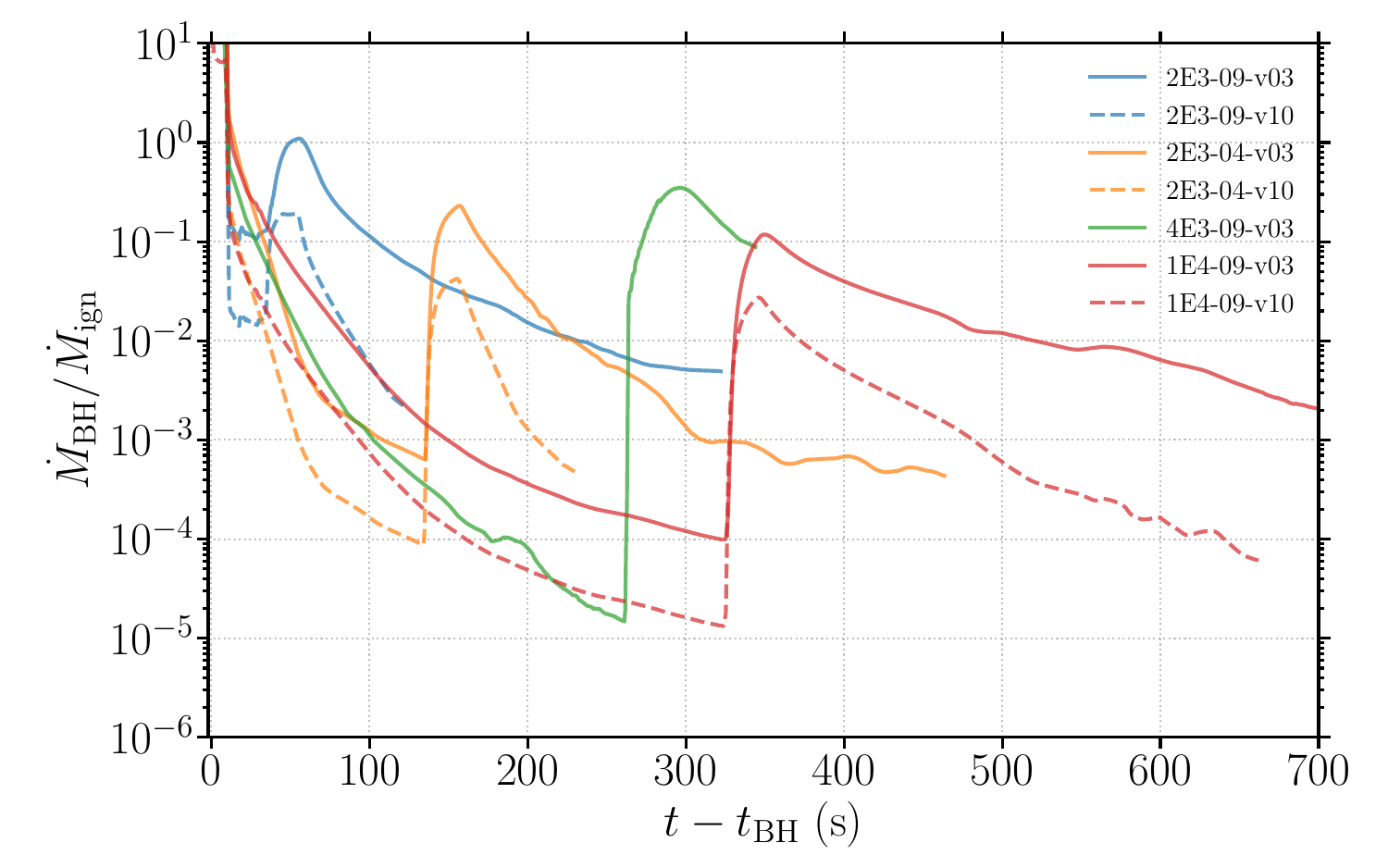}
    \caption{Time evolution of the mass accretion rate onto the black hole (top) and its ratio to the critical accretion rate for efficient neutrino cooling, $\dot{M}_\mathrm{ign}$ (bottom), for the viscous models. The solid and dashed curves denote models with $\alpha_\mathrm{vis}=0.03$ and 0.10, respectively. In the top panel, the thin lines show the corresponding values of $\dot{M}_\mathrm{ign}$. The sudden jumps occur when viscosity is switched on, and the disk begins to evolve viscously.}
    \label{fig:mdot-ign}
\end{figure}

In the present study, only alpha nuclei are included in the nuclear reaction network, and the evolution of the electron fraction, $Y_\mathrm{e}$, is not followed. The composition is therefore restricted to nuclei with $Y_\mathrm{e}=0.5$ because charged-current weak interactions, such as electron and positron captures, are not taken into account. These reactions can drive the matter away from $Y_\mathrm{e}=0.5$, either to neutron-rich or proton-rich sides, depending on the thermodynamic state and the relative importance of electron and positron captures. This effect is particularly important for lower-mass models, in which the temperature and density can become high enough for weak interactions to operate on a sufficiently short timescale. 

As discussed in \S~\ref{subsec:nuc-nu}, electron and positron captures on free nucleons can provide an additional neutrino-cooling channel. For the core mass below $\sim 10^4M_\odot$, the collapse trajectory enters the region in which the charged-current cooling rate is expected to exceed the cooling rate by pair processes. This suggests that weak interactions may further accelerate the runaway collapse of low-mass very massive stars. In addition, the change in $Y_\mathrm{e}$ caused by these reactions can affect the nucleosynthetic outcome \citep{Nagele2021nov}.

If the ejecta or disk matter deviates from $Y_\mathrm{e}=0.5$, the iron-group composition is not necessarily dominated by \isotope{56}{Ni}. Instead, neutron-rich or proton-rich iron-group nuclei may be synthesized. Therefore, the \isotope{56}{Ni} masses obtained in this study should be interpreted with caution, especially for lower-mass models whose disk matter reaches high density and temperature. In this sense, \isotope{56}{Ni} in our alpha-chain calculation should be regarded as representative of iron-group nuclei produced under nearly symmetric conditions, rather than as a robust prediction of the exact isotope composition.

On the other hand, if the disk matter attains a sufficiently low value of $Y_\mathrm{e}$, the disk ejecta may undergo neutron-capture nucleosynthesis. This possibility has been discussed in recent work in the context of very massive stars \citep[e.g.,][]{Siegel2022dec,Agarwal2026feb,Hernandez-Morales2026jan}.

To assess the possible deviation of $Y_\mathrm{e}$ from 0.5, we estimate the equilibrium electron fraction set by electron and positron captures, assuming that the matter is composed of free nucleons. The top panel of Fig.~\ref{fig:electron-fraction-2E3-09} shows the electron degeneracy parameter, $\eta_\mathrm{e}:=(\mu_\mathrm{e}-m_\mathrm{e}c^2)/kT$, at \SI{43}{s} after black hole formation for the 2E3-09 model \textit{without viscosity}. A region with positive electron degeneracy, $\eta_\mathrm{e}>0$, is present near the disk midplane and $x\lesssim \SI{1e9}{cm}\approx 3r_\mathrm{g,0}$. In this region, electron capture is favored, and the equilibrium value of $Y_\mathrm{e}$ can become smaller than 0.5.

The middle panel of Fig.~\ref{fig:electron-fraction-2E3-09} shows the equilibrium electron fraction determined by electron and positron captures. This quantity indicates the value toward which the matter would evolve if weak interactions were sufficiently fast. The disk contains regions where the equilibrium value is either below or above 0.5. In the region where electrons are degenerate, electron capture favors a neutron-rich equilibrium composition. For regions with $\eta_\mathrm{e}\sim 2$, the equilibrium electron fraction is $\approx 0.4$, and the region with $Y_\mathrm{e}^\mathrm{(eq)}\lesssim 0.2$ is present on the inner side. In less degenerate and sufficiently hot regions, on the other hand, electron-positron pairs are abundant, and positron capture on neutrons is kinematically favored by the neutron-proton mass difference, allowing the equilibrium electron fraction to exceed 0.5. This analysis suggests that charged-current weak interactions can produce both neutron-rich and proton-rich components.

The bottom panel of Fig.~\ref{fig:electron-fraction-2E3-09} shows the timescale for $Y_\mathrm{e}$ to approach its equilibrium value described above. In the region with $r\lesssim \SI{2e9}{cm}$, the equilibration timescale is shorter than the local viscous timescale,
\begin{align}
&t_\mathrm{vis} := \frac{r^2}{\nu} = \frac{1}{\alpha_\mathrm{vis} \Omega}\bigg(\frac{r}{H}\bigg)^2 \notag\\
&\approx \SI{22}{s}\,\bigg(\frac{\alpha_\mathrm{vis}}{0.03}\bigg)^{-1} \bigg(\frac{M_\mathrm{BH}}{2000M_\odot}\bigg)^{-1/2} \notag \\
&~~~~~~\times \bigg(\frac{r}{\SI{e9}{cm}}\bigg)^{3/2} \bigg(\frac{H/r}{0.3}\bigg)^{-2},
\end{align}
where we assumed the vertical hydrostatic structure of the disk at each radius. Thus, matter in this region can approach the weak-interaction equilibrium on a dynamical timescale. This suggests that charged-current weak interactions may significantly modify the composition of the disk and its ejecta. A quantitative prediction of the resulting nucleosynthesis, however, requires simulations that evolve $Y_\mathrm{e}$ consistently with charged-current neutrino interactions and a nuclear reaction network including neutron-rich and proton-rich nuclei.

The subsequent viscous evolution, however, weakens the possibility that such low-$Y_\mathrm{e}$ matter is ejected. After viscosity is switched on, the highly electron-degenerate matter near the disk midplane is rapidly accreted by the black hole. The remaining disk is less degenerate, making it difficult for electron captures to drive the matter substantially below $Y_\mathrm{e}=0.5$. To illustrate this point quantitatively, we compare in Fig.~\ref{fig:mdot-ign} the mass accretion rate onto the black hole with the threshold accretion rate above which the disk is expected to enter the neutrino-cooled regime \citep{Chen2007a},
\begin{align}
\dot{M}_\mathrm{ign} \approx 4.5 M_\odot\,\mathrm{s}^{-1} \bigg(\frac{\alpha_\mathrm{vis}}{0.03}\bigg)^{5/3}\bigg(\frac{M_\mathrm{BH}}{10^3M_\odot}\bigg)^{4/3}.
\end{align}
This criterion provides a broad condition for charged-current weak interactions to modify the electron fraction efficiently in the accretion flow. Figure~\ref{fig:mdot-ign} shows that, after the initial transient associated with the onset of viscosity, the accretion rate is generally well below $\dot{M}_\mathrm{ign}$. Thus, the viscously evolving disk studied in the current work is unlikely to become strongly neutrino-cooled or to develop substantially neutron-rich conditions. This suggests that strongly neutron-rich disk ejecta are unlikely in the present viscous models, although a definitive conclusion requires simulations that evolve $Y_\mathrm{e}$ and charged-current neutrino interactions self-consistently. 

For the 2E3-09 model with a lower viscosity $\alpha_\mathrm{vis}=0.03$, the black hole mass accretion rate is marginally close to the critical rate immediately after viscosity is switched on. As found by \citet{Hernandez-Morales2026jan}, disks with $\dot{M}_\mathrm{BH}\sim \dot{M}_\mathrm{ign}$ may instead develop proton-rich accretion flows. This suggests that, in the present viscous models, proton-rich disk ejecta may be more likely than strongly neutron-rich ejecta.

For progenitors with core masses below $10^3M_\odot$, low-$Y_\mathrm{e}$ regions may be more likely to appear in the accretion flow. However, it is not trivial whether matter whose $Y_\mathrm{e}$ is modified deep inside the disk can be ejected. Magnetohydrodynamic effects may provide a mechanism for ejecting such matter from the inner disk. We leave this issue for future work.

\section{Summary} \label{sec:summary}
We investigated the collapse of rotating very massive and supermassive stellar cores using numerical-relativity simulations. Our main survey covers initial core masses of $2 \times 10^3$--$5\times10^4M_\odot$. We also included selected higher-mass supermassive-star models from \cite{Fujibayashi2025mar} for comparison, with some of them recomputed using the present microphysical treatment. In addition to general-relativistic hydrodynamics, we included energy generation by an alpha-chain nuclear reaction network and energy loss by neutrino emission from pair processes. We focused on how the collapse dynamics, black hole formation, disk formation, mass ejection, and ejecta composition depend on the core mass and rotation.

We first found that the collapse profile differs systematically between the higher- and lower-mass models. The higher-mass models, whose collapse is triggered by general-relativistic instability, undergo a more homologous collapse. In contrast, the lower-mass models, which are unstable to the pair instability, show a more runaway-like collapse in the central region. This behavior is enhanced by neutrino cooling, which becomes more efficient because of the higher density and temperature of the lower-mass core collapse. As a result, the black hole formed in the lower-mass models initially contains a smaller fraction of the total core mass.

The lower-mass cores are less compact at the onset of collapse and can therefore have larger dimensionless spin. Consequently, they tend to form more rapidly rotating black holes and more massive disks for a given initial value of $T_\mathrm{rot}/|W|$. The disk forms earlier, in units of the initial gravitational timescale, in the lower-mass models due to a runaway nature of the collapse. The subsequent disk bounce drives an outgoing shock and produces mass ejection. For the high-mass models, the disk-bounce ejecta follow a relation corresponding to an average velocity of $V_\mathrm{ej}\approx0.2c$. Toward lower masses, the average velocity of the disk-bounce ejecta decreases because a larger amount of matter remains outside the black hole at the time of disk formation and dissipates a larger fraction of the kinetic energy of the shocks.

We also examined the nucleosynthetic properties of the ejecta. During collapse, oxygen burning starts at $T\approx3$ GK, and the composition approaches nuclear statistical equilibrium at $T\approx5$ GK. However, most of the newly synthesized nuclei in the collapsing phase are swallowed by the black hole, and the effects of nuclear burning on the infall dynamics are modest for the core-mass range considered in this study. The ejecta composition is nevertheless affected by nuclear burning. In the disk-bounce ejecta, significant \isotope{56}{Ni} production occurs only in the lower-mass models, for which the temperature becomes sufficiently high.

For selected models, we followed the subsequent viscous evolution of the disk after the disk-bounce ejecta saturated. Viscosity drives additional mass ejection from the disk and increases both the ejecta mass and kinetic energy. This effect is more pronounced in lower-mass models because they form more massive disks. For the slower-rotation cases, this component can dominate over the disk-bounce ejecta (e.g., the 2E3-04 model). The viscosity-driven ejecta is launched from high-temperature disk matter. In models with $M_0\lesssim10^4M_\odot$, the disk temperature exceeds $\approx5$ GK, and the disk matter approaches nuclear statistical equilibrium. As a result, the viscosity-driven ejecta can become rich in \isotope{56}{Ni}, substantially increasing the total \isotope{56}{Ni} mass.

We also discussed possible electromagnetic signals from these explosions. Because the circumstellar environment of the progenitor is uncertain, both interaction-powered emission and shock-cooling emission are possible. If the progenitor retains a hydrogen-rich envelope, the latter may appear as a long-lasting recombination-powered plateau. We find that \isotope{56}{Ni}-powered emission is unlikely to be the dominant component for the present models, unless the progenitor radius is much smaller than assumed in our estimate or an exceptionally large amount of \isotope{56}{Ni} is synthesized. Quantitative light-curve and spectral predictions require stellar-evolution models for the progenitor structure, together with detailed emission modeling.

Finally, we discussed the limitations associated with charged-current weak interactions. In the present simulations, the electron fraction is fixed to be 0.5 by the alpha-chain network, and charged-current weak interactions are not included. If these reactions operate efficiently, the disk and ejecta composition can deviate from $Y_\mathrm{e}=0.5$, producing either neutron-rich or proton-rich iron-group nuclei instead of \isotope{56}{Ni}. Thus, the \isotope{56}{Ni} masses obtained in this study should be interpreted as representative of iron-group nuclei synthesized under nearly symmetric conditions. Simulations that consistently evolve $Y_\mathrm{e}$ with charged-current neutrino interactions and a more complete nuclear reaction network are needed for a quantitative prediction of the nucleosynthetic yields.

\acknowledgements
SF thanks Tatsuya Matsumoto for a fruitful discussion. Numerical computation was performed on the clusters, Sakura and Momiji, at the Max Planck Computing and Data Facility. The authors benefited from discussions during the Yukawa Institute for Theoretical Physics (YITP) workshop YITP-T-25-02, ``Multi-Messenger Astrophysics in the Dynamic Universe''. 
This work was in part supported by Grant-in-Aid for Scientific Research of Japanese MEXT/JSPS (23H04900, 26K00732).
A.T.L.L. acknowledges support by NASA under award No. 80NSSC25K7213.

\appendix
\numberwithin{equation}{section}
\renewcommand{\theequation}{\Alph{section}\arabic{equation}}
\renewcommand{\theHequation}{\Alph{section}.\arabic{equation}}

\section{Steady-state treatment of \texorpdfstring{\((\alpha,p)(p,\gamma)\)}{} reaction paths} \label{app:alpha-network}
In addition to the standard $(\alpha,\gamma)$ and $(\gamma,\alpha)$ reactions, we include $(\alpha,p)(p,\gamma)$ and its inverse reactions between the species from \isotope{24}{Mg} to \isotope{56}{Ni} assuming steady-state abundance of the intermediate species (\isotope{27}{Al}, \isotope{31}{P}, ..., \isotope{55}{Co}) in the same manner as \cite{Timmes1999network, Navo2023jul}. Such reaction paths become important at high temperatures.
The key to reducing the size of the network is to only consider alpha nuclei, i.e., \isotope{4}{He}, \isotope{12}{C}, \isotope{16}{O}, ..., \isotope{52}{Fe}, and \isotope{56}{Ni}. They are connected by several reactions. For lighter nuclei, $\isotope{4}{He}(\alpha \alpha, \gamma)\isotope{12}{C}$ (triple-alpha), $\isotope{12}{C}(\isotope{12}{C},\gamma) \isotope{24}{Mg}$, $\isotope{16}{O}(\isotope{12}{C},\gamma) \isotope{28}{Si}$, $\isotope{16}{O}(\isotope{16}{O},\gamma) \isotope{32}{S}$ may be important. The main reaction is $(\alpha,\gamma)$ and its reverse reaction. For heavier nuclei, $(\alpha,p)(p,\gamma)$ may be important. Their rates are included by assuming a steady state of several minor nuclei. For example, for \isotope{24}{Mg}($\alpha,p$)\isotope{27}{Al}($p,\gamma$)\isotope{28}{Si} reaction chain, 
\begin{align}
0&=\frac{d\Yisotope{27}{Al}}{dt} \notag\\
&= \rho \lambda_{\isotope{24}{Mg}(\alpha,p)\isotope{27}{Al}} \Yisotope{24}{Mg} \Yisotope{4}{He} - \rho \lambda_{\isotope{27}{Al}(p,\gamma)\isotope{28}{Si}} \Yisotope{27}{Al} \Yisotope{1}{H} \notag\\
&- \rho \lambda_{\isotope{27}{Al}(p,\alpha)\isotope{24}{Mg}} \Yisotope{27}{Al} \Yisotope{1}{H} + \lambda_{\isotope{28}{Si}(\gamma,p)\isotope{27}{Al}}\Yisotope{28}{Si}
\end{align}
is assumed. Then the product of abundances $\Yisotope{27}{Al}\Yisotope{1}{H}$ is calculated as
\begin{align}
\Yisotope{27}{Al}\Yisotope{1}{H} = \frac{\rho \lambda_{\isotope{24}{Mg}(\alpha,p)\isotope{27}{Al}} \Yisotope{24}{Mg} \Yisotope{4}{He} + \lambda_{\isotope{28}{Si}(\gamma,p)\isotope{27}{Al}}\Yisotope{28}{Si}}{\rho (\lambda_{\isotope{27}{Al}(p,\gamma)\isotope{28}{Si}} + \lambda_{\isotope{27}{Al}(p,\alpha)\isotope{24}{Mg}})}
\end{align}
and thus the production rate of \isotope{28}{Si} through the chain reaction is calculated. \isotope{28}{Si} abundance is evolved as
\begin{align}
\frac{d\Yisotope{28}{Si}}{dt} = & + \rho \lambda_{\isotope{24}{Mg}(\alpha,\gamma)\isotope{28}{Si}} \Yisotope{24}{Mg} \Yisotope{4}{He} - \lambda_{\isotope{28}{Si}(\gamma,\alpha)\isotope{24}{Mg}} \Yisotope{28}{Si} \notag\\
& +\lambda_{\isotope{32}{S}(\gamma,\alpha)\isotope{28}{Si}}\Yisotope{32}{S} - \rho \lambda_{\isotope{28}{Si}(\alpha,\gamma)\isotope{32}{S}}\Yisotope{28}{Si} \Yisotope{4}{He}\notag\\
& + \rho \lambda_{\isotope{27}{Al}(p,\gamma)\isotope{28}{Si}} \Yisotope{27}{Al}\Yisotope{1}{H} - \lambda_{\isotope{28}{Si}(\gamma,p)\isotope{27}{Al}} \Yisotope{28}{Si}\notag\\
& +\rho \lambda_{\isotope{31}{P}(p,\alpha)\isotope{28}{Si}} \Yisotope{31}{P}\Yisotope{1}{H} - \rho \lambda_{\isotope{28}{Si}(\alpha,p)\isotope{31}{P}}\Yisotope{28}{Si} \Yisotope{4}{He}. \label{eq:evolution28Si}
\end{align}
By inserting the $\Yisotope{27}{Al}\Yisotope{1}{H}$ in Eq.~\eqref{eq:evolution28Si}, the differential equation is closed without solving an evolution equation for \isotope{27}{Al} or \isotope{1}{H}. The product $\Yisotope{31}{P}\Yisotope{1}{H}$ is also eliminated in the same manner.

\section{Charged-current weak interaction rates}\label{app:cc}
The specific cooling rate due to electron and positron captures by free nucleons is evaluated by
\begin{align}
q_\mathrm{cc} &= q_\mathrm{ec} + q_\mathrm{pc},\\
q_\mathrm{ec}&= X_\mathrm{p} K_\beta \int^\infty_0 d\omega \,\omega^3 \omega_+^2\sqrt{1-\bigg(\frac{m_\mathrm{e}c^2}{\omega_+}\bigg)^2}F_\mathrm{e^-}(\omega_+),\\
q_\mathrm{pc}&= X_\mathrm{n} K_\beta \int^\infty_{\omega_0} d\omega \,\omega^3 \omega_-^2\sqrt{1-\bigg(\frac{m_\mathrm{e}c^2}{\omega_-}\bigg)^2}F_\mathrm{e^+}(\omega_-),
\end{align}
where $\omega_\pm = \omega\pm\Delta$, $\omega_0 = \Delta + m_\mathrm{e}c^2$, ${K_\beta}^{-1} \approx \SI{1506}{s}\,(m_\mathrm{e}c^2)^5$, $\Delta\approx \SI{1.293}{MeV}$ is the mass difference between neutron and proton, $F_\mathrm{e^\pm}(\omega) = (e^{\omega/kT+\eta_\mathrm{e^\pm}}+1)^{-1}$ is the Fermi-Dirac distribution function for electrons and positrons. The mass fractions of free neutrons and protons, $X_\mathrm{n}$ and $X_\mathrm{p}$, are calculated in each density and temperature by assuming the NSE of neutrons, protons, and \isotope{4}{He} for symmetric matter $Y_\mathrm{e}=0.5$.

\bibliography{reference}

\end{document}